\documentclass[prd,preprint,superscriptaddress,nofootinbib]{revtex4}
\usepackage{hyperref}
\usepackage{amsmath}
\usepackage{graphicx}
\usepackage{epsfig}
\usepackage{latexsym}
\usepackage{float}
\usepackage[caption=false]{subfig}
\usepackage{cancel}
\usepackage{color}

\begin{document}

\title{Exploration of the Tensor Structure of the Higgs Boson Coupling
  to Weak Bosons in $e^+e^-$ Collisions} \date{\today}
  
\preprint{HRI-RECAPP-2014-011,WITS-CTP-135}

\author{Gilad Amar}
\email{gilad.amar@cern.ch}
\affiliation{School of Physics, University of the Witwatersrand, Wits 2050, South Africa}
\affiliation{National Institute for Theoretical Physics; School of Physics,
University of the Witwatersrand, Wits 2050, South Africa}

\author{Shankha Banerjee} \email{shankha@hri.res.in}
\affiliation{Regional Centre for Accelerator-based Particle Physics,
  Harish-Chandra Research Institute, Chhatnag Road, Jhusi, Allahabad -
  211019, India}

\author{Stefan von Buddenbrock}
\email{stefan.erich.von.buddenbrock@cern.ch}
\affiliation{School of Physics, University of the Witwatersrand, Wits 2050, South Africa}

\author{Alan S. Cornell}
\email{alan.cornell@wits.ac.za}
\affiliation{National Institute for Theoretical Physics; School of Physics,
University of the Witwatersrand, Wits 2050, South Africa}

\author{Tanumoy Mandal}
\email{tanumoymandal@hri.res.in}
\affiliation{Regional Centre for Accelerator-based Particle Physics, Harish-Chandra Research Institute, Chhatnag Road, Jhusi, Allahabad - 211019, India}

\author{Bruce Mellado}
\email{bmellado@mail.cern.ch}
\affiliation{School of Physics, University of the Witwatersrand, Wits 2050, South Africa}

\author{Biswarup Mukhopadhyaya}
\email{biswarup@hri.res.in}
\affiliation{Regional Centre for Accelerator-based Particle Physics, Harish-Chandra Research Institute, Chhatnag Road, Jhusi, Allahabad - 211019, India}

\begin{abstract}

Probing signatures of anomalous interactions of the Higgs boson with
pairs of weak vector bosons is an important goal of an $e^+ e^-$
collider commissioned as a Higgs factory. We perform a detailed
analysis of such potential of a collider operating at $250 - 300$
GeV. Mostly using higher dimensional operators in a gauge-invariant
framework, we show that substantial information on anomalous couplings 
can be extracted from the total rates of $s$-and $t$-channel Higgs 
production. The most obvious kinematic distributions, based on angular 
dependence of matrix elements, are relatively less sensitive with 
moderate coefficients of anomalous couplings, unless one goes to higher 
centre-of-mass energies. Some important quantities to use here,
apart from the total event rates, are the ratios of event rates at
different energies, ratios of $s$-and $t$-channel rates at fixed
energies, and under some fortunate circumstances, the correlated
changes in the rates for $W$-boson pair-production. A general scheme
of calculating rates with as many as four gauge-invariant operators is
also outlined.  At the end, we perform a likelihood analysis using
phenomenological parametrization of anomalous $HWW$ interaction, and
indicate their distinguishability for illustrative values of the
strength of such interactions.

\end{abstract}

\maketitle

\section{\label{sec:intro}Introduction}

Physicists are widely convinced now that they have discovered what
closely resembles the Higgs boson~\cite{ATLAS,CMS} postulated in the standard
electroweak model (SM)~\cite{GG2a,GG2b,GG2c,Higgs-originala,Higgs-originalb,
Higgs-originalc,Higgs-originald,Higgs-originale,Higgs-originalf}. Along with 
widespread exhilaration, such a
development brings in questions on whether this particle carries some
signature of physics beyond the standard model. Many studies in this
direction have appeared~\cite{before-discoverya,before-discoveryb,
before-discoveryc,before-discoveryd,before-discoverye,before-discoveryf,
before-discoveryg,before-discoveryh,before-discoveryi,before-discoveryj,
Desai-2011yj,before-discoveryk,before-discoveryl,higgs-phenoa,higgs-phenob,
higgs-phenoc,higgs-phenod,higgs-phenoe,higgs-phenof,higgs-phenog,higgs-phenoh,
higgs-phenoi,higgs-phenoj,higgs-phenok,higgs-phenol,higgs-phenom,higgs-phenon,
higgs-phenoo,higgs-phenop,higgs-phenoq,higgs-phenor,higgs-phenos,higgs-phenot,
higgs-phenou,higgs-phenov,higgs-phenow,higgs-phenox,higgs-phenoy,higgs-phenoz,
higgs-phenoaa,higgs-phenoab,higgs-phenoac,higgs-phenoad,higgs-phenoae,higgs-phenoaf} 
in the context of the Large Hadron Collider
(LHC) where the data available so far still allow some departure
from SM behaviour. Even a finite invisible branching
ratio ($BR$) for the Higgs cannot, at the moment, be ruled out~\cite{ATLAS-inv,CMS-inv}.
The issue can be probed through careful measurements of
the couplings of the Higgs (or Higgs-like scalar) to various pairs of
SM particles.  Among them, the couplings to pairs of vector bosons
($HVV$) are measured in a relatively more reliable manner. This possibility 
has been explained in the context of an $ep$ collider too~\cite{Han:2009pe,Biswal:2012mp}.

In view of the cumulative demand for a closer probe on the $HVV$ couplings (and of
course the couplings to other SM particles), the most desirable endeavour, however, is 
to build an electron-positron collider which provides a clean environment for
precise measurements of Higgs interaction strengths. The first step is
of course to develop a Higgs factory (at $\sqrt{s} \approx$ 250 - 300
GeV). Such a machine will not only produce the Higgs boson copiously
near resonance, but is also the first step before an $e^+ e^-$ machine
at even higher energies is developed. In this paper, we incorporate
some observations regarding the signatures of anomalous $HVV$ couplings,
manifest through higher dimensional operators (HDOs), at a Higgs factory. Other 
studies performed for an $e^+e^-$ machine can be found in~\cite{Biswal:2005fh}.

If the couplings arise through physics at a scale higher than that of
electroweak symmetry breaking, then the resulting higher-dimensional
effective interactions are expected to be gauge invariant. Such
interactions have not only been identified, but constraints on their
coefficients have also been obtained from the LHC data~\cite{higgs-phenoab,
HD-opsc,HD-opsa,HD-opsb,HD-opsd,HD-ops_eff}. In view of such
analyses, the coefficients are often restricted to such values where
many cherished kinematic distributions may fail to reveal their footprints.
In the current study, we point out some features which influence
the detectability (or otherwise) of the higher-dimensional couplings
at a Higgs factory. At the same time, we emphasise some possible
measurements that can elicit their signatures even for relatively
small coefficients of such operators.

We concentrate on two Higgs production channels, namely, $e^+ e^-
\longrightarrow ZH$ (the $s$-channel process) and $e^+ e^-
\longrightarrow \nu{\bar{\nu}}H$ (the $t$-channel process, which we
separate with the help of a simple kinematic cut around the Higgs
boson energy).  In principle, the HDOs that
will constitute our report can influence the rates in both
channels. In contrast, the most obvious kinematic distributions,
namely, those based on the angular dependence of matrix elements,
drawn with moderate values of their coefficients do not show a
perceptible difference with respect to the SM situation.  Keeping this
in view, we underscore the following points here:

\begin{enumerate}
\item The $s$-channel process has substantial rates at $\le$ 300 GeV
or thereabout. We show, through an analysis of the production amplitude
squared, why one cannot expect significantly different angular
distributions in this channel at such energies, if one uses moderate
values of the operator coefficients.

\item The $t$-channel process can have appreciable production rates at
  high energies ($\approx$ a TeV), too. Because of the production of two neutrinos in the final state, this process provides limited phase-space for the exploration of the tensor structure of the $HWW$ coupling. Here it is attempted to exploit the full kinematics of the Higgs boson by means of a correlated two-dimensional likelihood analysis.


\item We show that, given such impediment, it is possible to
uncover signatures of the aforementioned BSM operators through measurements
of rates at two different energies, which also cancels many
systematic uncertainties. In general, the energy dependence of the rates can be
sensitive to anomalous couplings.

\item The very fact that the additional operators should be electroweak
gauge invariant imply not only higher-dimensional $HVV$ interactions 
($V = W\,, Z\,, \gamma$) but also anomalous $WWV$ interactions ($V = Z, \gamma$) 
whose strengths are related to the former. We show that the concomitant variations
in Higgs production and W-pair production at Higgs factories may  
elicit the presence of such BSM interactions.

\item We also show that if the centre-of-mass energy (CME) of the colliding particles is 
$\approx 500$ GeV or more, then even moderate values of the operator coefficients 
can show some differences in the kinematic distributions. 

\item Lastly, we perform the analysis in a framework that allows one to
retain all the gauge-invariant operators at the same time.
\end{enumerate}

We summarise the gauge invariant couplings in the next section, and
subsequently point out the `phenomenological' anomalous couplings they
lead to. In section~\ref{sec:pheno}, we take up the $s$ and $t$-channel Higgs
production cross-sections in turn, and explain why one cannot expect too much
out of kinematic distributions at Higgs factory energies, so long as the
BSM coupling coefficients are subject to constraints imposed by the LHC data.
Their detectable signatures through event ratios at two energies, and
also via the simultaneous measurement of $W$-pair production are predicted
in section~\ref{sec:pheno}. A likelihood analysis and some related issues, mostly in terms
of the phenomenological forms to which all new couplings reduce, are
found in section~\ref{sec:likelihood}. We summarise our conclusions in section~\ref{sec:disc}.

\section{Effective Lagrangian Formalism}
\label{sec:ELF}

In this paper, we adopt two types of effective Lagrangian parametrizations which are commonly used
in the literature to probe the anomalous $HVV$ (where $V=W,Z,\gamma$) interactions. In one parametrization,
we take the most general set of dimension-6 gauge invariant operators which give rise to such
anomalous $HVV$ interactions. In the other one, we parametrize the $HVV$ vertices
with the most general Lorentz invariant structure.
Although, this formalism is not the most transparent one
from the viewpoint of the gauge structure of the theory, it is rather simple and more experiment-friendly.
Both formalisms modify the $HVV$ vertices
by introducing non-standard momentum-dependent terms.

We assume that the SM is a low-energy effective theory of a more complete perturbation theory valid below a 
cut-off scale $\Lambda$. In the present study, we are concerned mainly with the Higgs sector. The first order 
corrections to the Higgs sector will come from gauge invariant dimension 6 operators as there is only one 
dimension-5 operator which contributes to the neutrino masses. The relevant 
additional Lorentz structures in $HVV$ interactions are necessarily of dimensions higher than four. If they 
arise as a consequence of integrating out physics at a higher scale, all such operators will have to be 
invariant under $SU(2)_L\times U(1)_Y$. A general classification of such operators is found in the
literature~\cite{Buchmueller,min-basis,Hagiwara,Garcia}. The lowest order CP-conserving operators which are 
relevant for Higgs phenomenology are 

\begin{itemize}

\item
The operators containing the Higgs doublet $\Phi$ and its derivatives:
\begin{equation}
\mathcal{O}_{\Phi,1} = (D_{\mu}\Phi)^{\dagger}\Phi\Phi^{\dagger}(D^{\mu}\Phi);~~~
\mathcal{O}_{\Phi,2} = \frac{1}{2}\partial_{\mu}(\Phi^{\dagger}\Phi)\partial^{\mu}(\Phi^{\dagger}\Phi);~~~
\mathcal{O}_{\Phi,3} = \frac{1}{3}(\Phi^{\dagger}\Phi)^{3}
\end{equation}
\item
The operators containing the Higgs doublet $\Phi$ (or its derivatives) and bosonic field strengths :
\begin{equation}
\mathcal{O}_{GG} = \Phi^{\dagger}\Phi G_{\mu\nu}^a G^{a\,\mu\nu};~~~
\mathcal{O}_{BW} = \Phi^{\dagger}\hat{B}_{\mu \nu} \hat{W}^{\mu \nu} \Phi;~~~
\mathcal{O}_{WW} = \Phi^{\dagger}\hat{W}_{\mu \nu} \hat{W}^{\mu \nu} \Phi \nonumber
\end{equation}
\begin{equation}
\mathcal{O}_{W}  = (D_{\mu}\Phi)^{\dagger} \hat{W}^{\mu \nu} (D_\nu \Phi);~~~
\mathcal{O}_{BB} = \Phi^{\dagger}\hat{B}_{\mu \nu} \hat{B}^{\mu \nu} \Phi;~~~
\mathcal{O}_{B}  = (D_{\mu}\Phi)^{\dagger} \hat{B}^{\mu \nu} (D_\nu \Phi),
\end{equation}
\end{itemize}
where $\hat{W}^{\mu \nu}=i\,\frac{g}{2} \sigma_{a}W^{a \; \mu \nu}$ and $\hat{B}^{\mu \nu}=i\,\frac{g}{2}' B^{\mu \nu}$ and $g$, $g'$ are respectively the $SU(2)_L$ and $U(1)_Y$ gauge couplings. $W^a_{\mu \nu} = \partial_{\mu}W^a_{\nu}-\partial_{\nu}W^a_{\mu} - g \epsilon^{abc}W^b_{\mu} W^c_{\nu}$, $B_{\mu \nu} = \partial_{\mu}B_{\nu}-\partial_{\nu}B_{\mu}$ and $G^a_{\mu \nu} = \partial_{\mu}G^a_{\nu}-\partial_{\nu}G^a_{\mu} - g_s f^{abc}G^b_{\mu} G^c_{\nu}$. The Higgs doublet is denoted by $\Phi$ and its covariant derivative is given as
$D_{\mu}\Phi=(\partial_{\mu}+\frac{i}{2}g' B_{\mu} + i g \frac{\sigma_a}{2}W^a_{\mu})\Phi$.

Following are the properties of the aforementioned HDOs:

\begin{itemize}

\item $\mathcal{O}_{\Phi,1}$: Does not preserve custodial symmetry and is therefore severely constrained by 
the $T$-parameter (or equivalently the $\rho$ parameter). It modifies the SM $HZZ$ and $HWW$
couplings by unequal multiplicative factors.

\item $\mathcal{O}_{\Phi,2}$: Preserves custodial symmetry and modifies the SM $HZZ$ and $HWW$
couplings by multiplicative factors. This operator modifies the Higgs self-interaction as well.

\item $\mathcal{O}_{\Phi,3}$: Modifies only the Higgs self-interaction.

\item $\mathcal{O}_{GG}$: Introduces $HGG$ coupling which is same in structure as the SM effective $HGG$ coupling.
Since our discussion is limited to the context of an $e^+e^-$ collider and as we will also not consider the gluonic 
decay mode of the Higgs, we will not discuss this operator any further.

\item $\mathcal{O}_{BW}$: Drives the tree-level $Z\leftrightarrow\gamma$ mixing and
is therefore highly constrained by the electroweak precision test (EWPT) data~\cite{HD-opsc}.

\item $\mathcal{O}_{WW}$,
$\mathcal{O}_{W}$, $\mathcal{O}_{BB}$, $\mathcal{O}_{B}$: Modifies the $HVV$ couplings by
introducing new Lorentz structure in the Lagrangian. They are not severely constrained by the EWPT data\cite{HD-opsa,HD-opsb}.
\end{itemize}
Hence for the Higgs sector, we will choose our basis as $\mathcal{O}_i \in \{\mathcal{O}_{WW},\mathcal{O}_{W},\mathcal{O}_{BB},\mathcal{O}_B\}$.
In the presence of the above operators, the Lagrangian is parametrised as 
\begin{equation} 
\mathcal{L} = \kappa\left(\frac{2 m_W^2}{v} H W_{\mu}^+ W^{\mu -}+\frac{ m_Z^2}{v} H Z_{\mu} Z^{\mu } \right) + \sum_{i} \frac{f_{i}}{\Lambda^2}\mathcal{O}_{i}
\label{Lag}
\end{equation}

where $\kappa$ is the scale factor of the SM-like coupling, something
which needs to be accounted for when considering BSM physics. $f_i$
is a dimensionless coefficient which denotes the strength of the $i^{th}$
operator and $\Lambda$ is the cut-off scale above which new physics
must appear. We keep $\kappa$ to be the same for the $HWW$ and $HZZ$
couplings so that there is no unacceptable contribution to the
$\rho$-parameter. Another operator considered in this work is $\mathcal{O}_{WWW} =
Tr[\hat{W}_{\mu \nu} \hat{W}^{\nu \rho} \hat{W}^{\mu}_{\rho}]$.  This
only affects the triple gauge boson couplings and does not affect the
Higgs sector. \\
The effective Lagrangian which affects the Higgs sector is 
\begin{align}
\label{eq:lagHVV}
\mathcal{L}_{eff} &= 
g_{HWW}^{(1)}~(W_{\mu\nu}^{+}W^{-\mu}\partial^{\nu}H + h.c.) +
g_{HWW}^{(2)}~HW_{\mu\nu}^{+}W^{-\mu\nu} \nonumber \\
&+ g_{HZZ}^{(1)}~Z_{\mu\nu}Z^{\mu}\partial^{\nu}H +
g_{HZZ}^{(2)}~HZ_{\mu\nu}Z^{\mu\nu} \nonumber \\
&+ g_{HZ\gamma}^{(1)}~A_{\mu\nu}Z^{\mu}\partial^{\nu}H +
g_{HZ\gamma}^{(2)}~HA_{\mu\nu}Z^{\mu\nu}+g_{H\gamma\gamma}H A_{\mu \nu} A^{\mu \nu},
\end{align}
where
\begin{align}
\label{eq:lagHVVcoeff}
g^{(1)}_{HWW}&=\left(\frac{g M_W}{\Lambda^2}\right) \frac{f_W}{2};~~~ g^{(2)}_{HWW}=-\left(\frac{g M_W}{\Lambda^2}\right)f_{WW} \nonumber \\
g^{(1)}_{HZZ}&=\left(\frac{g M_W}{\Lambda^2}\right) \frac{c^2 f_W + s^2 f_B}{2 c^2};~~~g^{(2)}_{HZZ}=-\left(\frac{g M_W}{\Lambda^2}\right) \frac{s^4 f_{BB} + c^4 f_{WW}}{2 c^2} \nonumber \\
g^{(1)}_{HZ\gamma}&=\left(\frac{g M_W}{\Lambda^2}\right)\frac{s(f_W-f_B)}{2 c};~~~g^{(2)}_{HZ\gamma}=\left(\frac{g M_W}{\Lambda^2}\right)\frac{s(s^2 f_{BB}-c^2 f_{WW})}{c} \nonumber \\
g_{H\gamma\gamma}&=-\left(\frac{g M_W}{\Lambda^2}\right)\frac{s^2(f_{BB}+f_{WW})}{2}
\end{align} with $s\,(c)$ being the sine (cosine) of the Weinberg angle.
The operators $\mathcal{O}_W$, $\mathcal{O}_B$ and $\mathcal{O}_{WWW}$ contribute to the anomalous triple gauge boson interactions. 
The interactions can be summarised as 
\begin{align}
\label{eq:lagWWV}
\mathcal{L}_{WWV}=-i g_{WWV}\left\{g_1^V\left(W_{\mu\nu}^+W^{-\mu}V^{\nu}-W_{\mu}^+V_{\nu}W^{-\mu \nu}\right)+\kappa_V W_{\mu}^+W_{\nu}^-V^{\mu \nu} + \frac{\lambda_V}{M_W^2}W_{\mu \nu}^+ W^{-\nu \rho} V_{\rho}^{\mu}\right\},
\end{align}
where $g_{WW\gamma}=g \, s$, $g_{WWZ} = g \, c$, $\kappa_V=1+\Delta\kappa_V$ and $g_1^Z=1+\Delta g_1^Z$ with 
\begin{align}
\label{eq:lagWWVcoeff}
\Delta \kappa_{\gamma}&=\frac{M_W^2}{2 \Lambda^2}\left(f_W+f_B\right);~~~\lambda_{\gamma}=\lambda_Z=\frac{3g^2M_W^2}{2\Lambda^2} f_{WWW} \nonumber \\
\Delta g_1^Z&=\frac{M_W^2}{2 c^2 \Lambda^2} f_W;~~~\Delta \kappa_Z=\frac{M_W^2}{2 c^2 \Lambda^2}\left(c^2 f_W - s^2 f_B\right)
\end{align}

The limits on these operators have been derived in many references.
The most comprehensive of these are listed in references~\cite{higgs-phenoab,HD-opsc,HD-opsa,HD-opsb,HD-opsd}. 
These operators, even within their current limits, 
have been shown to modify the efficiencies of the various selection cuts
for the relevant final states in the context of the LHC~\cite{HD-ops_eff}. 

All of the aforementioned HDOs lead essentially to one effective coupling (each for $HWW$ and $HZZ$), when $CP$-violation is neglected. 
These can be alternatively used in a phenomenological way for example, the $H(k)W_\mu^+(p)W_\nu^-(q)$ vertex can be parametrised as
\cite{probingSpinParity}:
\begin{equation}
i\Gamma^{\mu\nu}(p,q) \epsilon_\mu(p)\epsilon^*_\nu(q),
\end{equation}
where deviations from the SM form of
$\Gamma^{\mu\nu}_{SM}(p,q)=-gM_Wg^{\mu\nu} $ would indicate the
presence of BSM physics. These BSM deviations, including $CP$-violating ones (not considered among the gauge invariant operators), can be specified as 
\begin{equation}
\label{eq:LIP}
\Gamma^{BSM}_{\mu\nu}(p,q)=\frac{g}{M_W}[\lambda(p.q g_{\mu\nu} - p_\nu q_\mu) + 
\lambda^\prime \epsilon_{\mu\nu\rho\sigma}p^\rho q^\sigma],
\end{equation}
where $\lambda $ and $\lambda^\prime $
are the effective strengths for the anomalous CP-conserving and
CP-violating operators respectively. 

Precise identification of the non-vanishing nature of $\lambda,
\lambda^\prime$ is a challenging task. If ever accomplished, it can
tell us whether the modification in $HVV$-couplings are $CP$-conserving
or $CP$-violating in nature and, if both are present, what their
relative proportion is. Here we analyse the process $e^+ e^- \to
H \nu \bar{\nu}$ and see if there is any BSM physics
involved by incorporating a likelihood analysis of the SM hypothesis
tested against BSM hypotheses. 

A few comments are in order on the two ways of parametrizing the anomalous Higgs couplings. The latter, of course, encapsulates all possible modified Lorentz invariant couplings in the
lowest possible order, including both $CP$-conserving and $CP$-violating
ones, in the coefficients $\lambda$ and $\lambda^\prime$
respectively. All of the anomalous $HWW$ and $HZZ$ couplings listed in
the gauge-invariant formulation reduce basically to one term if one confines
oneself to a $CP$-conserving scenario. Thus we can say that the latter parametrization shows us a
rather `economic' way of relating the anomalous $HVV$ interactions to
collider phenomenology. On the other hand, the process of relating the anomalous couplings to specific effective 
interactions is more transparent from the viewpoint of gauge
structures when one uses the gauge invariant HDOs. It paves an easier path towards understanding the ultraviolet 
completion of the scenario. In addition to this, the formulation in terms of gauge-invariant operators relates the 
anomalous $HWW$ and $HZZ$
interactions.  One finds, in this way, a pattern in the departure of
the $ZH$ and $\nu\bar{\nu}H$ final state production rates from the
corresponding SM prediction. Finally, some of the
gauge-invariant operators lead simultaneously to anomalous triple
gauge boson interactions.  There is thus an associated variation in
the $ZH$, $\nu\bar{\nu}H$ and $W^+W^-$ production rates as well as in the
kinematic distributions associated with each final state. Such an
association enables one to use various pieces of data to determine
each new operator.

\section{Phenomenology at an $e^+e^-$ Collider}
\label{sec:pheno}

In this section, we discuss various important Higgs production mechanisms through $HVV$ vertices at
an $e^+e^-$ collider. For the collider phenomenology, we have implemented the Lagrangians of 
Eqs.~(\ref{eq:lagHVV})~and~(\ref{eq:lagWWV}) in \texttt{FeynRules}~\cite{Alloul:2013bka} to generate \texttt{Universal FeynRules
Model (UFO)}~\cite{Degrande:2011ua} files suitable for interfacing with \texttt{MadGraph}~\cite{Alwall:2011uj}. We also use 
\texttt{FORM}~\cite{Vermaseren:2000nd} to compute many cross-sections analytically.   

\subsection{Higgs production at an $e^+e^-$ collider}

\begin{figure}
\centering
\subfloat{
\begin{tabular}{ccc}
\resizebox{55mm}{!}{\includegraphics{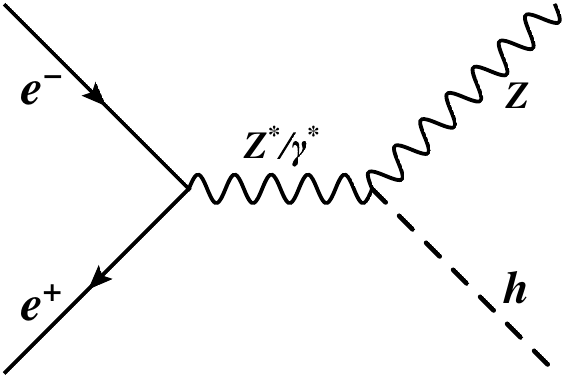}} &&~~~
\resizebox{55mm}{!}{\includegraphics{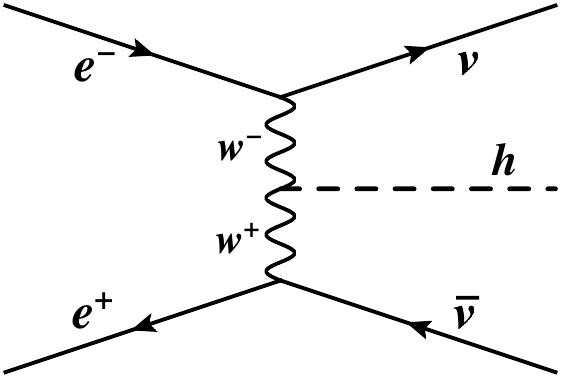}} \\
\hspace{0mm}(a)&&\hspace{8mm}(b)
\end{tabular}}
\caption{(a) $s$-channel Feynman diagrams (b) $t$-channel Feynman Diagram.}
\label{fig:FD}
\end{figure}

We concentrate on two main Higgs production mechanisms {\it viz.} $e^+e^-\to ZH$ and $e^+e^- \to \nu\bar{\nu}H$, at 
an $e^+e^-$ collider with energies ranging from 250 GeV to 500 GeV. The $e^+e^-\to ZH$ channel includes only the $s$-channel 
processes -- $e^+e^-\to Z^*/\gamma^*\to ZH$ (shown in Fig.~\ref{fig:FD}(a)). Whereas $e^+e^- \to \nu \bar{\nu} H$ includes both the $s$-channel processes, $e^+e^-\to Z^*/\gamma^* \to ZH\to \nu \bar{\nu} H$ as well as the $t$-channel process $e^+e^-\to \nu\bar{\nu}W^*W^*\to \nu\bar{\nu}H$ ($WW$ fusion process as shown in Fig.~\ref{fig:FD}(b)).

The $s$ and $t$-channel
processes have different kinematics and hence are affected differently by the inclusion of the HDOs. Moreover,
the $t$-channel process allows us to explore the tensor structure of the $HWW$ vertex alone, free from any
contamination from the $HZZ$ and $HZ\gamma$ vertices. On the other hand, the $s$-channel process is free from
any contamination due to the $HWW$ vertex. Hence, the measurement of the $s$-channel 
contribution will shed light on the tensorial nature of the $HZZ$ and $HZ\gamma$ vertices. We, therefore, analyse the $s$ and 
$t$-channel processes separately to shed more light on the anomalous behaviour of the $HVV$ vertices. We separate the $s$-channel 
($t$-channel) contribution from the $e^+e^-\to \nu\bar{\nu}H$ events by applying a simple kinematic cut on the Higgs energy 
($E_H$) as follows:
\begin{equation}
\label{eq:stcut}
E_H\textrm{-cut:}~~
\Big\vert E_H-\frac{s+M_H^2-M_Z^2}{2\sqrt{s}}\Big\vert \leq \Delta~~~
\left(E_H^c\textrm{-cut:}~~
\Big\vert E_H-\frac{s+M_H^2-M_Z^2}{2\sqrt{s}}\Big\vert \geq \Delta\right),
\end{equation}
where $\sqrt{s}$ is the CME of the two colliding $e^+e^-$ beams and $\Delta$ is an 
energy-window around $E_H$. Here, $E_H^c$-cut is complementary to the $E_H$-cut. We use $\Delta=5$ GeV 
throughout our analysis~\footnote{Typical values of $\Delta$ can be estimated from the energy uncertainties 
of the $b$-jets coming from the Higgs decay. The jet energy uncertainty $\Delta E_{jet}$ (1$\sigma$) of a 
jet having energy $E_{jet}$ are related as, $\Delta E_{jet}/E_{jet} \lesssim 0.3/\sqrt{E_{jet}}$ at the 
ILC~\cite{ILC}. For example, if there are two $b$-jets each with energy 100 GeV, the total 
uncertainty in their energy measurement is $\sqrt{2\times (0.3\times \sqrt{100})^2}\sim 4$ GeV 
(added in quadrature).}.
We must mention here that for the rest of this paper the $s$-channel process will be studied at the $ZH$ 
level without any cuts, unless otherwise specified. 
One can easily get an estimate of the cross-section for any decay modes of $Z$ by multiplying the appropriate BR.
This is because for the $e^+e^-\to l^+l^-H$ channel, a simple 
invariant mass cut on the two leptons about the $Z$ boson mass will separate the $s$-channel to a very high degree. 
For $e^+e^-\to \nu \bar{\nu}H$, on the other hand, the cut on $E_H$ separates the $s$ and $t$-channels. The $s$-channel 
contribution surviving the cut is found to be very close to what one would have found from the rate for $l^+l^-H$, through a 
scaling of BRs. One is thus confident that the $E_H$-cut is effective in minimising mutual contamination of the $s$ and $t$-channel 
contributions.

It should also be mentioned here that the effects of
  beam energy spread are not taken into account in
  Eq.~\ref{eq:stcut} for simplification. While we present the
  basic ideas of distinguishing anomalous interactions of the Higgs,
  the relevant energy window for precision studies has to factor in
  the effects of bremsstrahlung as well as beamstrahlung (depending on
  whether the Higgs factory is a circular or a linear collider).

\begin{table}[t]
\centering
\begin{tabular}{|c|c|c|c|c|c|c|c|}
\hline
$\sqrt{s}$ & Benchmark & $\sigma^{tot}_{\nu\bar{\nu}H}$ & $\sigma^s_{\nu\bar{\nu}H}$ & $\sigma^t_{\nu\bar{\nu}H}$ & $\sigma^{int}_{\nu\bar{\nu}H}$ & $\sigma^{s,ac}_{\nu\bar{\nu}H}$ & $\sigma^{t,ac}_{\nu\bar{\nu}H}$ \\
(GeV) & point & (fb) & (fb) & (fb) & (fb) & (fb) & (fb) \\
\hline    
300 & SM & 52.43 & 36.35 & 17.83 & -1.75 & 37.24 & 15.19 \\
    & BP1 & 52.11 & 35.29 & 18.83 & -2.01 & 36.76 & 15.35 \\    
\hline    
500 & SM & 84.80 & 11.64 & 74.07 & -1.11 & 11.93 & 72.83 \\
    & BP1 & 87.38 & 7.37 & 81.50 & -1.49 & 7.83 & 79.55 \\   
\hline
\end{tabular}
\caption{\label{tab:sigst} We show the total $\nu\bar{\nu}H$
  cross-section ($\sigma_{\nu\bar{\nu}H}^{tot}$), only $s$-channel
  cross-section ($\sigma_{\nu\bar{\nu}H}^{s}$), only $t$-channel
  cross-section ($\sigma_{\nu\bar{\nu}H}^{t}$) and their interference
  contribution ($\sigma_{\nu\bar{\nu}H}^{int}$) for the SM
  ($\kappa=1,f_{WW}=0,f_W=0,f_{BB}=0,f_B=0$) and for HDO benchmark
  point BP1 ($\kappa=1,f_{WW}=-3,f_W=8,f_{BB}=-4,f_B=3$) for two
  different CMEs. We also present the $s$
  ($\sigma^{s,ac}_{\nu\bar{\nu}H}$) and $t$-channel
  ($\sigma^{t,ac}_{\nu\bar{\nu}H}$) cross-sections separated from the
  $\nu\bar{\nu}H$ events after applying the cut defined in
  Eq.~\ref{eq:stcut}.  The superscript $ac$ means after cut.}
\end{table}

In Table~\ref{tab:sigst}, we show the effect of the $E_H$-cut on the $\nu\bar{\nu}H$ channel in the SM and in presence 
of HDOs for one benchmark point, BP1 ($\kappa=1,f_{WW}=-3,f_W=8,f_{BB}=-4,f_B=3$) which closely mimics the SM cross-section. The 
$E_H$-cut keeps almost all the $s$-channel contribution but the $E_H^c$-cut cuts out a small portion around $E_H$ from the 
$t$-channel contribution. Therefore, the $s$-channel cross-sections after this cut increase slightly from their without-cut values due 
to this small $t$-channel contamination. On the other hand, the $t$-channel cross-sections after cut decrease slightly from their 
without-cut values. We also estimate the interference between the $s$ and $t$-channel diagrams and present the numbers 
in Table~\ref{tab:sigst}. Interference contribution is expected to be tiny in the $\sqrt{s}$ region sufficiently away from the 
$s$-channel threshold energy $(M_H+M_Z) \approx 226$ GeV. We find that the interference contribution is only $\sim 3.5$\% of 
the total cross-section for $\sqrt{s}=300$ GeV, in the SM. This re-affirms the statement at the end of the previous paragraph. 
We also note that the inclusion of HDOs with moderate values of coefficients 
does not affect this contribution much. Hence, by neglecting the interference term, we approximate the total $\nu\bar{\nu}H$ 
cross-section as
\begin{equation}
\sigma^{tot}_{\nu \bar{\nu}H}\approx \sigma_{ZH}\times BR_{Z\to \nu \bar{\nu}}+\sigma_{\nu\bar{\nu}H}^t,
\end{equation} 
where $\sigma_{ZH}$ is the $s$-channel cross section and $BR_{Z\to \nu \bar{\nu}}$ is the invisible branching fraction 
($\approx 20\%$) of the $Z$-boson.

\begin{figure}
\includegraphics{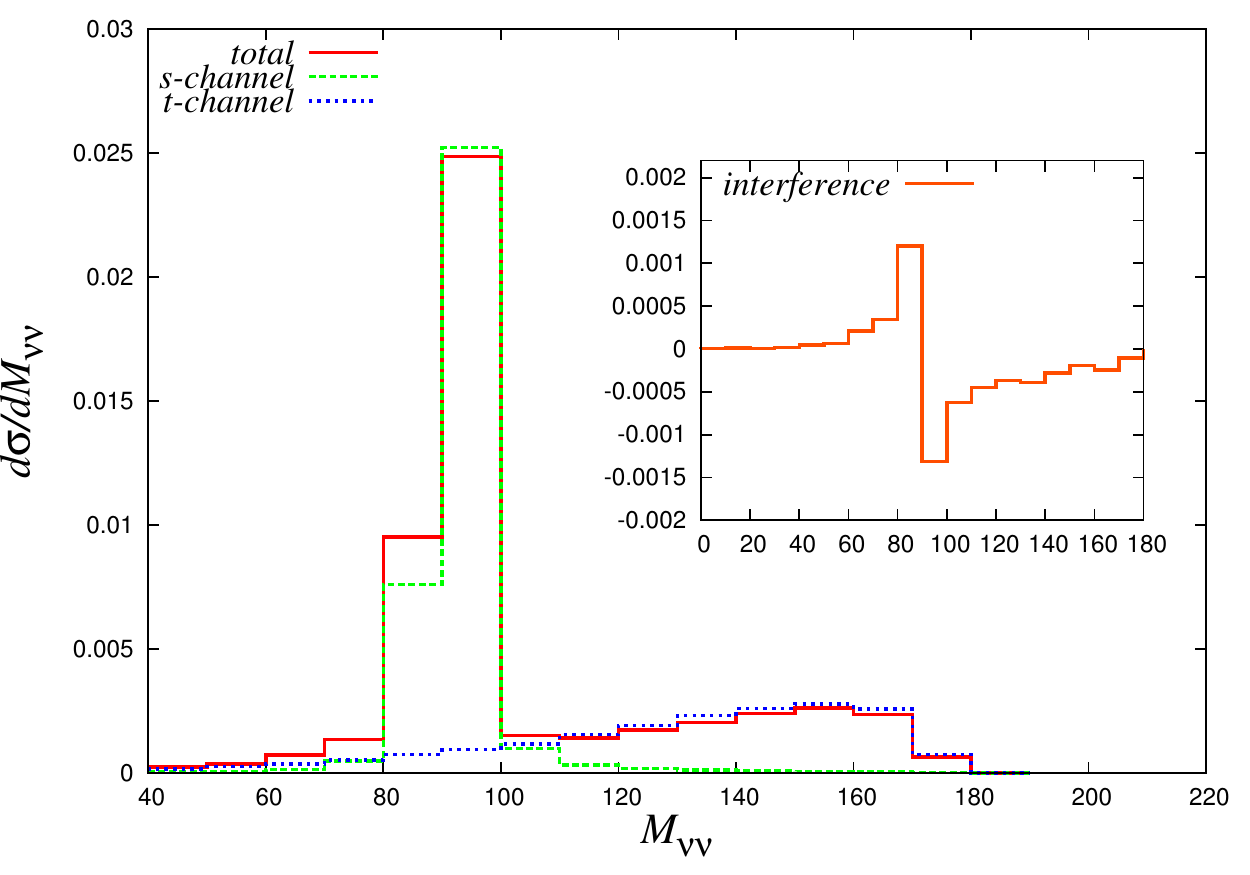}
\caption{Invariant mass distributions of $\nu\bar{\nu}$ of the process $e^+e^-\to\nu\bar{\nu}H$ at $\sqrt{s}=300$ GeV and for the benchmark point BP1 ($\kappa=1,f_{WW}=-3,f_W=8,f_{BB}=-4,f_B=3$). The red, green, blue histograms are for the total ($s+t+ interference$), $s$ and $t$ channels respectively. The inset (orange) plot shows the interference ($total-s-t$) contribution.}
\label{fig:Mvv}
\end{figure}

Fig.~\ref{fig:Mvv} shows the invariant mass distribution of the neutrino pair for the process $e^+e^-\to\nu\bar{\nu}H$ at $\sqrt{s}=300$ GeV and for the benchmark point BP1. We separately show the distributions for the total process (which includes the $s$ and $t$ channels as well as the interference) and also the $s$ and $t$ channels separately. In an inset plot we show the distribution due to this interference. This clearly shows that it is negligible when compared to the $s$ and $t$ channel contributions. This nature generally holds for the parameter space under consideration.

\subsection{A general expression for the cross-sections}

In this analysis, we keep $\kappa$, $f_{WW}/\textrm{TeV}^2$, $f_{W}/\textrm{TeV}^2$, $f_{BB}/\textrm{TeV}^2$ and $f_{B}/\textrm{TeV}^2$ as free
parameters. 
The $HWW$ vertex depends on three parameters ($\kappa$, $f_{WW}$ and $f_{W}$) whereas the 
$HZZ$ and the $HZ\gamma$ vertices depend on five parameters ($\kappa$, $f_{WW}$, $f_{W}$, $f_{BB}$ and $f_{B}$). The
$\kappa$ dependence enters the $HZ\gamma$ vertex through the $W$-loop in the effective $HZ\gamma$ vertex.
The amplitude for the process $e^+e^-\to ZH/\nu\bar{\nu}H$ is a linear combination of $x_i\in\{\kappa,f_{WW},f_{W},f_{BB},f_B\}$ 
and therefore, the cross-section can always be expressed as a bi-linear form,
$\sigma(S,x_i)=\displaystyle\sum_{i,j=1}^{5}x_i C_{ij}(S) x_j$, where $C_{ij}(S)$ is the $ij^{th}$ element of the coefficient matrix 
$\mathcal{M}(\sqrt{s})$ at a CME of $\sqrt{s}$. Hence, the cross-section can be written in the following closed form
\begin{equation}
\sigma(\sqrt{s})=\mathcal{X}\cdot\mathcal{M}(\sqrt{s})\cdot\mathcal{X}^T,
\end{equation}
where $\mathcal{X}=(\kappa,f_{WW},f_W,f_{BB},f_B)$ is a row vector.

The matrices of coefficients for the $e^+e^-\to Z H$ process at $\sqrt{s}=250$ GeV and $300$ GeV are

\begin{equation}
\label{Ms}
\footnotesize
\mathcal{M}^{s,ZH}_{250}= 
\begin{pmatrix}
 241.32 & -7.11 & -2.29 & -0.55 & -0.51 \\
 -7.11 & 0.35 & 0.13 & -0.02 & -0.05 \\
 -2.29 & 0.13 & 0.06 & -0.01 & -0.03 \\
 -0.55 & -0.02 & -0.01 & 0.01 & 0.02 \\
 -0.51 & -0.05 & -0.03 & 0.02 & 0.04
\end{pmatrix};
\mathcal{M}^{s,ZH}_{300}= 
\begin{pmatrix}
 181.67 & -6.43 & -2.99 & -0.51 & -0.71 \\
 -6.43 & 0.46 & 0.18 & -0.03 & -0.08 \\
 -2.99 & 0.18 & 0.14 & -0.02 & -0.06 \\
 -0.51 & -0.03 & -0.02 & 0.02 & 0.03 \\
 -0.71 & -0.08 & -0.06 & 0.03 & 0.08
\end{pmatrix}
\end{equation}

Similar matrices for the $t$-channel process (after the $E_H^c$-cut) for the channel 
$e^+e^-\to \nu \bar{\nu}H$ at $\sqrt{s}=250$ GeV and $300$ GeV are

\begin{equation}
\label{Mt}
\footnotesize
\mathcal{M}^{t,\nu\bar{\nu}H}_{250}=
\begin{pmatrix}
 4.63 & 5.2\times 10^{-3} & 0.02 \\
 5.2\times 10^{-3} & 2.9\times 10^{-4} & -1.2 \times 10^{-4} \\
 0.02 & -1.2 \times 10^{-4} & 1.6 \times 10^{-4}
\end{pmatrix};
\mathcal{M}^{t,\nu\bar{\nu}H}_{300}=
\begin{pmatrix}
  15.36 & 0.04 & 0.07 \\
  0.04 & 1.2\times10^{-3} & -7.7\times10^{-4} \\
  0.07 & -7.7\times10^{-4} & 4.6\times10^{-4}.
\end{pmatrix}
\end{equation}

We must mention here that the matrices in Eq.~\ref{Mt} are
three-dimensional compared to the five-dimensional matrices in
Eq.~\ref{Ms} because the $t$-channel only involves the $HWW$ vertex
which is not affected by the operators $\mathcal{O}_{BB}$ and
$\mathcal{O}_B$ (Eqs.~\ref{eq:lagHVV},~\ref{eq:lagHVVcoeff}).  We also
observe that in Eq.~\ref{Ms}, the coefficients of the matrix related
to either $f_{BB}$ or $f_B$ are much less pronounced compared to the
coefficients involving the other three parameters, {\textit{viz.}}
$\kappa$, $f_{WW}$ and $f_W$.  Also from Eq.~\ref{Mt} we see that
barring the (1,1) entry in the matrices, all the other coefficients
are small implying that the HDOs will have small but non-negligible
effects on the $t$-channel cross-sections for energies at the Higgs
factories.

\begin{figure}
\includegraphics{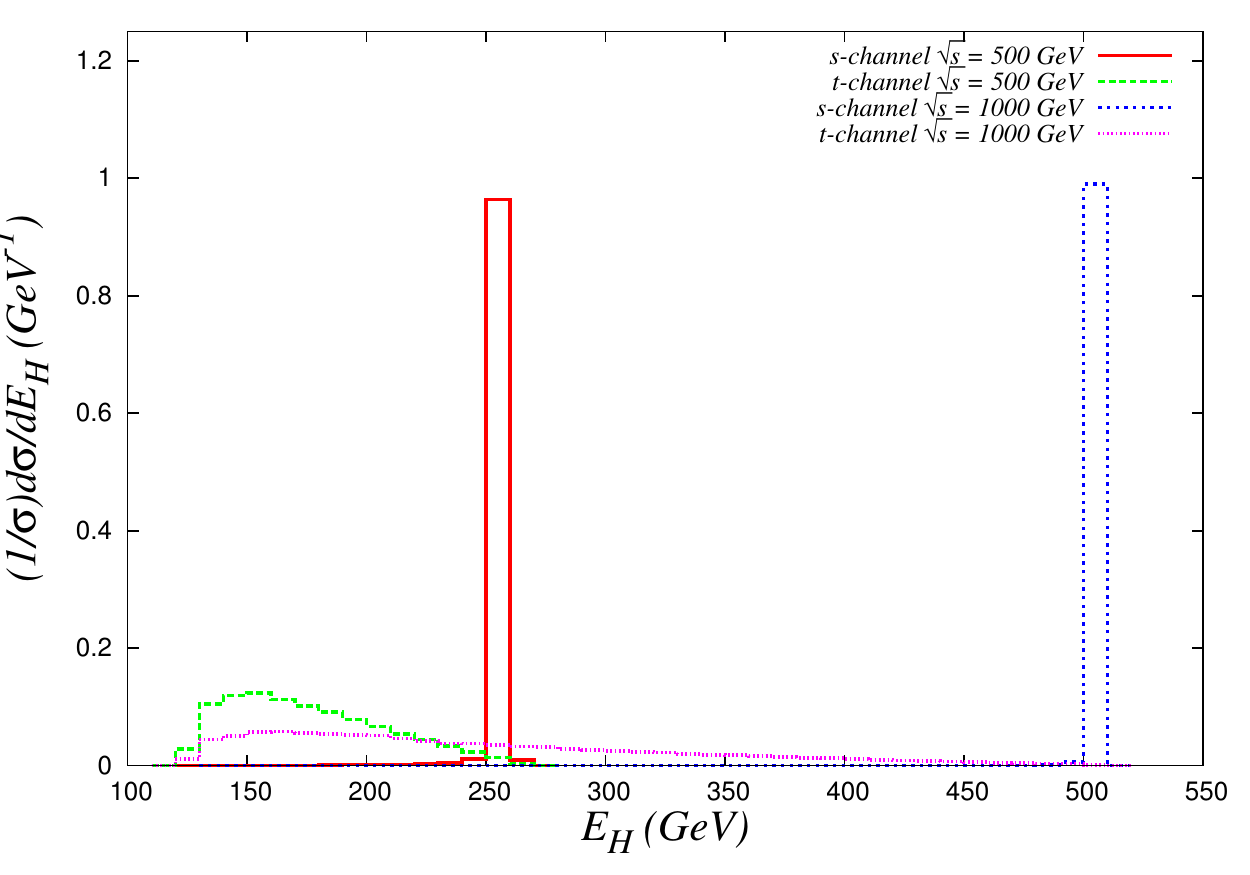}
\caption{Normalised distributions of the Higgs energy ($E_H$) for the $s$-channel (red : $\sqrt{s}=500$ GeV and blue : $\sqrt{s}=1$ TeV) and $t$-channel (green : $\sqrt{s}=500$ GeV and magenta : $\sqrt{s}=1$ TeV) for the benchmark point BP1.}
\label{fig:EH}
\end{figure}

An explanation of relatively less dependence of the $t$-channel
cross-section compared to the $s$-channel on the anomalous operators
can also be understood from Fig.~\ref{fig:EH}. The plots reveal that, for
the former process (essentially a vector boson fusion channel), the
Higgs emerges with much smaller energy. The higher-dimensional
couplings, on the other hand, contain derivatives which translate
into a direct dependence on the energy of the Higgs, thus putting the
$t$-channel process at a relative disadvantage. The Higgs energy
distribution shows a longer tail for higher centre-of-mass energies,
thus offering a partial recompense to the $t$-channel process for
an energy as high as a TeV.

In this study we also consider the process $e^+e^-\to W^+ W^-$ which
involves the triple-gauge boson vertices $WW\gamma$ and $WWZ$. These
are concomitantly affected by the operators $\mathcal{O}_W$ and
$\mathcal{O}_B$. Besides, as mentioned in section~\ref{sec:ELF}, such
vertices are also affected by the operator $\mathcal{O}_{WWW}$ which
does not affect the Higgs sector. In the basis of
$x_{i}^{WW}\in\{1,f_W,f_B,f_{WWW}\}$, the coefficient matrix at
$\sqrt{s}=300$ GeV is given by
\begin{equation}
\label{MWW}
\footnotesize
\mathcal{M}^{WW}_{300}=
\begin{pmatrix}
 13.48 & 1.10\times 10^{-2} & 5.65\times 10^{-3} & 4.24\times 10^{-3} \\
 1.10\times 10^{-2} & 4.98\times 10^{-4} & 5.27\times 10^{-5} & 2.02\times 10^{-4} \\
 5.65\times 10^{-3} & 5.27\times 10^{-5} & 1.17\times 10^{-4} & 1.96\times 10^{-5} \\
 4.24\times 10^{-3} & 2.02\times 10^{-4} & 1.96\times 10^{-5} & 8.18\times 10^{-4}
 \end{pmatrix}.
\end{equation} 
As we can see above, all the $C_{ij}s$ are very small when compared to
$C_{11}$, which gives us the SM cross-section. We will discuss this
channel in more details later in this paper.

\subsection{Energy dependence of $s$ and $t$-channel cross-sections}

\begin{figure}[!h]
\centering
\subfloat{
\begin{tabular}{ccc}
\resizebox{70mm}{!}{\includegraphics{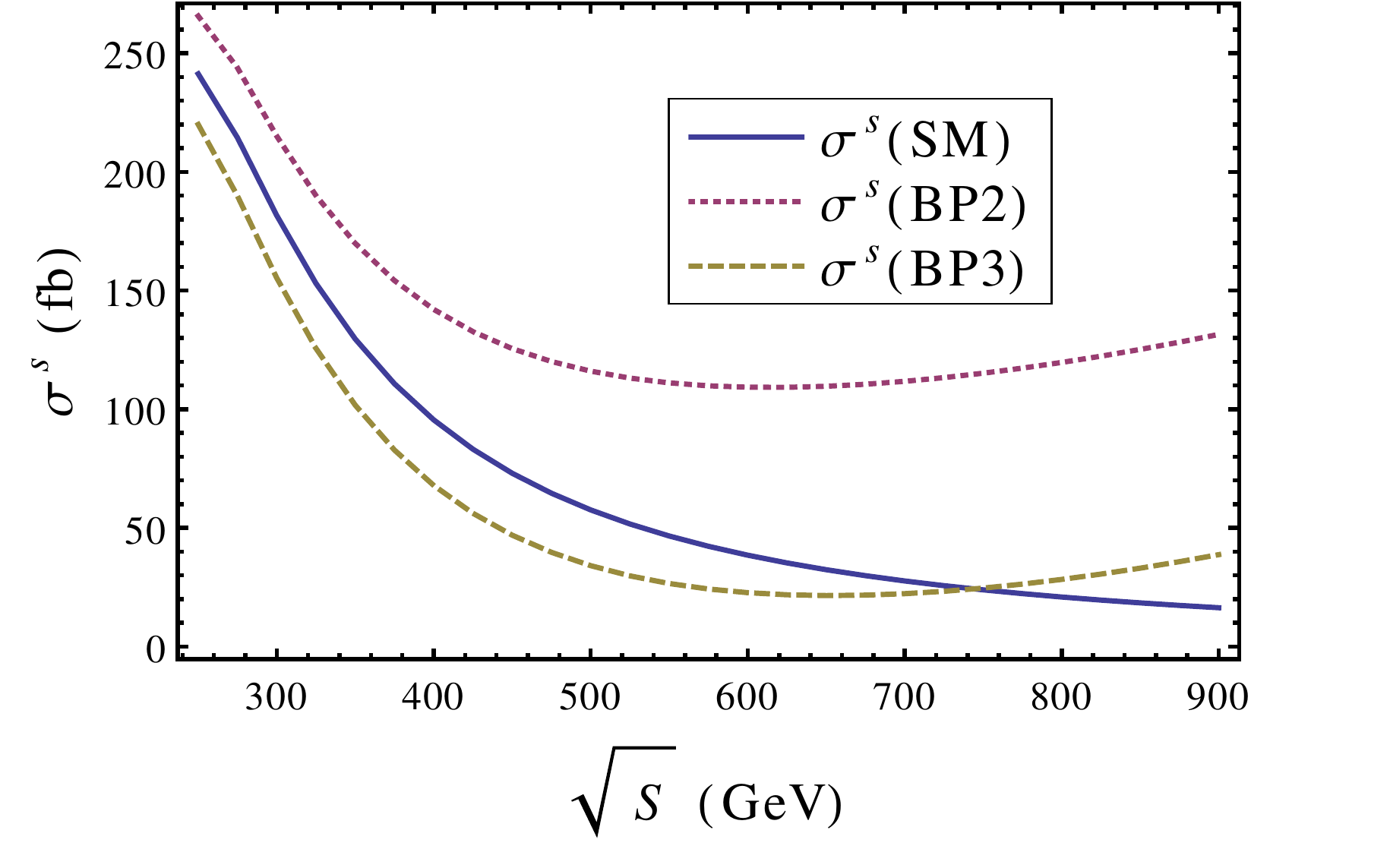}} &&
\resizebox{70mm}{!}{\includegraphics{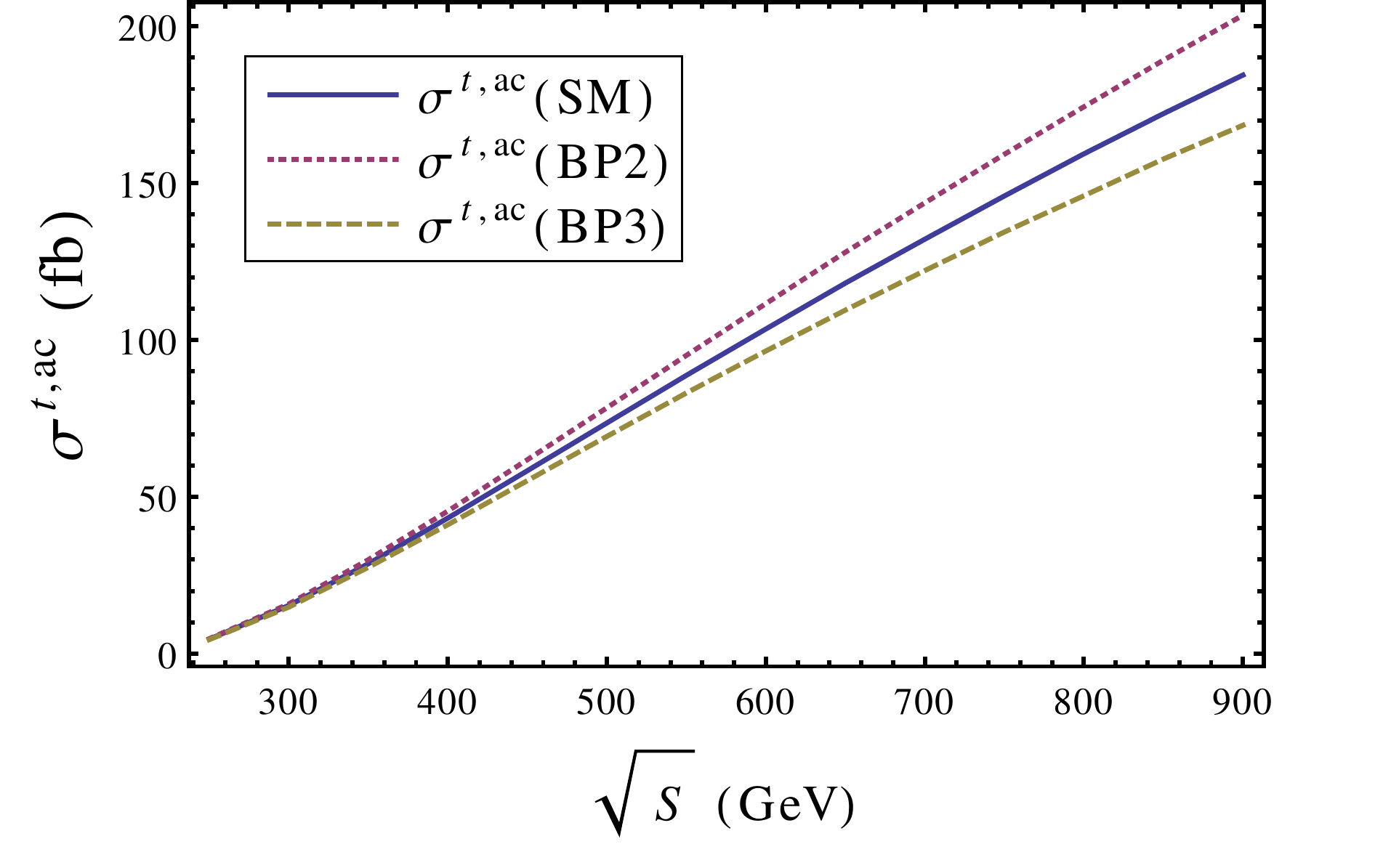}} \\
\hspace{8mm}(a)&&\hspace{10mm}(b)
\end{tabular}}
\caption{(a) : $\sigma^s$ (in fb) for the channel $e^+e^-\to Z H$ and (b) : $\sigma^{t,ac}$ (in fb) for the channel 
$e^+e^-\to \nu \bar{\nu} H$ as functions of the CME, $\sqrt{s}$. The cross-sections have been computed for three 
benchmark points, viz. SM ($x_i\in \{1,0,0,0,0\}$), BP2 ($x_i\in \{1,0,5,0,0\}$) and BP3 ($x_i\in \{1,0,-5,0,0\}$). The 
superscript $ac$ denotes the after cut scenario.}
\label{stenergy}
\end{figure}

It is well-known that in SM, the cross-section for the $s$-channel
falls with the CME as $1/S$ and that for the $t$-channel, rises as
$\ell n{S}$~\cite{Altarelli}. However, for sets of values of our
parameters, different from the SM, the nature of the $s$-channel curve
can be completely different from its SM-counterpart. The $t$-channel
cross-section however is not affected so significantly on the
introduction of HDOs as has been discussed in detail in the previous
sub-section. We show the variation of the $s$ and $t$-channel
processes for $\sqrt{s}$ ranging from $250$ GeV to $900$ GeV. In
contrast to the SM nature of a fall in the $s$-channel cross-section
with energy, the introduction of HDOs does in no way ensure such a
nature which can be seen in Fig.\ref{stenergy} (a) for two benchmark
points (BP2 ($x_i\in \{1,0,5,0,0\}$) and BP3 ($x_i\in
\{1,0,-5,0,0\}$)) alongside the SM. The above two benchmark points
have been chosen as the cross-sections are quite sensitive to $f_W$
and the two points are allowed from EWPT constraints. On the whole it
is clear from the diagrams that the ratio of the $s$ and $t$-channel
cross-sections in some channel at a particular energy can be an
important probe to the nature of new Higgs couplings\footnote{The
  visible rise with $\sqrt{s}$ (in Fig.\ref{stenergy}(a) for the
  benchmark points BP2 and BP3) does not threaten unitarity, since the
  additional degrees of freedom responsible for the effective
  operators take care of it when $\sqrt{s}$ approaches $\Lambda$. The
  rise is not noticeable if one has the operators
  $\mathcal{O}_{WW}/\mathcal{O}_{BB}$ instead of
  $\mathcal{O}_W/\mathcal{O}_B$. The different momentum dependence in
  the former case tames the rise with $\sqrt{s}$ as can be verified
  from the corresponding Feynman rules in\cite{Garcia}.}

\subsection{More information from the total rates}

The total rates and their ratios at different CMEs can be important
probes to identify the tensor structure of the $HVV$ couplings. We
show how the total rates for the $s$ and $t$-channel processes are
affected on the introduction of the effective operators (Eqs.~\ref{Ms}
and~\ref{Mt}).

We must make a statement about the values of the
  coefficients, $f_i/\Lambda^2$ ($i$ is the index of the operator
  under consideration) chosen in the
  rest of the paper. In most cases, $f_i/\Lambda^2$ is allowed to vary
  in the range $[-20,20]$ TeV$^{-2}$. Now, a reasonable criterion for
  the validity of the effective field theory~\cite{effval} is $f_i x(g) E^2/\Lambda^2 < 1$,
  where $x(g)$ are the $SU(2)_L/U(1)_Y$ factors for the operators
  under study and $E$ is the scale of the process. For the production
  case, it is the centre of mass energy of the $e^+e^-$ colliding
  beams, which is $250-300$ GeV, while for  decays, it is the mass of
  the Higgs boson. For the production case, we perform a rough calculation
  taking $g \approx 0.65$, $g'\approx 0.74$ and the cut-off scale $\Lambda=1$ TeV.
  Hence, for the operator $\mathcal{O}_W$, $f_W x(g) E^2/\Lambda^2 \approx f_W
  \frac{0.65}{2} 300^2/1000^2 \approx 0.029 f_W$, which can take $f_W$
  to values $\simeq 34$. Similarly, for $\mathcal{O}_B$, the reach will be around $f_B \simeq 30$. For $\mathcal{O}_{WW}$, we have two factors of $g$ and two
  factors of $\frac{1}{2}$, which can take $f_{WW}$ to an even larger
  value. Thus the values chosen in our scan approximately conforms to
  the requirement of a valid effective theory.

\subsubsection{One parameter at a time}
\label{sec:1param}

\begin{figure}[!h]
\centering
\subfloat{
\begin{tabular}{ccc}
\resizebox{70mm}{!}{\includegraphics{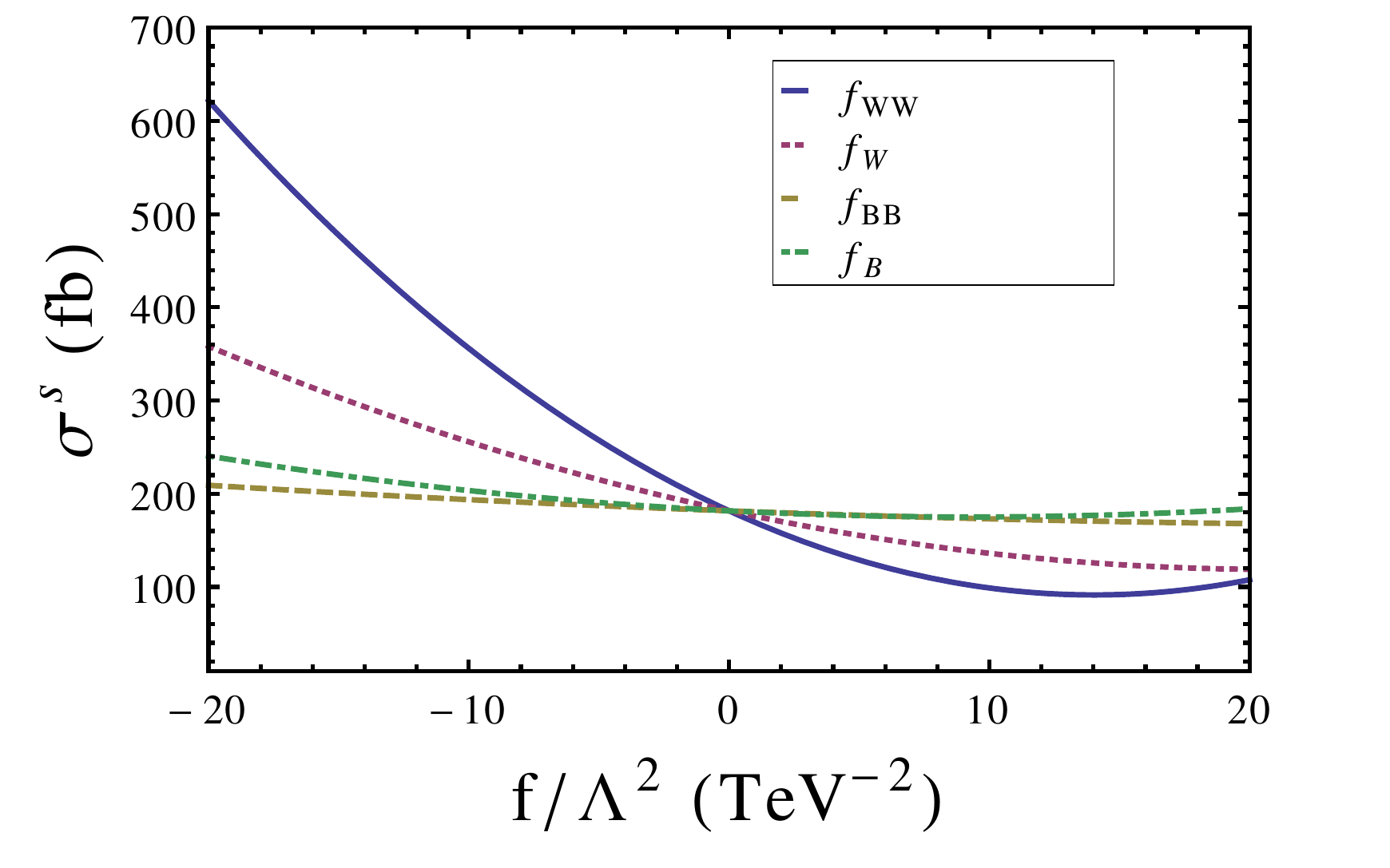}} &&
\resizebox{70mm}{!}{\includegraphics{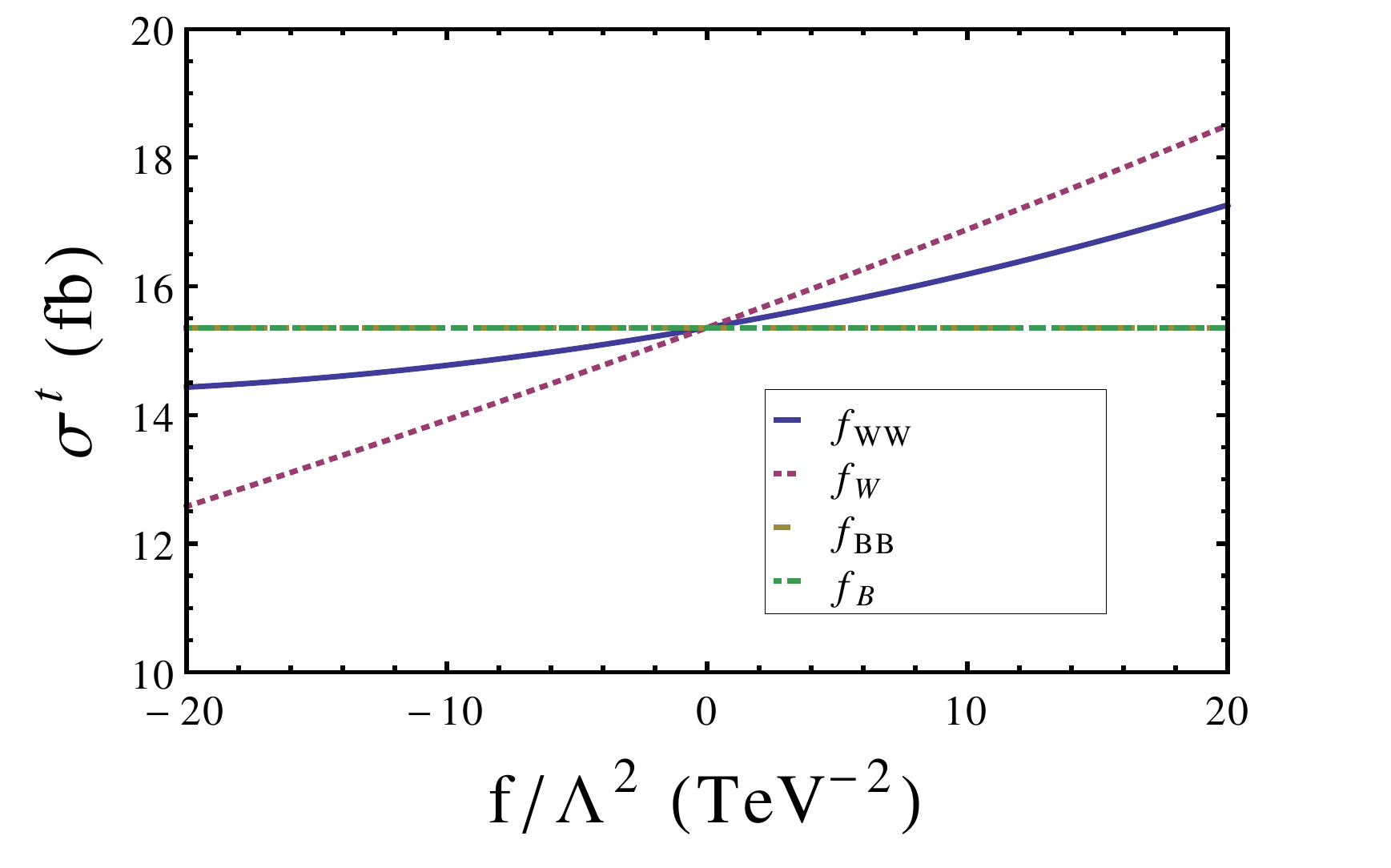}} \\
\hspace{8mm}(a)&&\hspace{20mm}(b) \\
\resizebox{70mm}{!}{\includegraphics{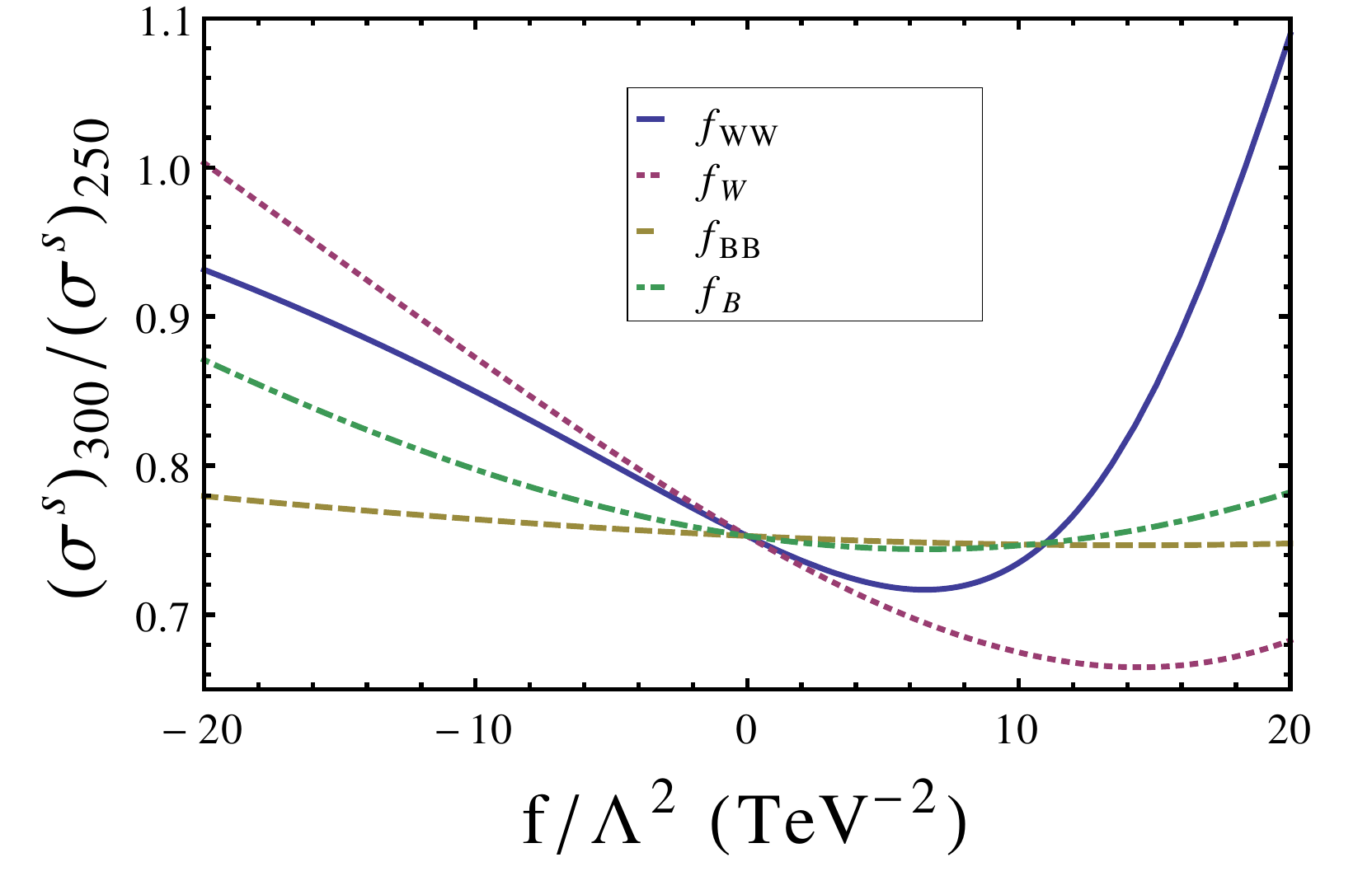}} &&
\resizebox{70mm}{!}{\includegraphics{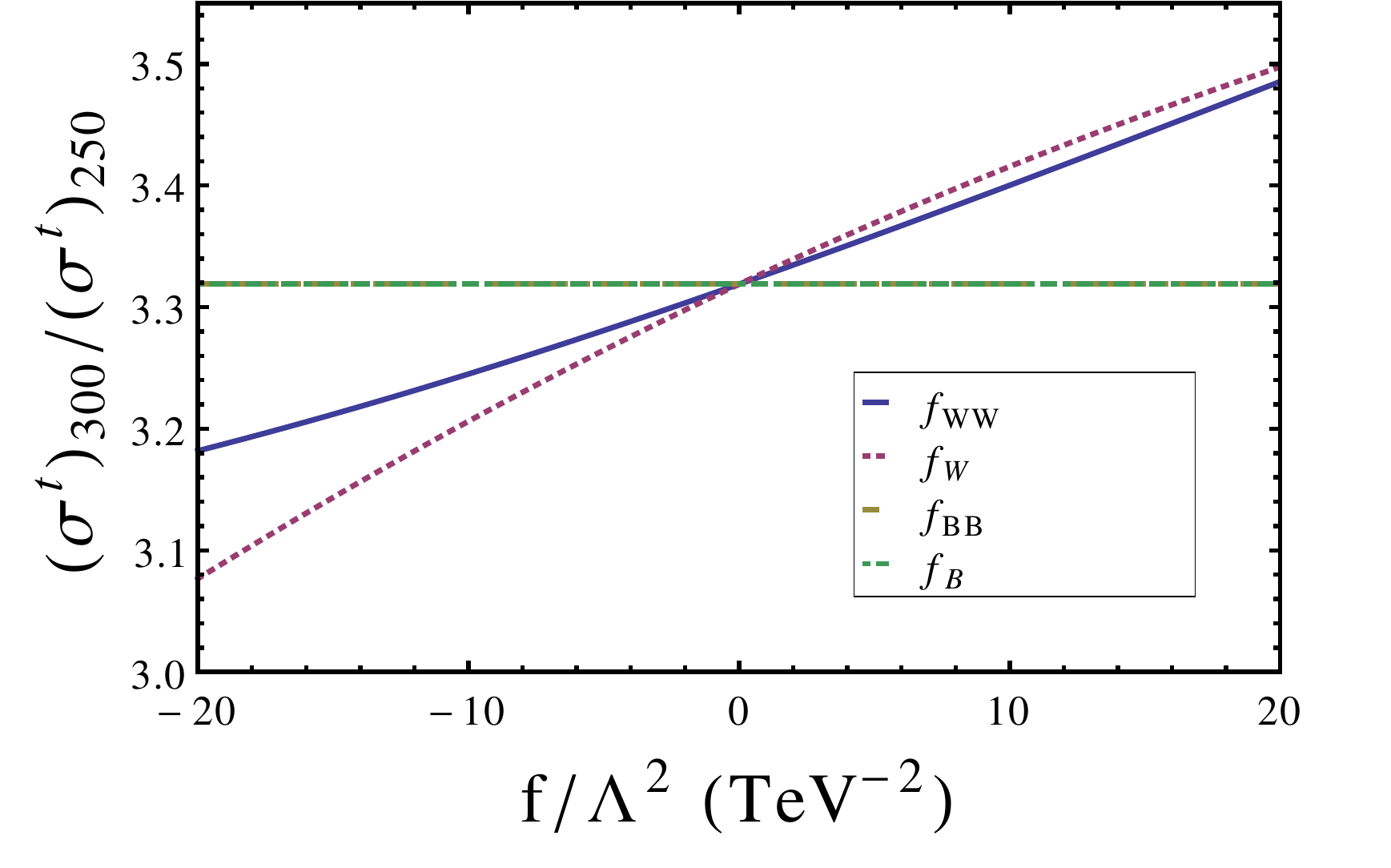}} \\
\hspace{8mm}(c)&&\hspace{20mm}(d)
\end{tabular}}
\caption{Variations of (a) $\sigma^s_{ZH}(300)$ (fb) and (c) $\sigma^s_{ZH}(300)/\sigma^s_{ZH}(250)$ for $e^+e^-\to Z H$ and of 
(b) $\sigma^{t,ac}_{\nu\bar{\nu}H}(300)$ (fb) and (d) $\sigma^t_{\nu\bar{\nu}H}(300)/\sigma^{t,ac}_{\nu\bar{\nu}H}(250)$ for 
$e^+e^-\to \nu \bar{\nu} H$ with $f_{WW}$, $f_W$, $f_{BB}$, $f_B$. $\kappa=1$ for all the cases. The superscript $ac$ denotes 
the cut in Eq.\ref{eq:stcut}. The numbers in the brackets are the CMEs.}
\label{fig:1d}
\end{figure}

\begin{figure}[!h]
\centering
\subfloat{
\begin{tabular}{ccc}
\resizebox{70mm}{!}{\includegraphics{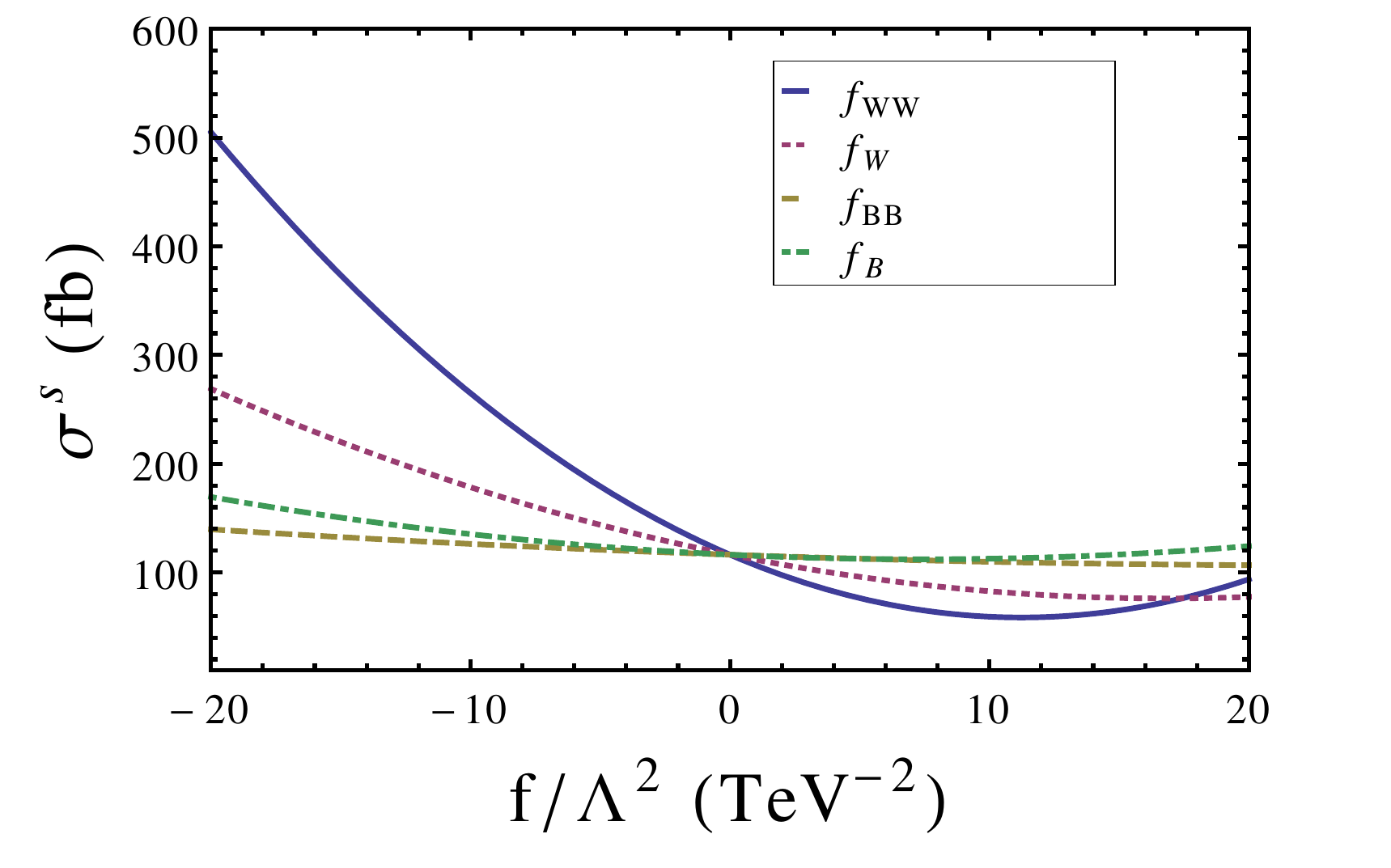}} &&
\resizebox{70mm}{!}{\includegraphics{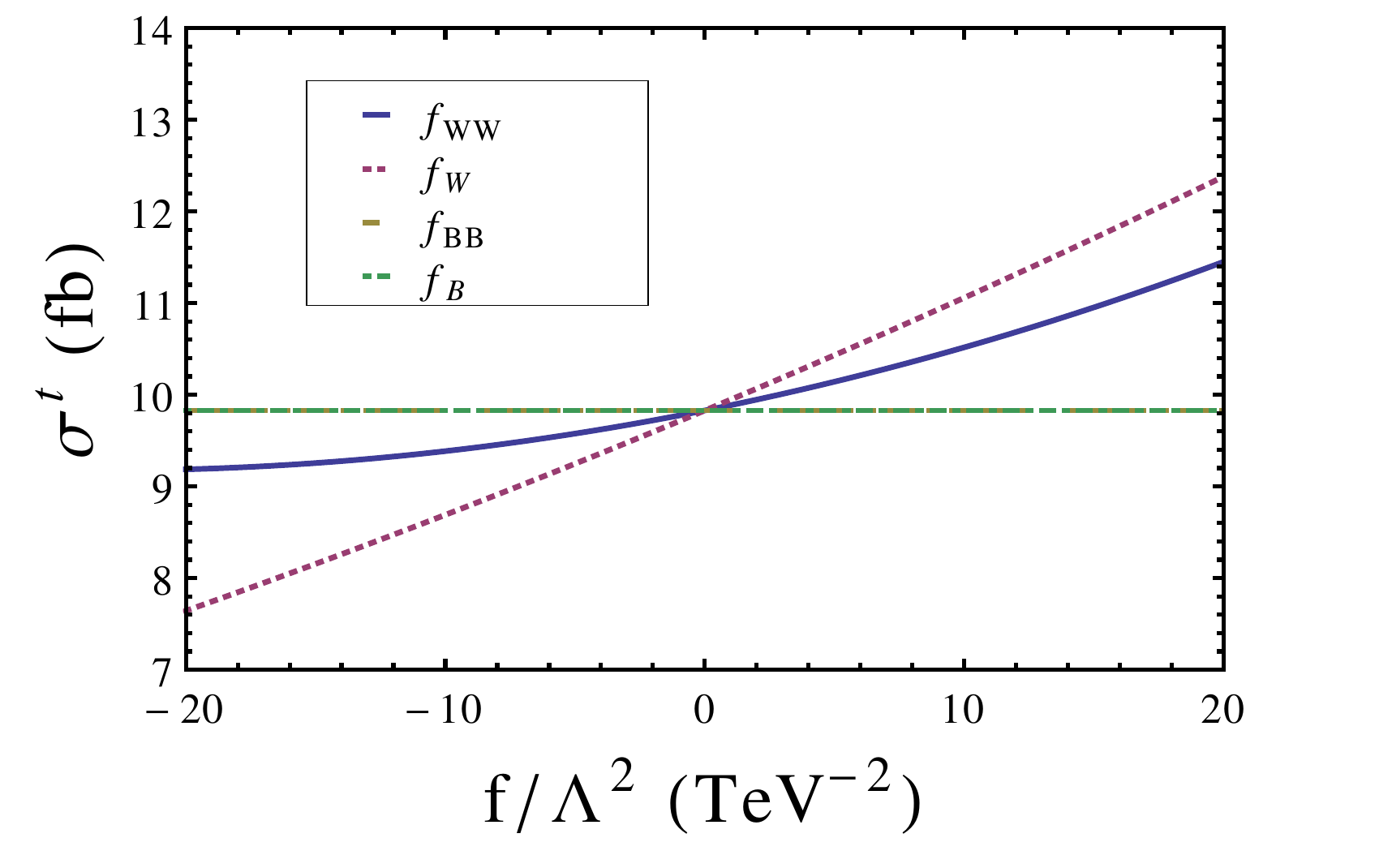}} \\
\hspace{8mm}(a)&&\hspace{20mm}(b) \\
\resizebox{70mm}{!}{\includegraphics{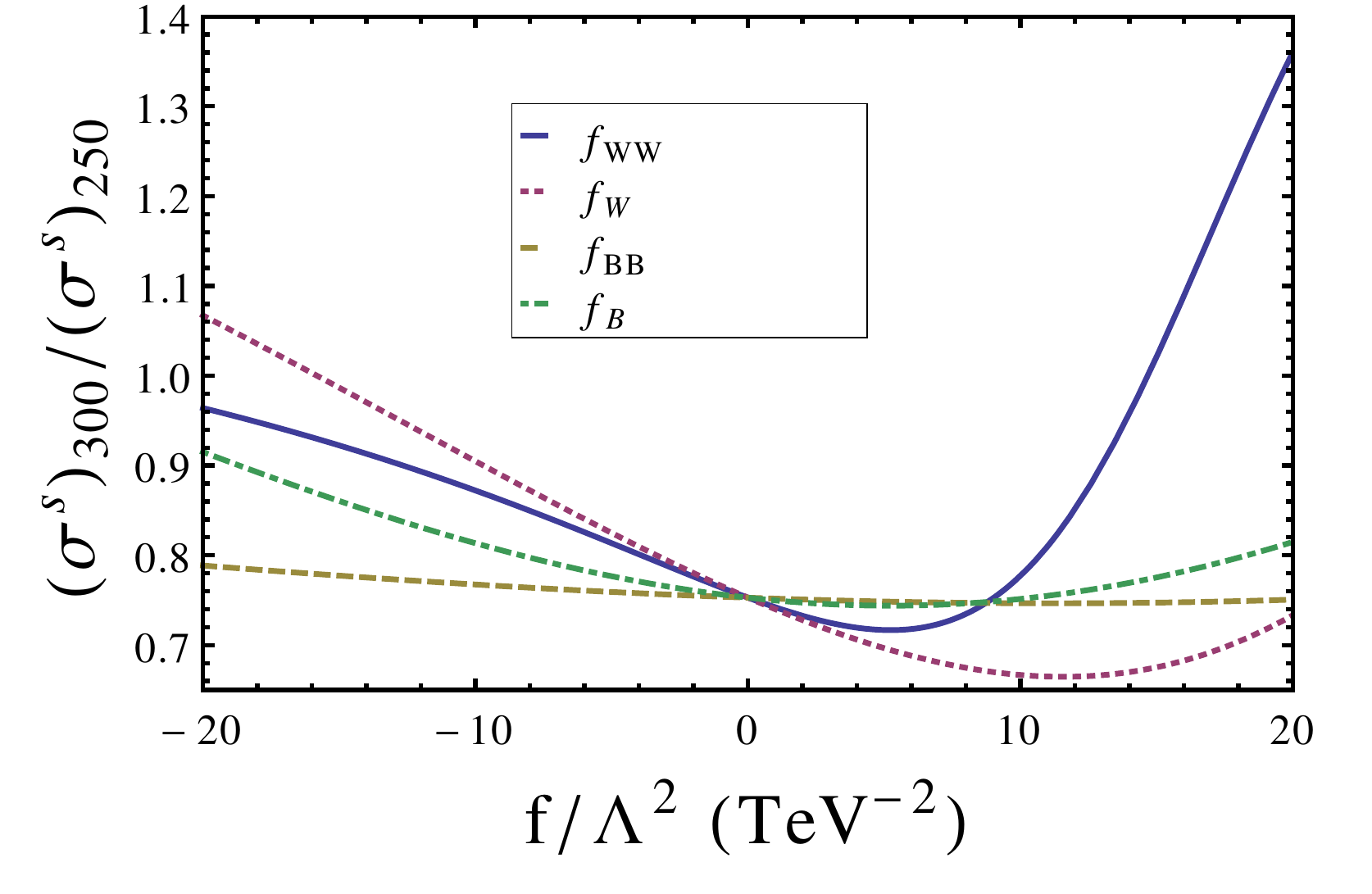}} &&
\resizebox{70mm}{!}{\includegraphics{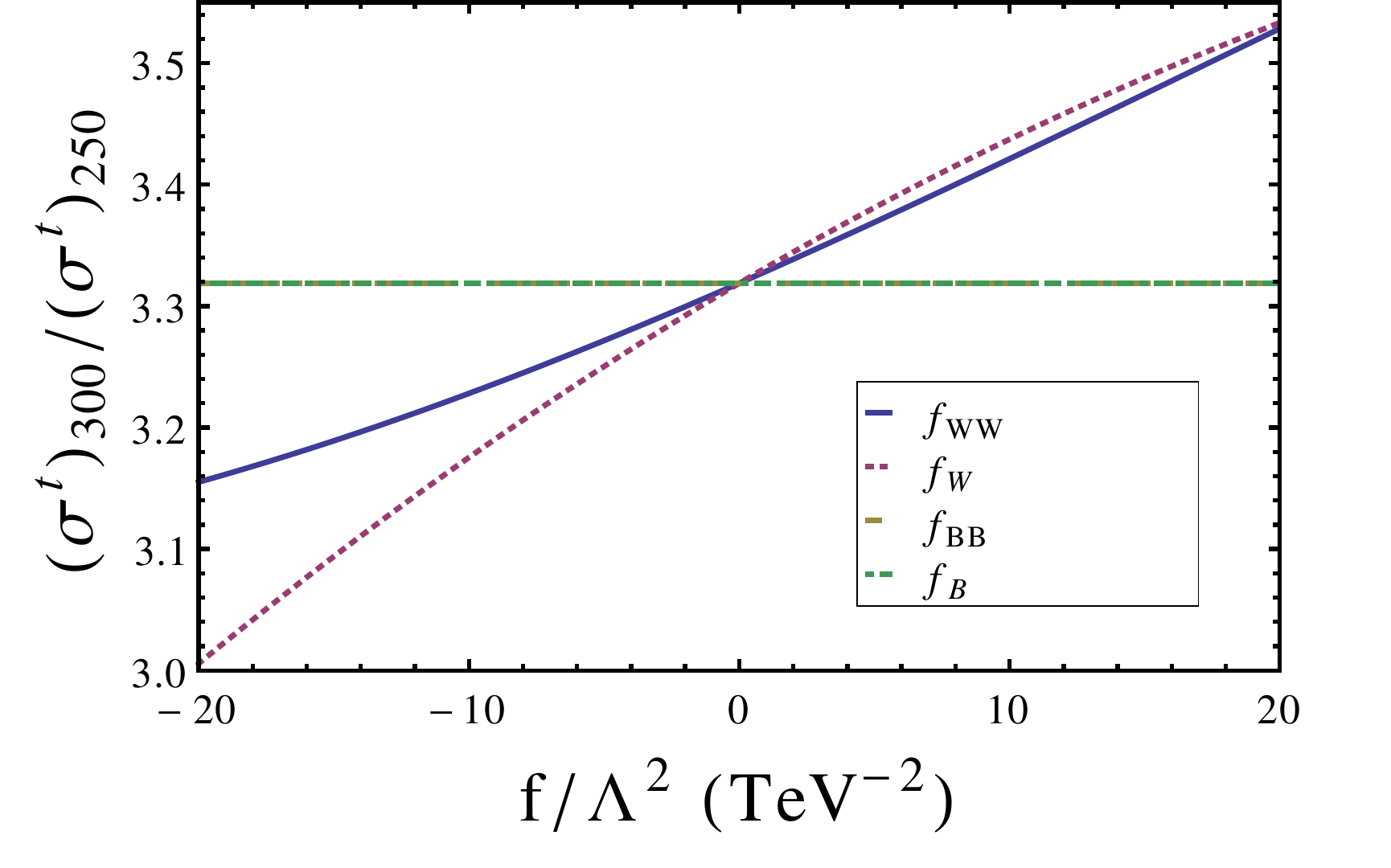}} \\
\hspace{8mm}(c)&&\hspace{20mm}(d)
\end{tabular}}
\caption{Variations of (a) $\sigma^s_{ZH}(300)$ (fb) and (c) $\sigma^s_{ZH}(300)/\sigma^s_{ZH}(250)$ for $e^+e^-\to Z H$ and of 
(b) $\sigma^{t,ac}_{\nu\bar{\nu}H}(300)$ (fb) and (d) $\sigma^t_{\nu\bar{\nu}H}(300)/\sigma^{t,ac}_{\nu\bar{\nu}H}(250)$ for 
$e^+e^-\to \nu \bar{\nu} H$ with $f_{WW}$, $f_W$, $f_{BB}$, $f_B$. $\kappa=0.8$ for all the cases. The superscript $ac$ denotes 
the cut in Eq.\ref{eq:stcut}. The numbers in the brackets are the CMEs.}
\label{fig:1dpt8}
\end{figure}

In Figs.~\ref{fig:1d} and~\ref{fig:1dpt8}, we show the variations of the $e^+e^-\to Z H$ and 
$e^+ e^- \to \nu \bar{\nu} H$ ($t$-channel) cross-sections as functions of 
a single parameter by keeping all other parameters fixed at their SM values. We show that even for small values of the operator 
coefficients, the cross-sections can vary significantly from the SM expectations. We also show that the ratios of the cross sections 
at two different energies can vary non-trivially with these parameters. If there is no new tensor structure in the $HVV$ couplings, 
the ratio plots will be flat horizontal curves. Any departure from a horizontal nature of such curves will shed light on new tensor 
structure  in such $HVV$ vertices. The main sources of departure are the interference terms between the SM and HDO contributions. 
Such terms, occurring in both the numerator and the denominator of the ratio, carry the dependence on $f$ as well as $\sqrt{s}$.

\begin{figure}[!h]
\centering
\subfloat{
\begin{tabular}{ccc}
\resizebox{70mm}{!}{\includegraphics{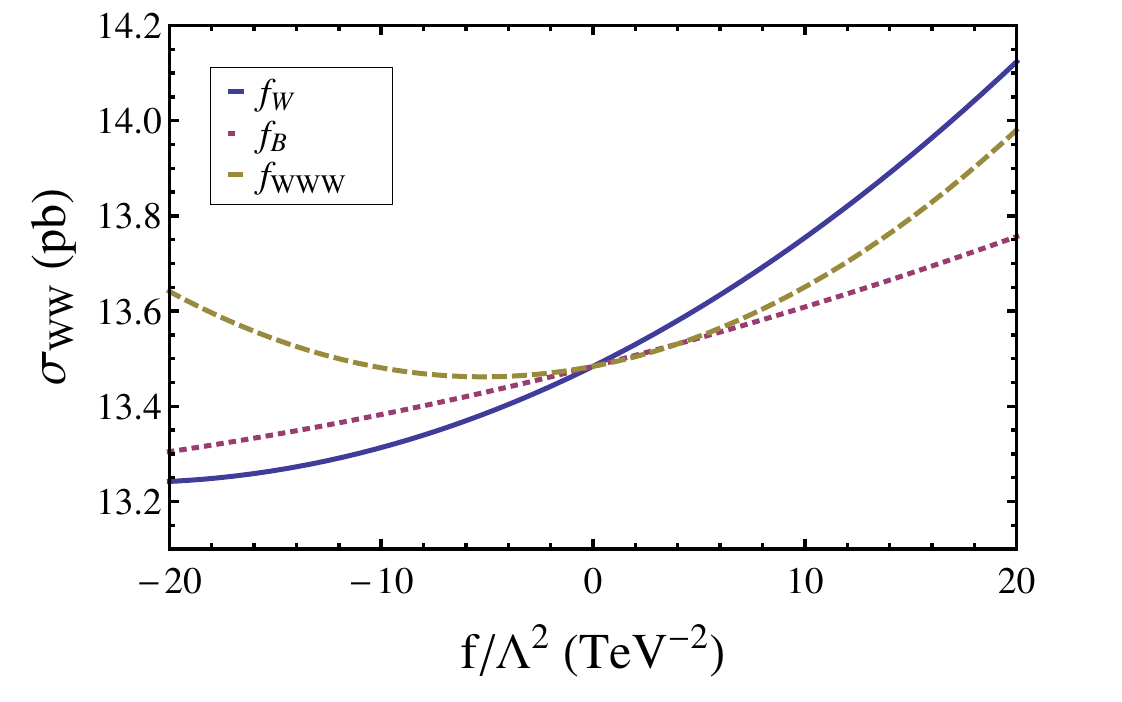}} &&
\resizebox{70mm}{!}{\includegraphics{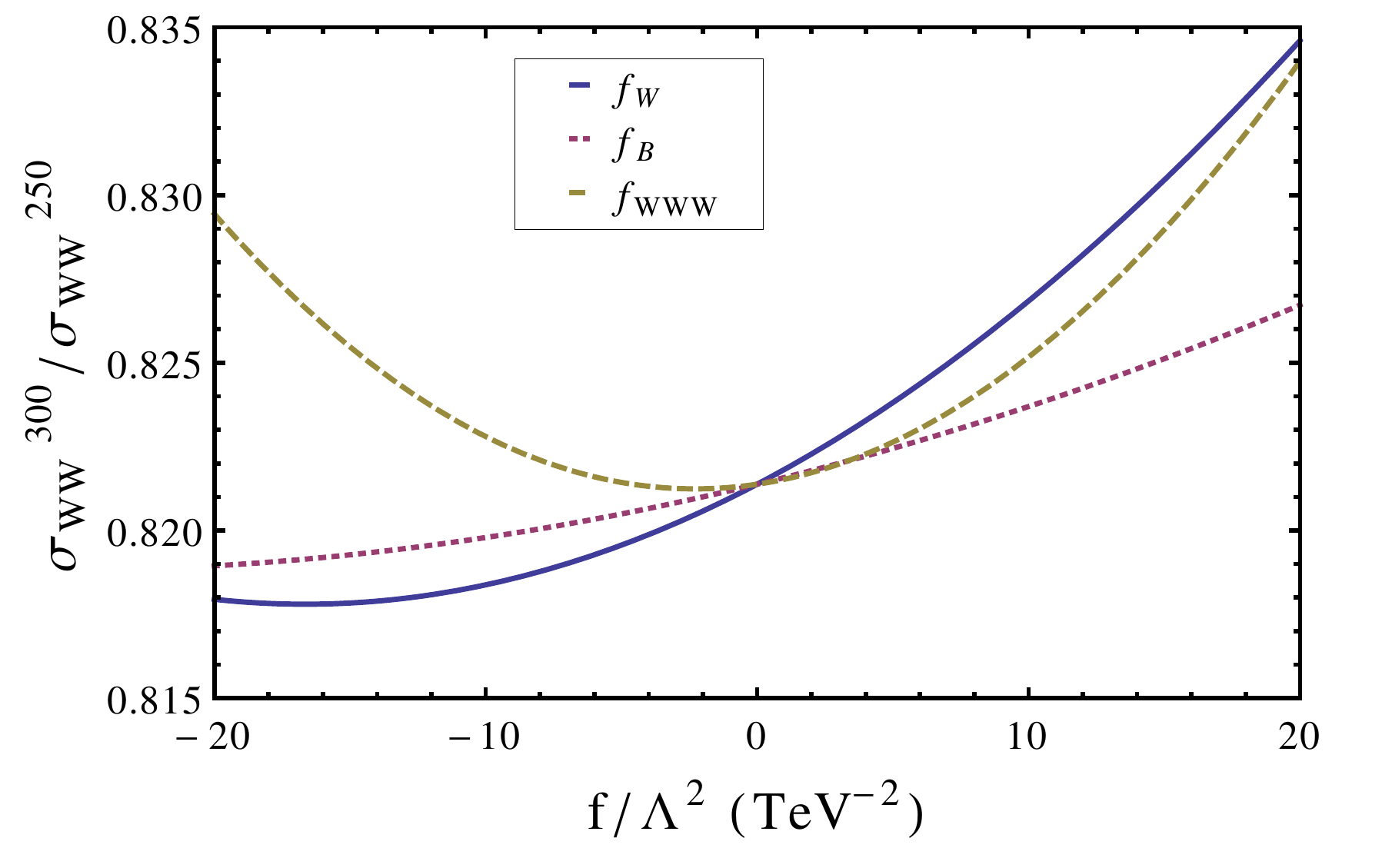}} \\
\hspace{8mm}(a)&&\hspace{20mm}(b) \\
\end{tabular}}
\caption{(a) Cross section ($\sigma$ (in pb)) for the process $e^{+}e^{-}\to W^{+} W^{-}$ for $\sqrt{s} = 300$ GeV and (b) ratio of cross sections
($\sigma_{300}/\sigma_{250}$) for the same process as functions of $f$'s.}
\label{fig:ee2ww}
\end{figure}

We also remind the reader that the use of gauge invariant
higher-dimensional operators implies a correlated modification in
triple gauge boson couplings(Eqs.~\ref{eq:lagWWV},~\ref{eq:lagWWVcoeff}). 
$f_W$ and $f_B$ are thus responsible for
altering the rates of $e^+ e^- \rightarrow W^+ W^-$ concomitantly with
those for Higgs boson production. Such a concomitance, if verified in
an $e^+ e^-$ collision experiment, should point rather unmistakably at
one or the other of the gauge invariant operators mentioned here.  We
show the modified rates of the $WW$ final state in Fig.~\ref{fig:ee2ww} where we also show the effects of the operator driven
by $f_{WWW}$ (which does not affect the Higgs couplings).

It should however be mentioned that the actual presence of anomalous couplings in $e^+e^-\to W^+W^-$ is best reflected 
in a detailed study of various kinematic regions~\cite{Hagiwara:1986vm}. Such a study, however is not the subject of the present paper.

The main conclusion emerging from Figs.~\ref{fig:1d},~\ref{fig:1dpt8} and~\ref{fig:ee2ww} are as follows :

\begin{itemize}
 \item In Figs.~\ref{fig:1d}(a) and~\ref{fig:1dpt8}(a), for the process $e^+ e^- \rightarrow
   Z H$, we find that the operator $\mathcal{O}_{WW}$ changes the
   cross section from its SM expectation by $\sim 30\%$ even in the
   range $-5 < f_{WW} < 5$.  The major contribution to the cross
   section modification comes from the operators $\mathcal{O}_{WW}$
   and $\mathcal{O}_{W}$. $\mathcal{O}_{B}$ and $\mathcal{O}_{BB}$
   have lesser contributions to the cross section.
 \item In Figs.~\ref{fig:1d}(b) and~\ref{fig:1dpt8}(b), for the cut-applied $t$-channel contribution in the 
   process $e^+ e^- \to \nu \bar{\nu} H$, the operator $\mathcal{O}_W$ 
   maximally affects the cross-section. The effect of $\mathcal{O}_{WW}$ 
   is comparatively less pronounced. $\mathcal{O}_{BB}$ and $\mathcal{O}_B$ 
   does not change this cross-section as the $HWW$ vertex is unaffected by 
   these operators. Most importantly, it should be noted that the effect 
   of these operators on the $t$-channel process is much less pronounced 
   than its $s$-channel counterpart (Eqs.~\ref{Ms},~\ref{Mt}).

 \item In Figs.~\ref{fig:1d}(c) and~\ref{fig:1dpt8}(c), the ratio of the cross sections for the 
   $e^+ e^- \rightarrow Z H$ channel at $\sqrt{s}=300$ GeV and $\sqrt{s}=250$ 
   GeV shows a different nature. In the range $-20 < f_i < 20$ for the four 
   operators discussed above, the ratio changes by $\sim 33 \%$ for
   $\mathcal{O}_W$. The effect of $\mathcal{O}_{WW}$ is less than this. 
   The change in the ratio is the least for $\mathcal{O}_{BB}$.
 \item In Figs.~\ref{fig:1d}(d) and~\ref{fig:1dpt8}(d), the ratio of cross-sections for the cut-applied 
   $t$-channel process varies in the range $\sim[3.1,3.5]$ for $-20 < f_i < 20$.
 \item We see that in Fig.~\ref{fig:ee2ww}, the cross-sections do not vary significantly with 
   the operator coefficients. This is because the $e^+e^-\to W^+ W^-$ channel has 
   a strong $\nu_e$ mediated $t$-channel contribution which does not involve the triple-gauge boson vertex. 
   This has a significant interference with the  $s$-channel. In order to bring out the feature of 
   the triple gauge boson vertices, we need to devise some strategy which will tame down the $t$-channel 
   effect, such as using right-polarised electrons if one uses a linear collider.
\end{itemize}

\subsubsection{Two parameters at the same time}

\begin{figure}[H]
\centering
\subfloat{
\begin{tabular}{ccc}
\resizebox{65mm}{!}{\includegraphics{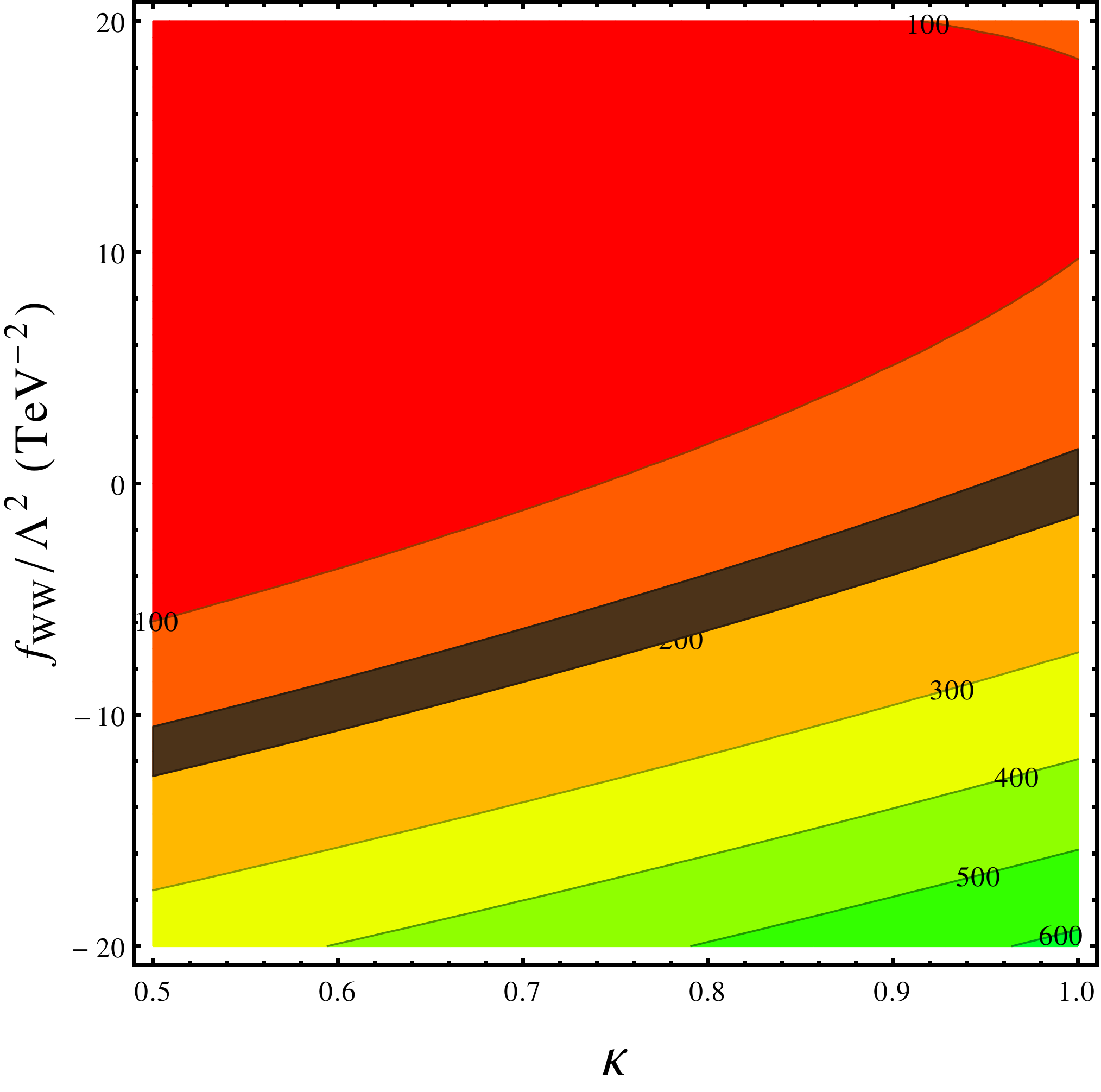}} &&
\resizebox{65mm}{!}{\includegraphics{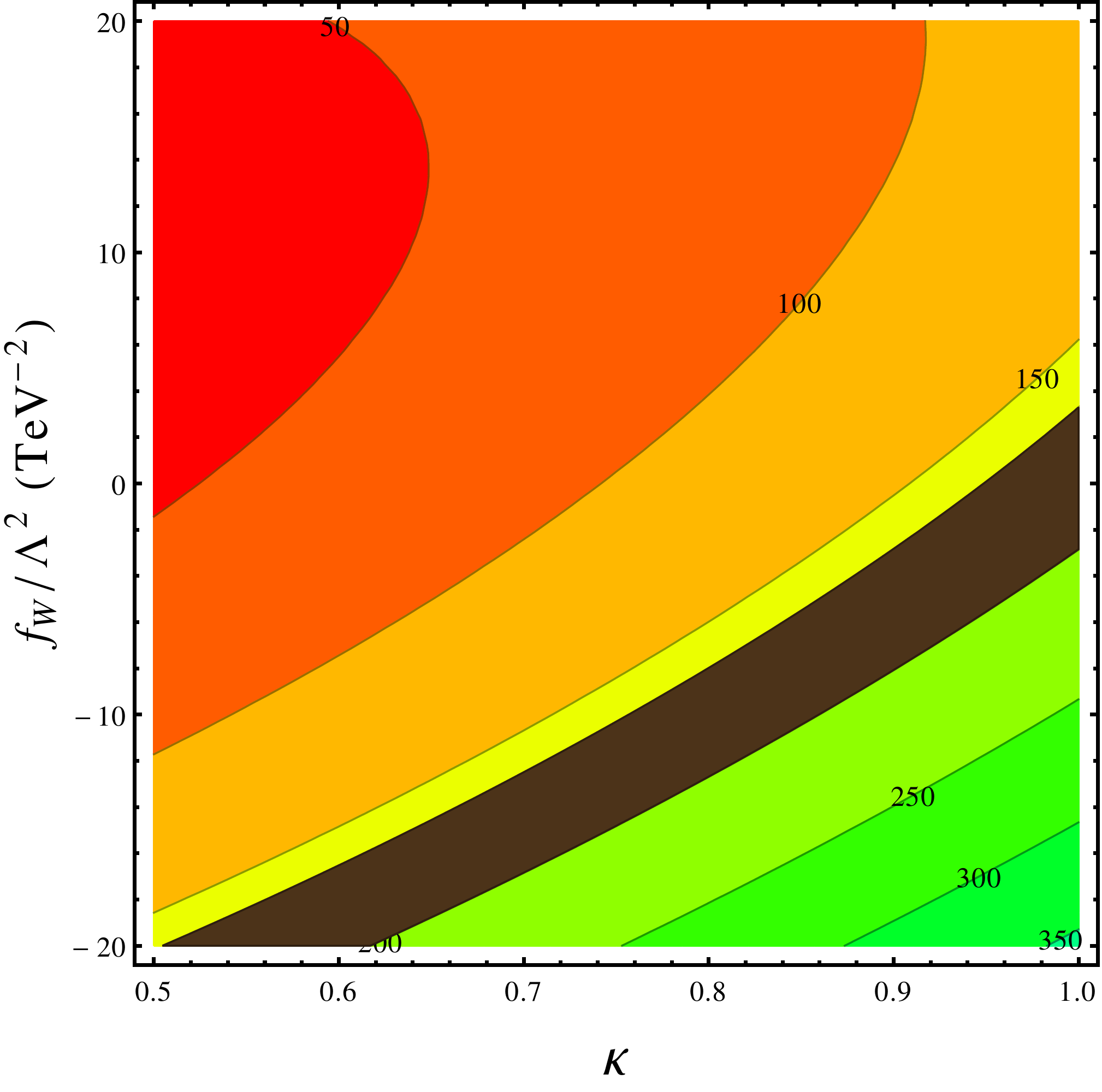}} \\
\hspace{8mm}(a)&&\hspace{15mm}(b) \\
\resizebox{65mm}{!}{\includegraphics{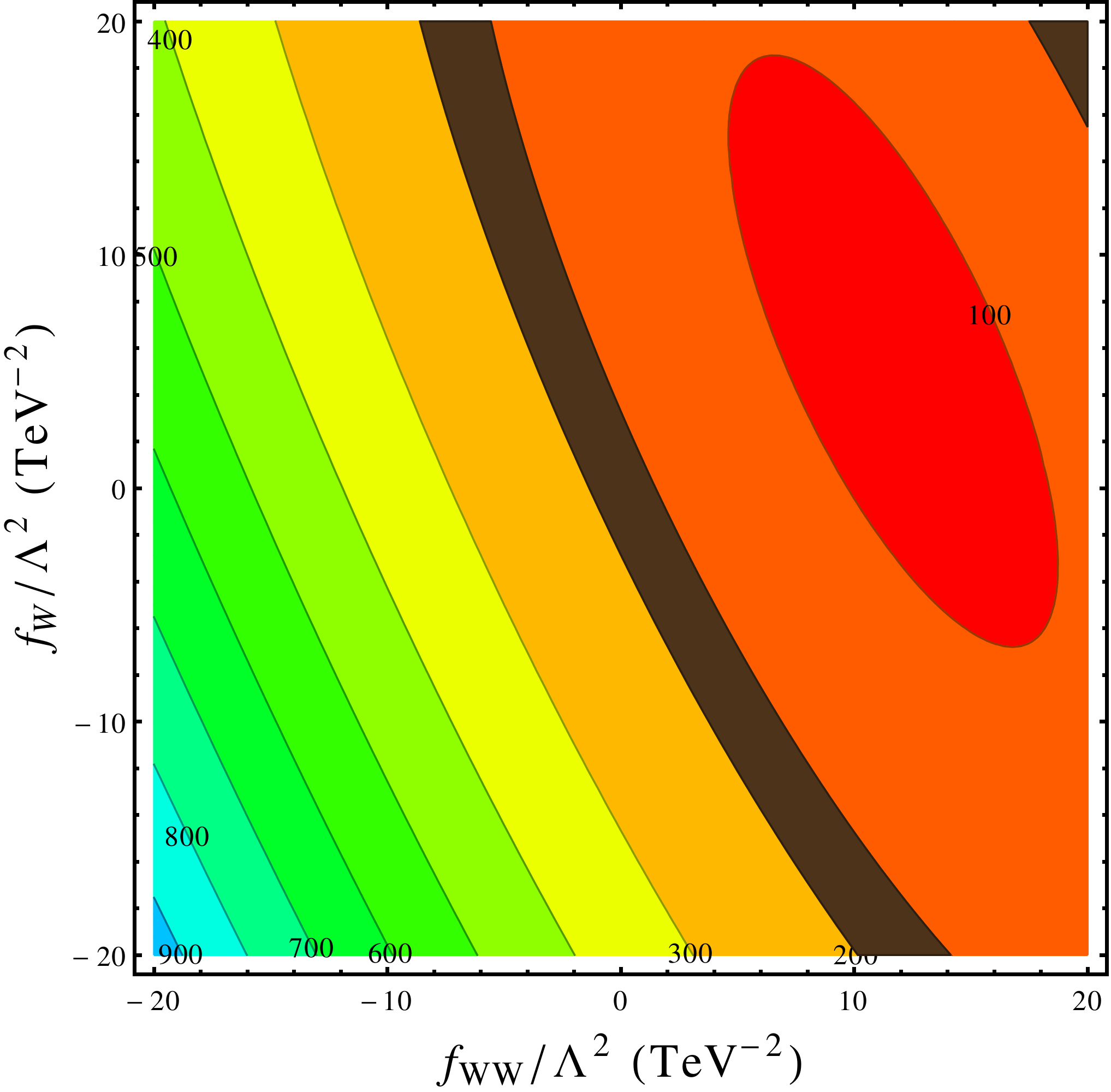}} &&
\resizebox{65mm}{!}{\includegraphics{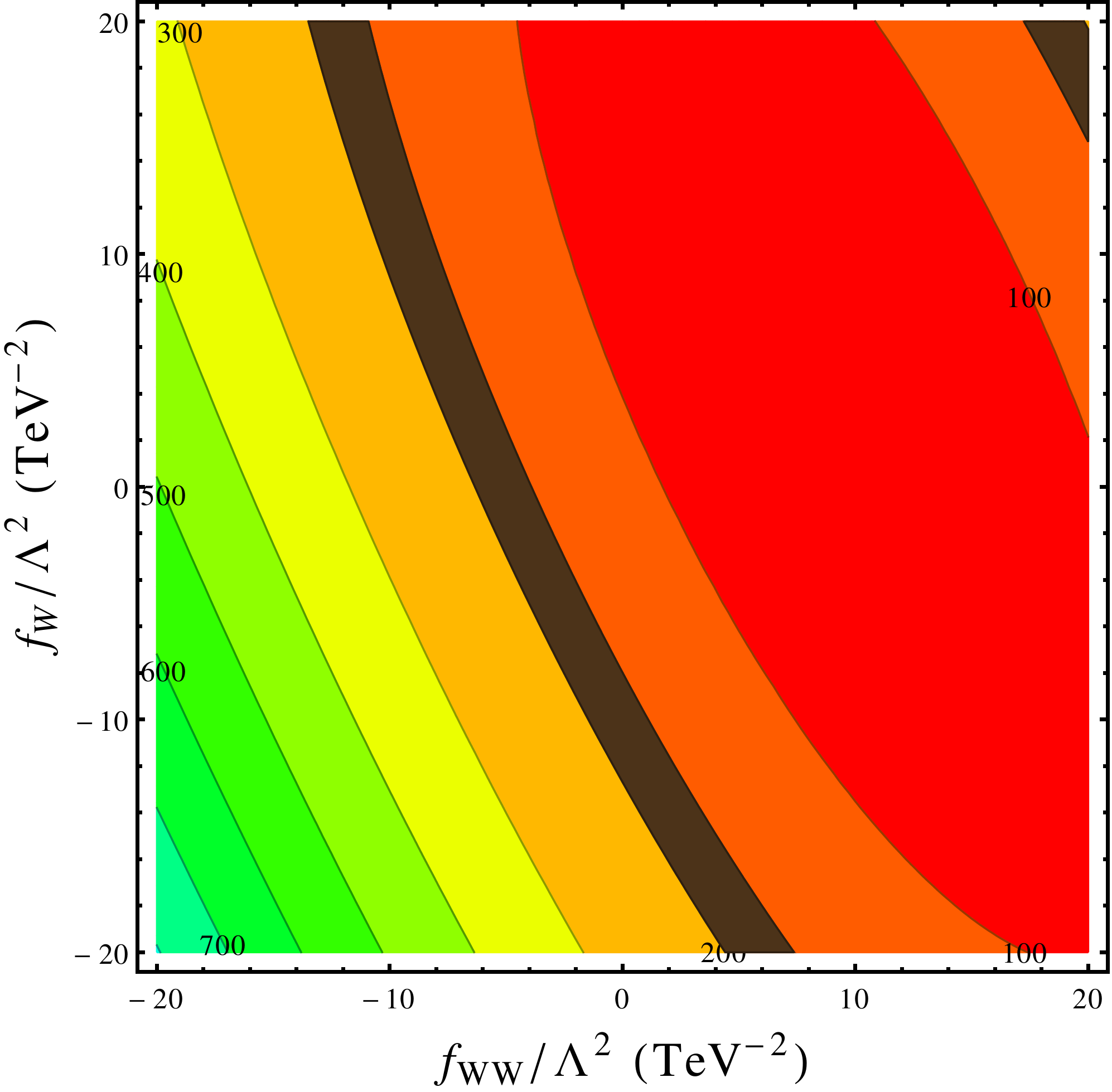}} \\
\hspace{8mm}(c)&&\hspace{15mm}(d) \\
\resizebox{65mm}{!}{\includegraphics{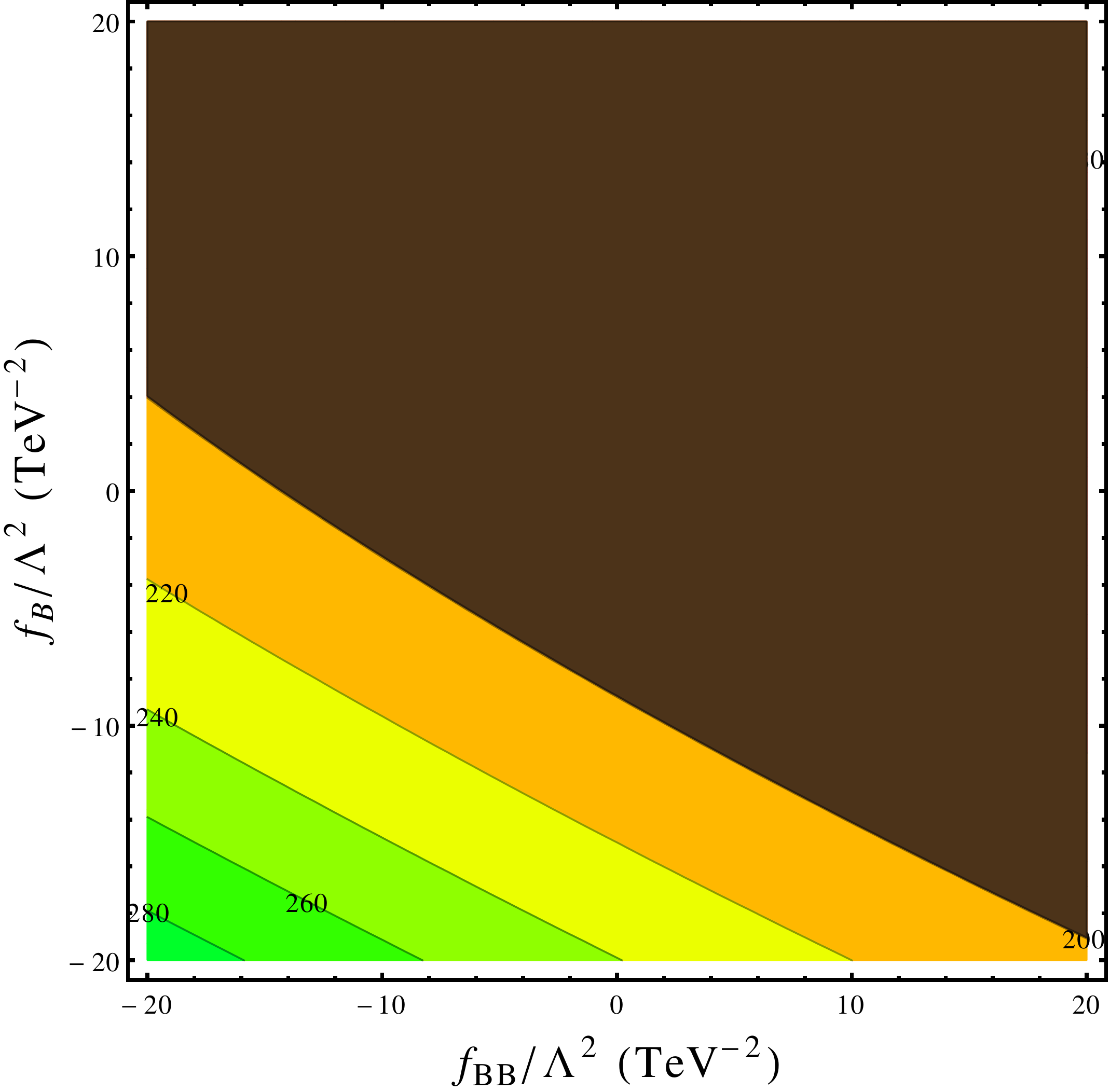}} \\
\hspace{8mm}(e)
\end{tabular}}
\caption{Variations of $\sigma_s^{300}$ for $e^+e^-\to Z h$ with (a) $\kappa$ and $f_{WW}$, (b) $\kappa$ and $f_W$,
(c) $f_{WW}$ and $f_W$ for $\kappa=1$, (d) $f_{WW}$ and $f_W$ for $\kappa=0.8$ and (e) $f_{BB}$ and $f_B$ for $\kappa=1$. For each 
case all the other $f$s are set to zeroes. Brown patches signify cross-sections within $\pm 10$\% of the SM expectation.}
\label{fig:2ds}
\end{figure}

\begin{figure}[H]
\centering
\subfloat{
\begin{tabular}{ccc}
\resizebox{65mm}{!}{\includegraphics{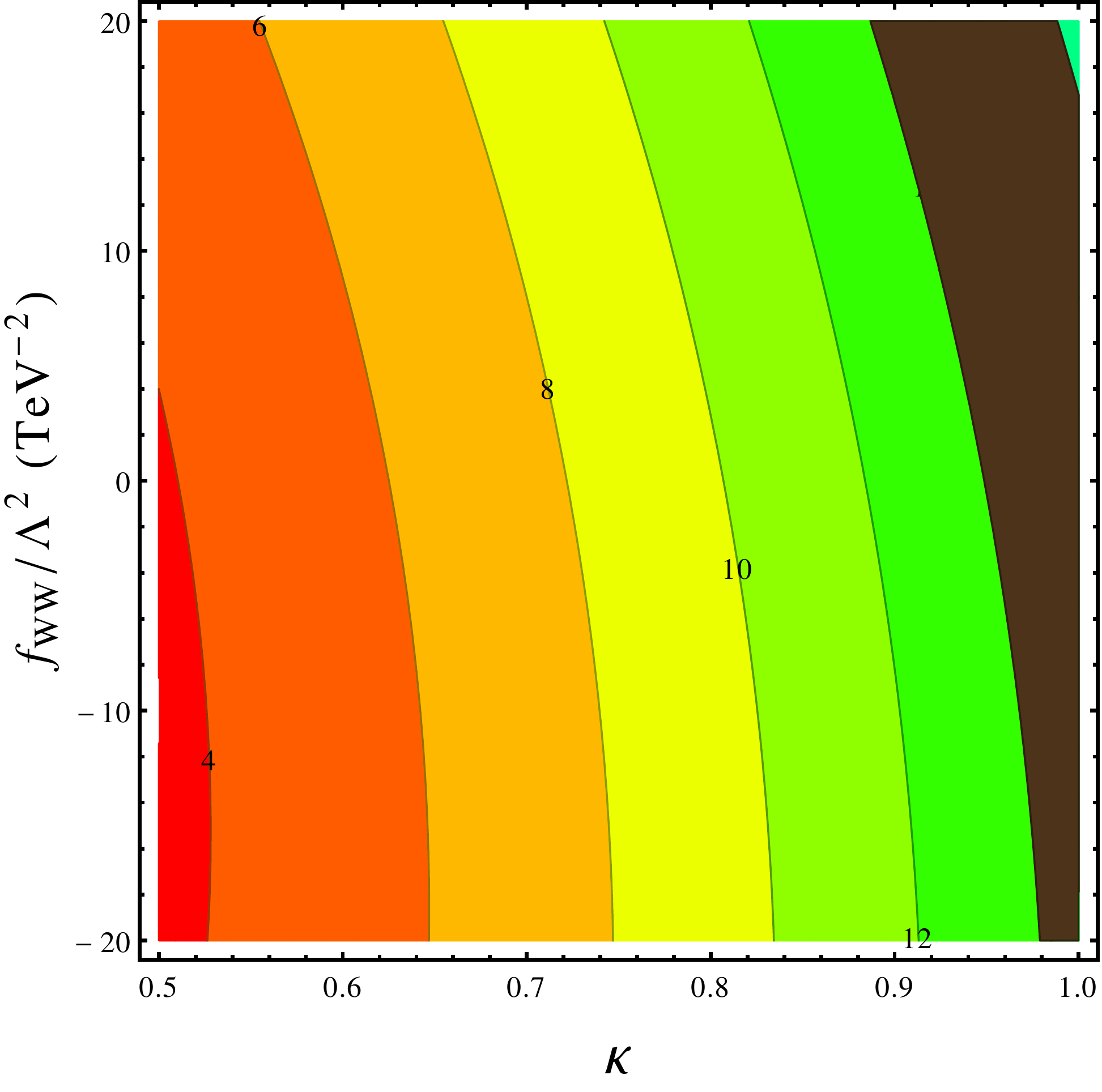}} &&
\resizebox{65mm}{!}{\includegraphics{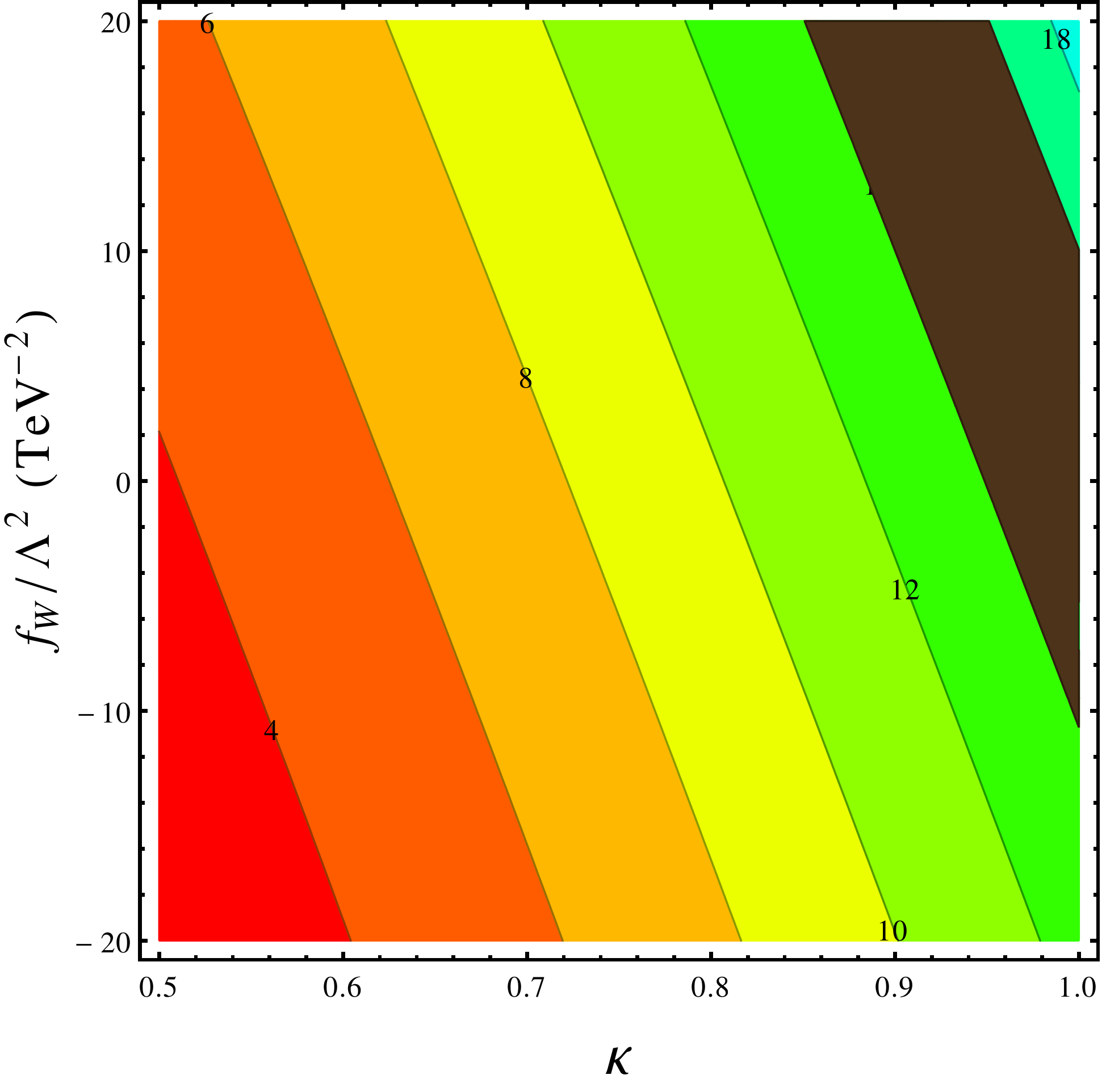}} \\
\hspace{8mm}(a)&&\hspace{20mm}(b) \\
\resizebox{65mm}{!}{\includegraphics{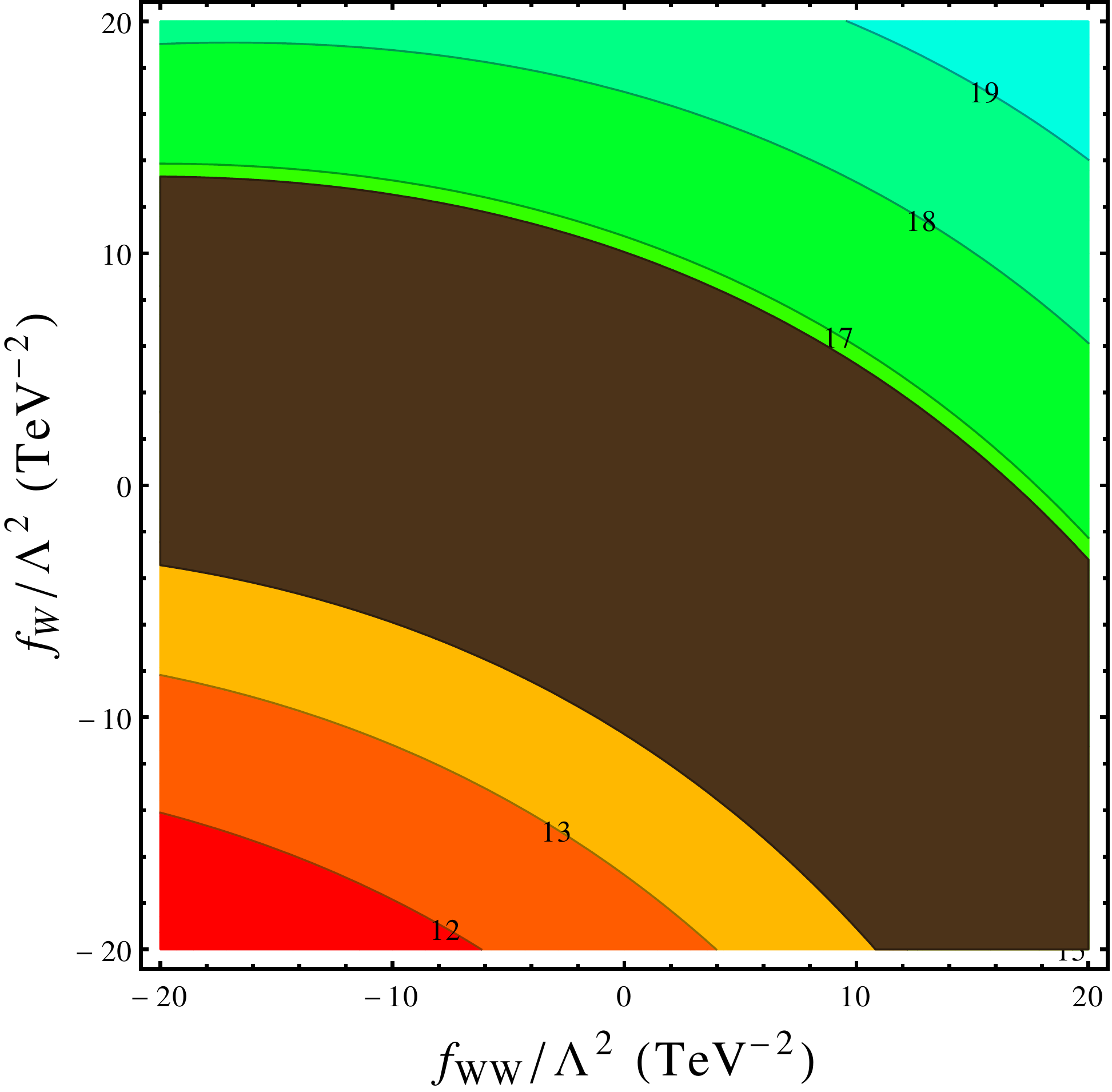}} &&
\resizebox{65mm}{!}{\includegraphics{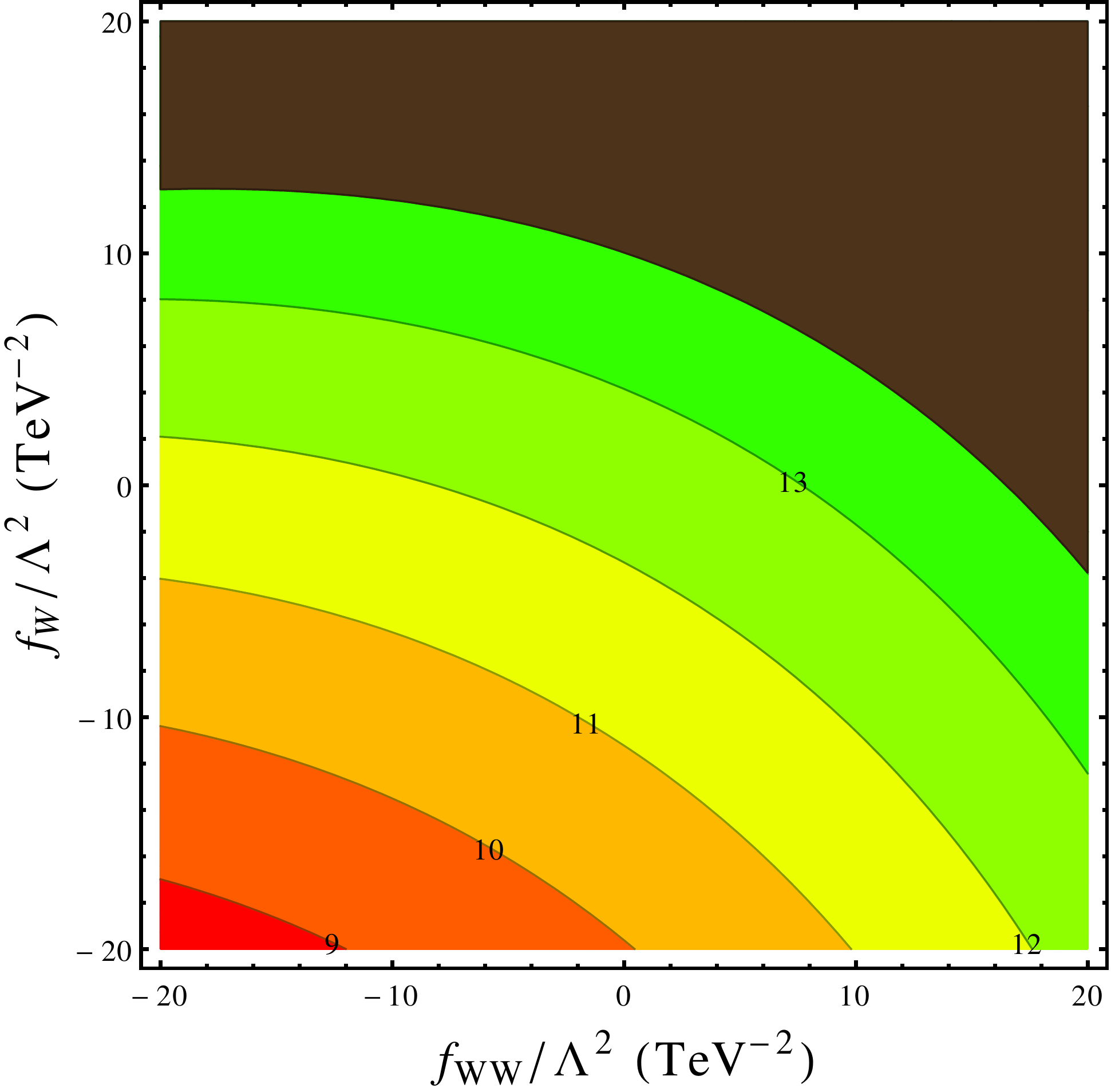}} \\
\hspace{8mm}(c)&&\hspace{20mm}(d)
\end{tabular}}
\caption{Variations of $\sigma_t^{300,ac}$ for $e^+e^-\to Z h$ with (a) $\kappa$ and $f_{WW}$, (b) $\kappa$ and $f_W$,
(c) $f_{WW}$ and $f_W$ for $\kappa=1$, (d) $f_{WW}$ and $f_W$ for $\kappa=0.9$. For each 
case all the other $f$s are set to zeroes. Brown patches signify cross-sections within $\pm 10$\% of the SM expectation.}
\label{fig:2dt}
\end{figure}

In Figs.~\ref{fig:2ds} and~\ref{fig:2dt}, we show some fixed cross-section contours in the planes of 
two parameters varied at the same time. In Figs.\ref{fig:2ds} and \ref{fig:2dt}, all the parameters apart from the ones shown in 
the axes, are kept fixed. In each of these figures, we have marked regions in brown where the cross-section is 
$\sigma(SM)\pm10\%\times\sigma(SM)$. Hence, we see that for each of these plots, some regions even with large values of the parameters 
can closely mimic the SM cross-section. The above statement for the ranges of the coefficients of the HDOs will be somewhat modified 
if we consider the Higgs decays. This is because then we will have branching ratios depending on the effects of the HDOs. Even 
for fermionic decays of the Higgs, which are independent of the operators under study, the $BR$ will have non-trivial effects on the 
operator couplings through the total decay width. But, we must mention here that unless we go to very high values of the operator 
coefficients, the total decay width remains close to the SM expectation and hence fermionic decay channels would show similar features 
as these plots. Of course, when we study the effects of all the operators in the basis that we have considered by  
considering every possible decay mode of the Higgs, then the higher-dimensional operators will come to play at the $HVV$ decay vertices 
also. Hence, we will get modified bounds on the operator coefficients from a similar approach. We should mention that these operators 
are also constrained by the electroweak precision observables, {\textit viz.} $S$, $T$ and $U$ parameters. An important observation which is carried forward from Fig.~\ref{fig:1d} (a) is that the $HZZ$ and 
$H\gamma Z$ vertices are very less affected by the operators $\mathcal{O}_{BB}$ and $\mathcal{O}_B$. This fact is corroborated in 
Fig.\ref{fig:2ds} (e). The above mentioned pair of operators thus allow a wide 
region of parameter space which has cross-sections within $10\%$ of the SM value. 

Some salient features of Figs.~\ref{fig:2ds} and ~\ref{fig:2dt} are :
\begin{itemize}
 \item Fig.~\ref{fig:2ds} shows the variation of the total rate for the channel $e^+e^-\to Z H$ as functions of two parameters taken 
 together. All the other parameters are fixed for these plots. In Figs.~\ref{fig:2ds}(a)-(d), the cross-section varies significantly from the 
 SM value for the allowed ranges of the parameters. However, Fig.~\ref{fig:2ds}(e) shows a large region of the parameter space to have 
 cross-sections similar to the SM (within $10\%$).
 \item Fig.~\ref{fig:2dt} shows the variation of the cross-sections for the $t$-channel process in $e^+e^-\to \nu \bar{\nu} H$ as 
 functions of two parameters varied at the same time. Figs.~\ref{fig:2dt}(c) and~\ref{fig:2dt}(d) shows a substantial amount of parameter space agreeing with the SM cross-section.
\end{itemize}

\subsubsection{All parameters at the same time}

\begin{figure}[H]
\centering
\subfloat{
\begin{tabular}{ccc}
\resizebox{65mm}{!}{\includegraphics{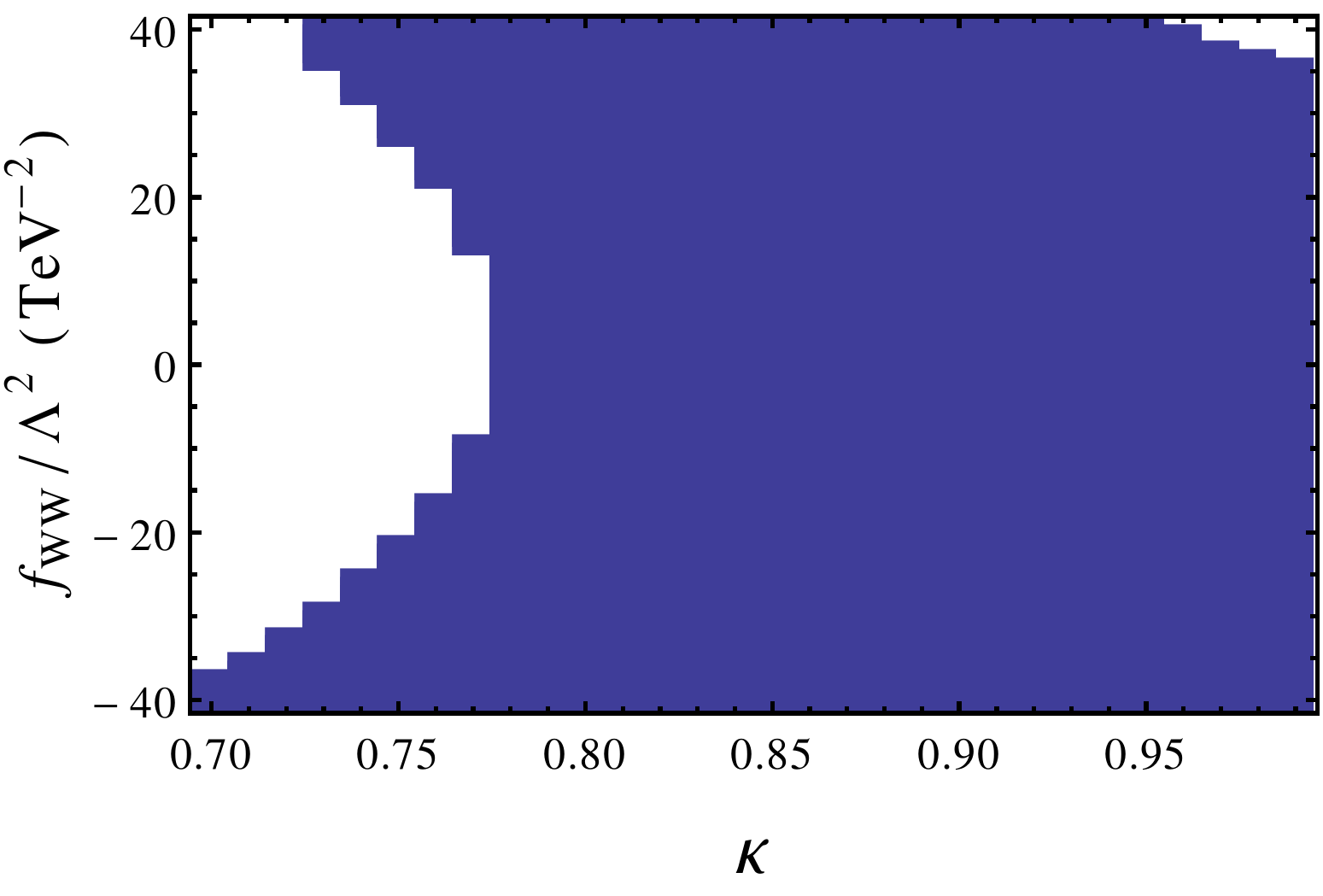}} &&
\resizebox{65mm}{!}{\includegraphics{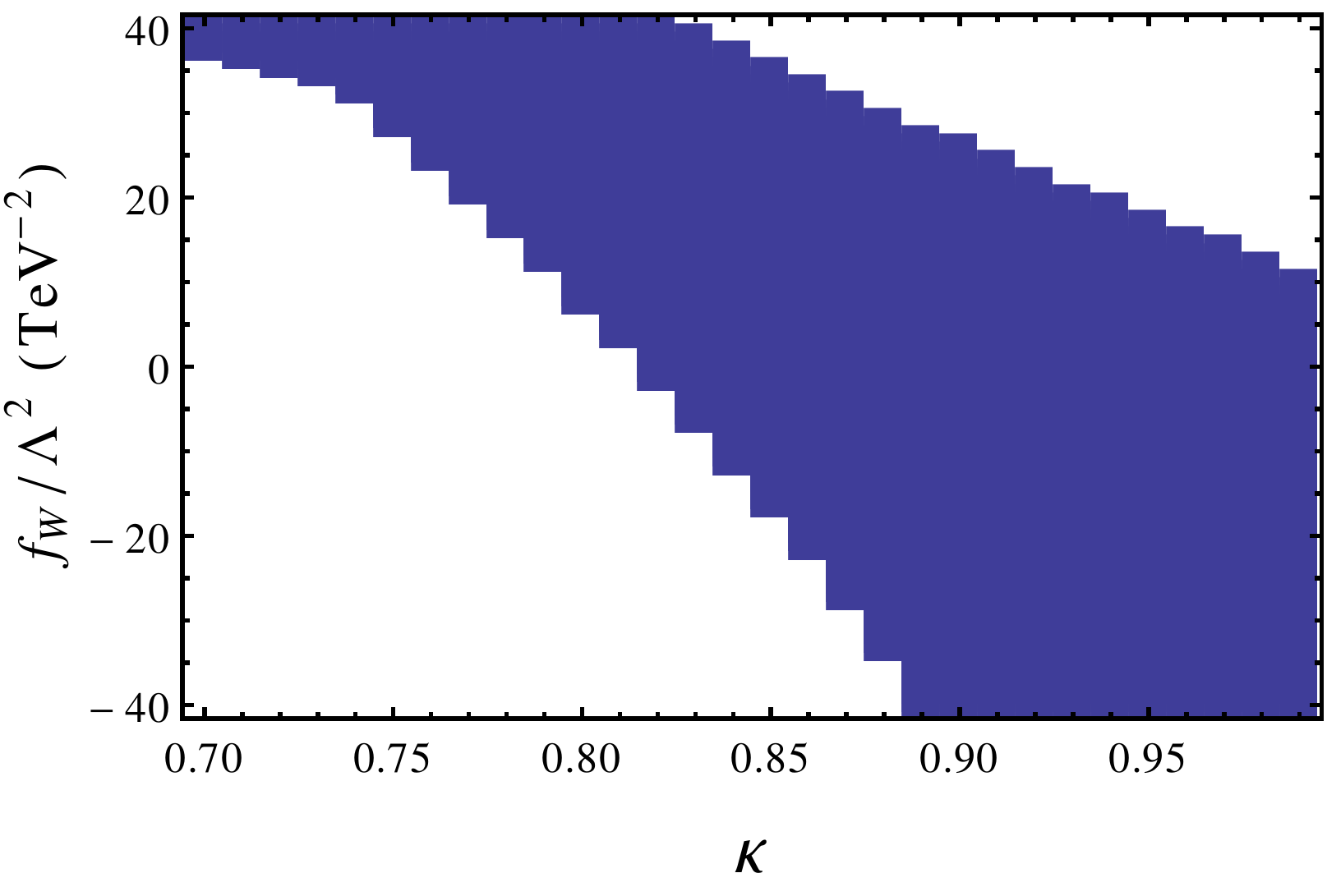}} \\
\hspace{5mm}(a)&&\hspace{15mm}(b) \\
\resizebox{65mm}{!}{\includegraphics{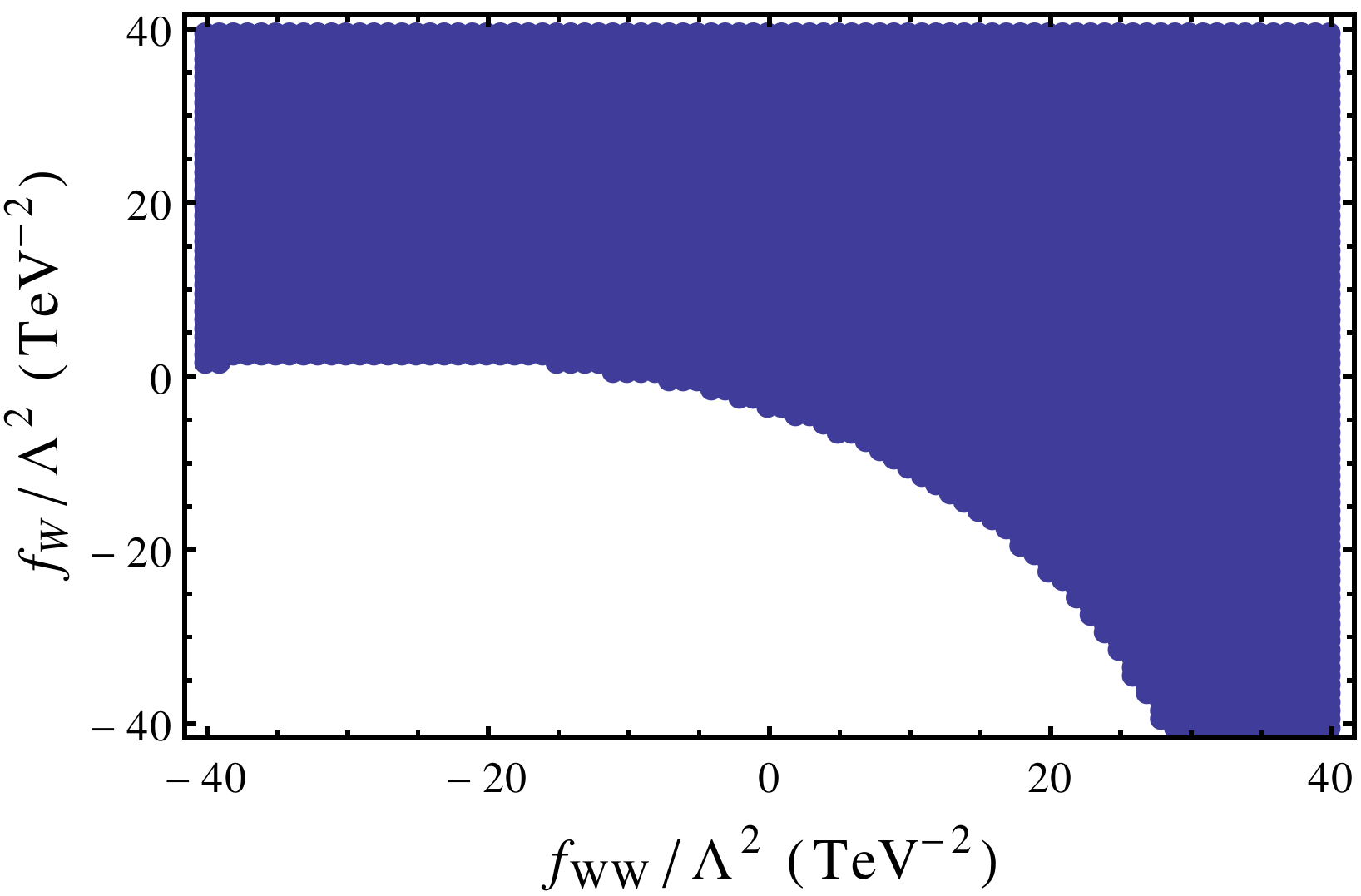}} &&
\resizebox{65mm}{!}{\includegraphics{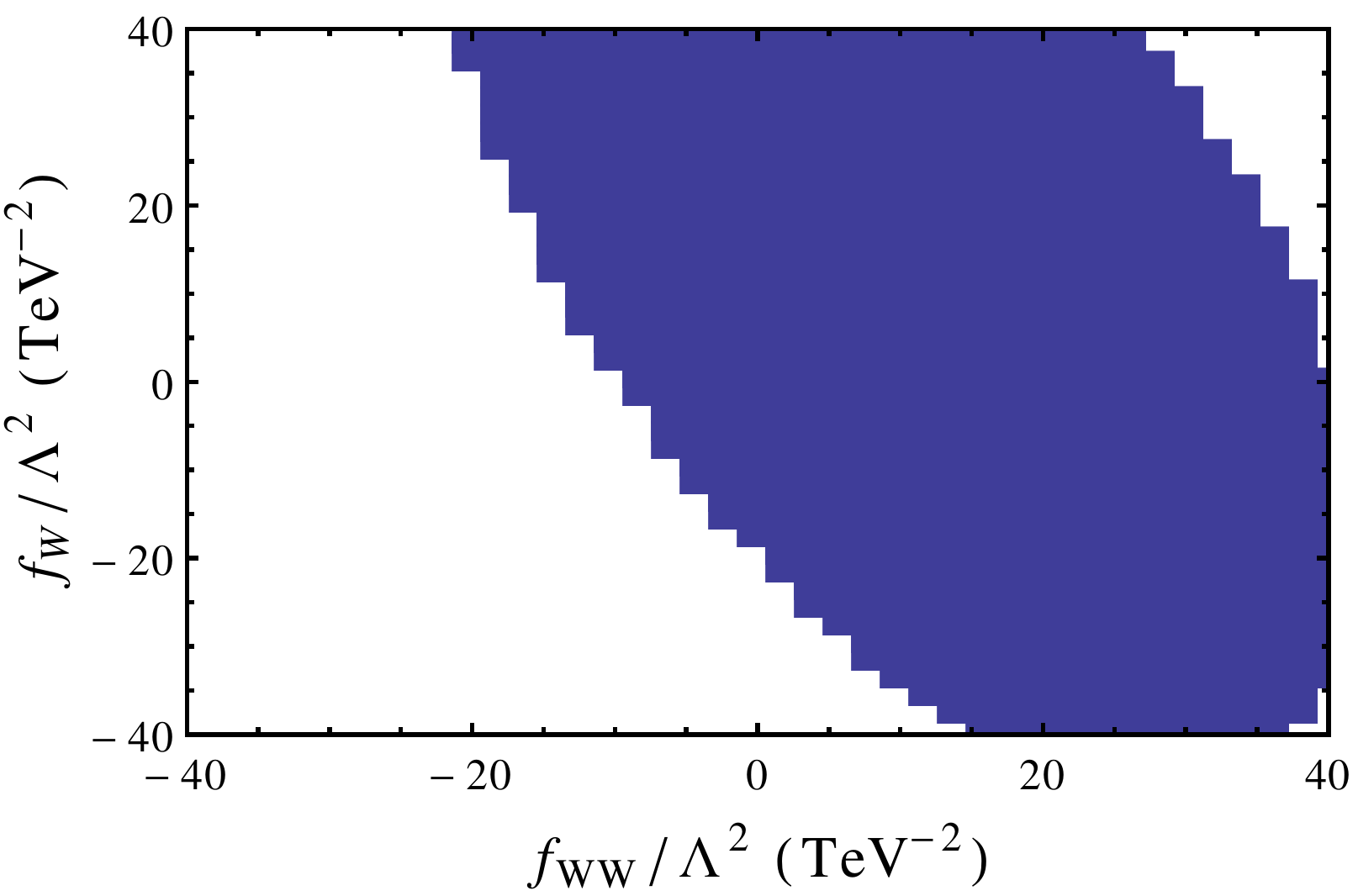}}\\
\hspace{5mm}(c)&&\hspace{15mm}(d)
\end{tabular}}
\caption{Allowed parameter space for $\sigma^{t,ac}_{\nu\bar{\nu}H}$ within $10\%$ of its SM value : (a) 
$f_{WW}$ vs $\kappa$ ($f_W$ varied) , (b) $f_W$ vs $\kappa$ ($f_{WW}$ varied),  
(c) $f_W$ vs $f_{WW}$ ($\kappa$ varied) and for $\sigma^{s}_{Z H}$ within $10\%$ of its SM value : 
(d) $f_W$ vs $f_{WW}$ ($\kappa$ , $f_{BB}$ and $f_B$ varied).
$\sqrt{s}=300$ GeV.}
\label{fig:allpar}
\end{figure}

The most general case will be to vary all the parameters simultaneously to obtain the most realistic parameter space. Here, we 
demonstrate this scenario for the cut-applied $t$-channel cross section in the $e^+e^-\to \nu \bar{\nu} H$ channel. In 
Figs.\ref{fig:allpar} (a), (b) and (c) we present three slices of the 3-dimensional hyper-surface. For each of these plots, there is a third parameter 
which has been varied. We see that a very large parameter space is allowed which can mimic the SM cross section within its $10\%$ value. Of 
course these plots are for illustrative purposes only. In Fig.~\ref{fig:allpar} (d), we have shown one such slice of the five-dimensional hyper-surface 
in the space of ($\kappa$, $f_{WW}$, $f_W$, $f_{BB}$ and $f_B$) for the $s$-channel process. 

\underline{\textbf{Discussion on EWPT constraints}} : All the benchmark points chosen throughout this
paper are consistent with all constraints available till date \cite{HD-opsc,HD-opsa}.
However, if one looks at the contour plots in Figs.~\ref{fig:2ds},~\ref{fig:2dt} and~\ref{fig:allpar}, there may exist certain points which are disfavoured by the precision constraints.


\subsection{The effects on kinematic distributions}

\begin{figure}[H]
\centering
\subfloat{
\begin{tabular}{ccc}
\resizebox{70mm}{!}{\includegraphics{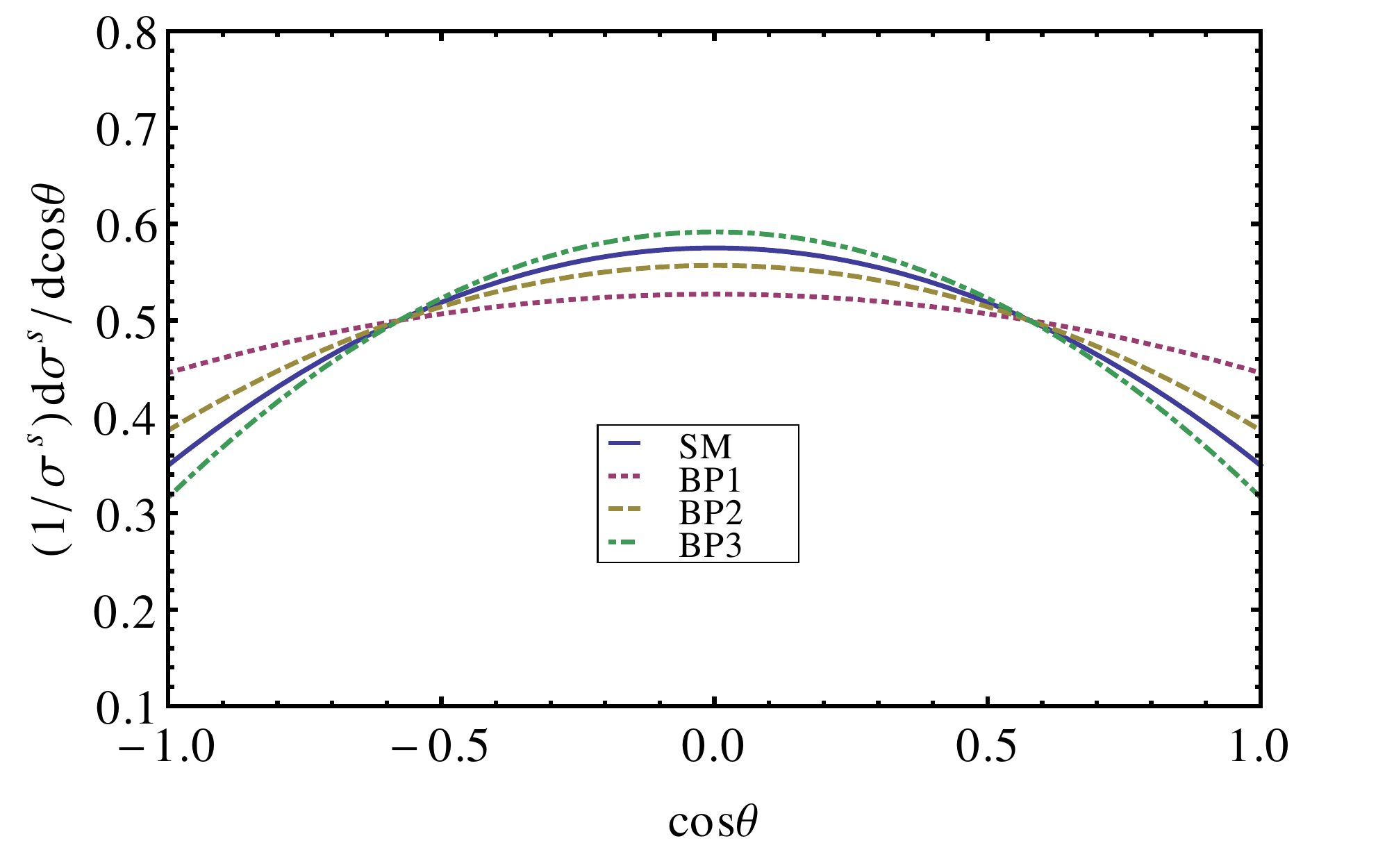}} &&
\resizebox{70mm}{!}{\includegraphics{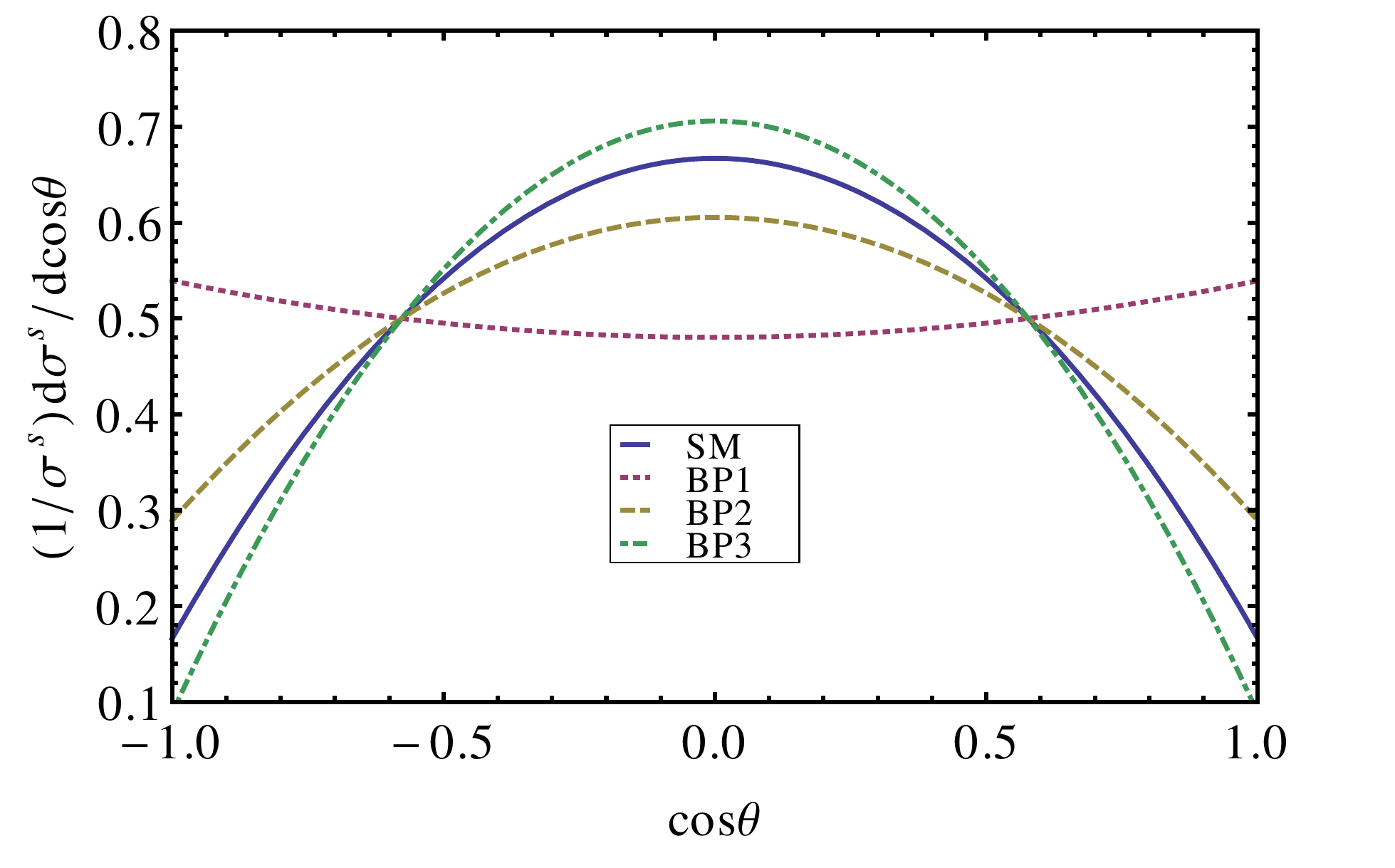}} \\
\hspace{2mm}(a)&&\hspace{8mm}(b) \\
\resizebox{70mm}{!}{\includegraphics{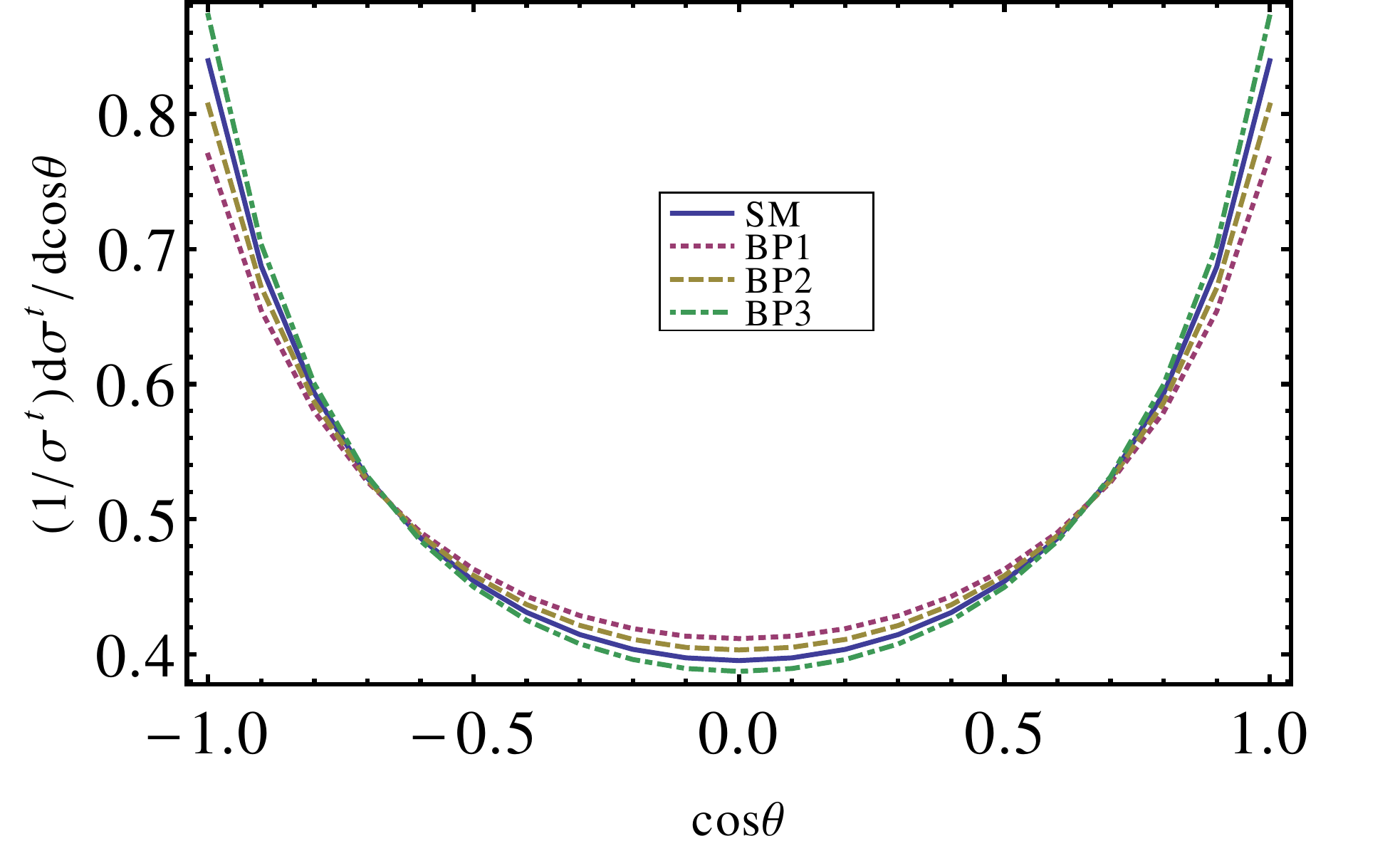}} &&
\resizebox{70mm}{!}{\includegraphics{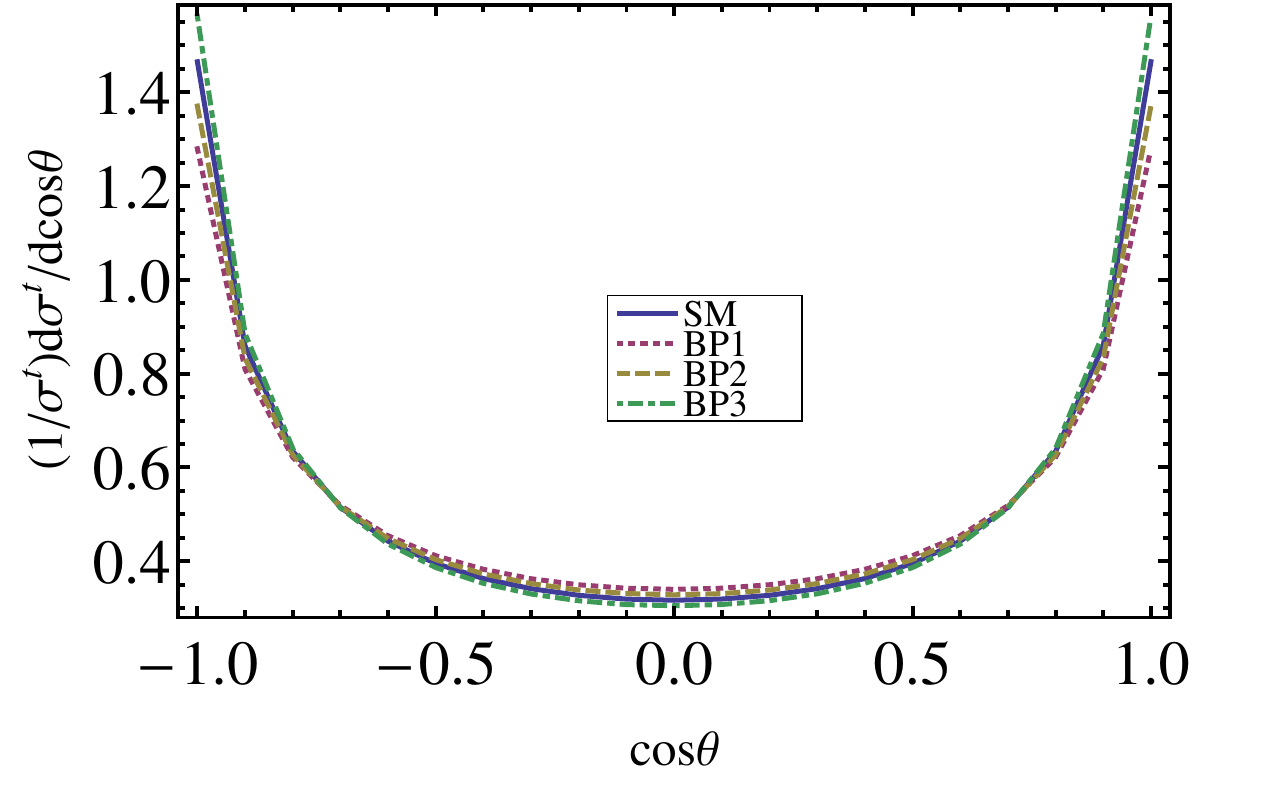}}\\
\hspace{2mm}(c)&&\hspace{10mm}(d)\\
\resizebox{70mm}{!}{\includegraphics{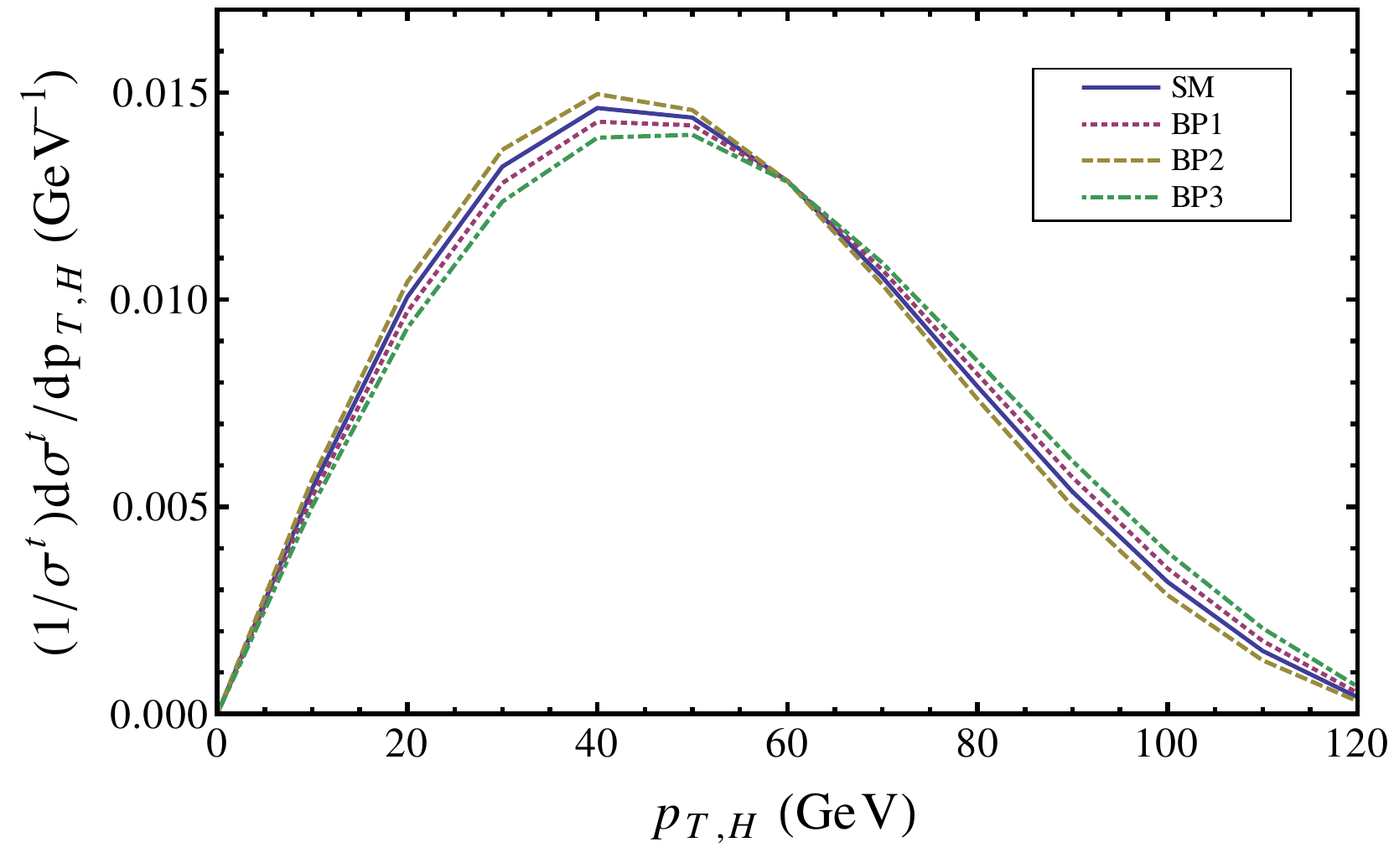}} &&
\resizebox{70mm}{!}{\includegraphics{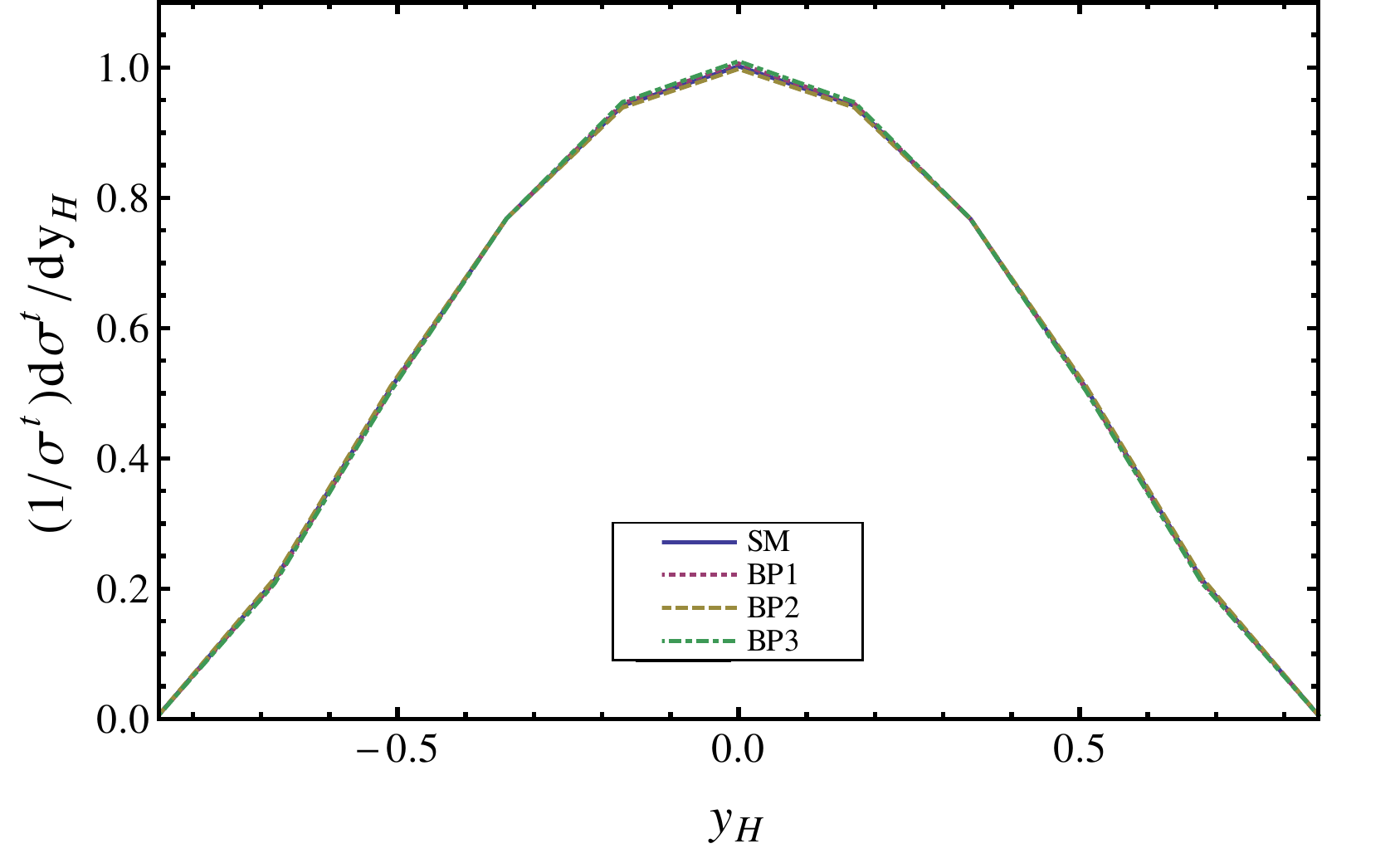}}\\
\hspace{2mm}(e)&&\hspace{10mm}(f)
\end{tabular}}
\caption{Normalised kinematic distributions $(1/\sigma^s)d\sigma^s/d\cos\theta$ for the channel $e^+e^-\to ZH$ for (a) $\sqrt{s}=300$ GeV and (b) $\sqrt{s}=500$ GeV. Normalised kinematic distributions $(1/\sigma^t)d\sigma^t/d\cos\theta$ for the $t$-channel process in $e^+e^-\to \nu \bar{\nu} H$ for (c) $\sqrt{s}=300$ GeV and (d) $\sqrt{s}=500$ GeV. Distributions for (e) $(1/\sigma^t)d\sigma^t/d p_{T,H}$ and (f) $(1/\sigma^t)d\sigma^t/d y_H$ for the $t$-channel process in $e^+e^-\to \nu \bar{\nu}H$ at $\sqrt{s}=300$ GeV. Benchmark points, {\textit viz.} SM ($x_i\in \{1,0,0,0,0\}$), BP1 ($x_i\in \{1,-3,8,-4,3\}$), 
BP2 ($x_i\in \{1,0,5,0,0\}$) and BP3 ($x_i\in \{1,0,-5,0,0\}$).}
\label{fig:KD}
\end{figure}
The presence of anomalous $HVV$ vertex can in principle also affect the
shapes of various kinematic distributions. In Figs.~\ref{fig:KD}(a)~and~\ref{fig:KD}(b)
[Figs.~\ref{fig:KD}(c) and (d)], we show the normalised angular (angle
of Higgs with the $z$-axis) distributions for the $s$-channel
($t$-channel) processes for $\sqrt{s}=300$ GeV and 500 GeV
respectively. We find that the angular dependence for the $s$-channel
is very sensitive in some regions of the parameter space allowed by
the EWPT constraints and the LHC data. We also find the $\cos\theta$
dependence can be completely opposite as we increase the CME. This can
be seen in Figs.~\ref{fig:KD}(a)~and~\ref{fig:KD}(b), if we compare
the curves for BP1. In contrast, the $t$-channel is not significantly
affected by the inclusion of HDOs. The angular dependence of the
differential cross-sections can be expressed as
\begin{equation}
\frac{d\sigma(\sqrt{s},x_i)}{d\cos\theta} = a(\sqrt{s},x_i) + b(\sqrt{s},x_i)\cos^2\theta
\end{equation}

It is found that, between coefficients $a$ and $b$ above, $a$ is more
affected by the anomalous couplings rather than $b$, unless $\sqrt{s}$
is 500 GeV or well above that. As a result, angular distributions are insensitive
to the new interactions at the proposed energy scale of a Higgs
factory.

In Figs.~\ref{fig:KD}(e)~and~\ref{fig:KD}(f), we show the normalised
$d\sigma/d p_{T,h}$ and $d\sigma/d y_h$ distributions respectively for
the $t$-channel where $p_{T,h}$ is the transverse momentum of the
Higgs and $y_h$ is its rapidity. We want to emphasise that it is very
difficult to see any significant differences in the various kinematic
distributions in most of the parameter space allowed by the LHC and
EWPT constraints while performing experiments with smaller CME. In
both the channels, we do not consider the final decay products of the
Higgs. If we consider the Higgs boson decaying to fermionic final
states, then the HDOs under consideration will not affect these decay
vertices and the above normalised distributions will remain
intact. However, if we consider the bosonic decay modes of the Higgs,
then the HDOs will affect these distributions non-trivially. 

We end this subsection with the following admission. Various
kinematical distributions are canonically emphasized as the best
places to find the signature of non-standard Lorentz structures in
interaction terms. While this expectation is not completely belied in the
present case as well, we note that the anomalous couplings are
reflected in distributions {\em at relatively high CMEs.} The reason behind this has already been explained
above. While this prospect is encouraging, electron-positron
colliders, especially those designed as Higgs factories, are likely to
start operating at energies as low as $250-300$ GeV. Our observation
is that the imprint of anomalous couplings can be found even at such
low energies at the level of total rates and their ratios. A detailed
study involving all possible decay products and their various
correlations can in principle go further in revealing traces of
anomalous couplings. We will take up such a study in a subsequent
work.

\subsection{Discussion on relevant backgrounds}

We wish to see the effects of anomalous $HVV$ couplings on the Higgs
production alone.  Therefore, we do not look at bosonic decay modes of
Higgs and limit our discussion only to those signal processes where
$H$ decays maximally to a $b\bar{b}$ pair. For the $e^+e^-\to ZH$
process, the $Z$ can either decay visibly to $b\bar{b}$, $jj$,
$\ell^+\ell^-$ (here $j=g,u,d,c,s$ and $\ell=e,\mu$) modes or
invisibly to a $\nu\bar{\nu}$ pair.  So the dominant backgrounds
relevant for these final states are the non-Higgs $e^+e^-\to
b\bar{b}b\bar{b},b\bar{b}jj,b\bar{b}\ell^+\ell^-,b\bar{b}+\cancel{E}$.
The non-Higgs $e^+e^-\to b\bar{b}+\cancel{E}$ process can also act as
the dominant background for the $e^+e^-\to \nu\bar{\nu}H$ channel. We
select events after the following kinematic cuts:

Trigger cuts : $p_T(b,j) > 20$ GeV, $p_T(\ell) > 10$ GeV, $|y(b,j)|<
5.0$, $|y(\ell)|< 2.5$, $\Delta R(bb,bj,jj,b\ell,j\ell) > 0.4$,
$\Delta R(\ell\ell) > 0.2$.

Finally we estimate two of the aforementioned backgrounds by applying the cuts below:
\begin{itemize}

\item \underline{Non-Higgs $e^+e^-\to bb\ell\ell$}

We demand the two $b$'s to fall within the Higgs-mass window and the two $\ell$'s to 
fall within the $Z$-mass window as follows:
\begin{equation}
|M(bb) - M_h| < 10~\textrm{GeV~~AND}~~|M(\ell\ell) - M_Z| < 10~\textrm{GeV}
\end{equation}
Finally the total background cross-section for the $bb\ell\ell$ final
state is defined as, $\mathcal{B}_{bb\ell\ell} =
\eta_b^2~\sigma_{bb\ell\ell}$ where $\eta_b$ is the $b$-tagging
efficiency which we take as $0.6$ for our analysis. The signal is also
scaled by the same factor, $\eta_b^2$.

\item \underline{Non-Higgs $e^+e^-\to bb+\cancel{E}$}

We demand the two $b$'s to fall within the Higgs-mass window, $|M(bb) - M_h| < 10$ GeV. Here the background is 
$\mathcal{B}_{bb+\cancel{E}} = \eta_b^2~\sigma_{bb+\cancel{E}}$. The signal\footnote{The channel $e^+e^-\to H + \cancel{E} \to b\bar{b} + \cancel{E}$ also includes diagrams involving the triple-gauge boson vertices. These 
effects are almost nullified when the selection cuts for this channel are employed.} has also been scaled by the $b$-tagging 
efficiency.
\end{itemize}

\begin{table}[H]
\centering
\begin{tabular}{|c|c|c|c|c|c|c|c|}
\hline
Final states       & $\sqrt{s}$ & $\sigma_{SM,tc}^{sig}$ & $\sigma_{SM,ac}^{sig}$ & $\sigma_{BP1,tc}^{sig}$ & $\sigma_{BP1,ac}^{sig}$ & $\sigma_{tc}^{bkg}$ & $\sigma_{ac}^{bkg}$ \\
                   & (GeV)      & (fb)                   & (fb)                   & (fb)                    & (fb)                    & (fb)                & (fb) \\
\hline
$b\bar{b}l^+l^-$   & 250        & 2.68                   & 2.46                   & 2.76                    & 2.52                    & 10.33               & 0.09 \\
                   & 300        & 2.33                   & 1.91                   & 2.31                    & 1.83                    & 9.17                & 0.07 \\
\hline
$b\bar{b}+\cancel{E}$ & 250     & 12.25                  & 10.31                  & 12.36                   & 10.53                   & 20.53               & 0.33 \\
                      & 300     & 13.67                  & 9.79                   & 13.26                   & 9.62                    & 18.00               & 0.29 \\ 
\hline
\end{tabular}
\caption{\label{tab:bkg} We show the signal and backgrounds for two
  different final states, {\textit{viz.}} $b\bar{b}l^+l^-$ and
  $b\bar{b}+\cancel{E}$. $\sigma_{tc}$'s are the cross-sections after
  the basic trigger cuts mentioned above and $\sigma_{ac}$'s are the
  cross-sections after the channel-specific cuts. The analysis has
  been done for the SM and the benchmark point BP1 ($x_i\in
  \{1,-3,8,-4,3\}$).}
\end{table}

\begin{figure}
\centering
\subfloat{
\begin{tabular}{ccc}
\resizebox{70mm}{!}{\includegraphics{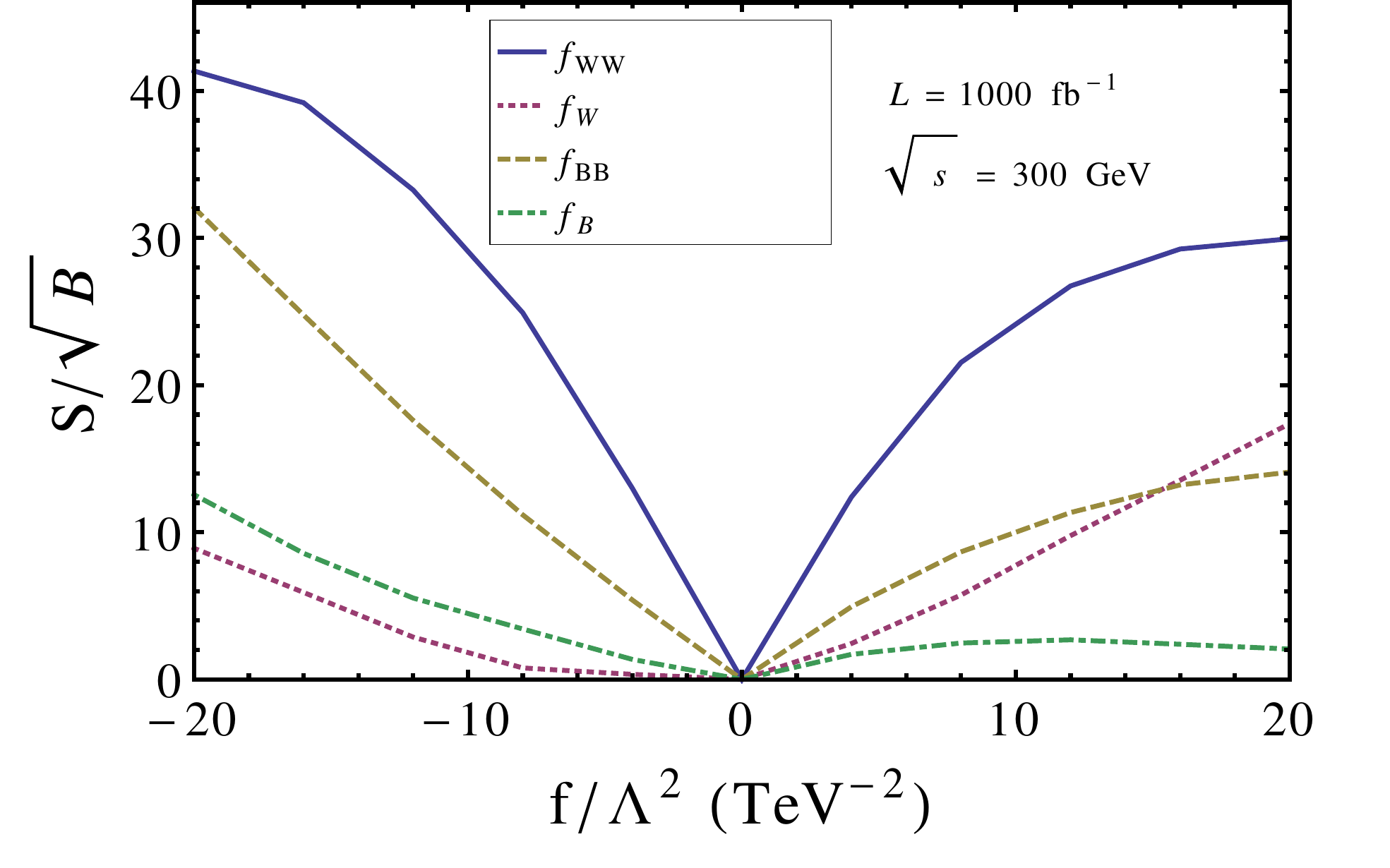}} &&
\resizebox{70mm}{!}{\includegraphics{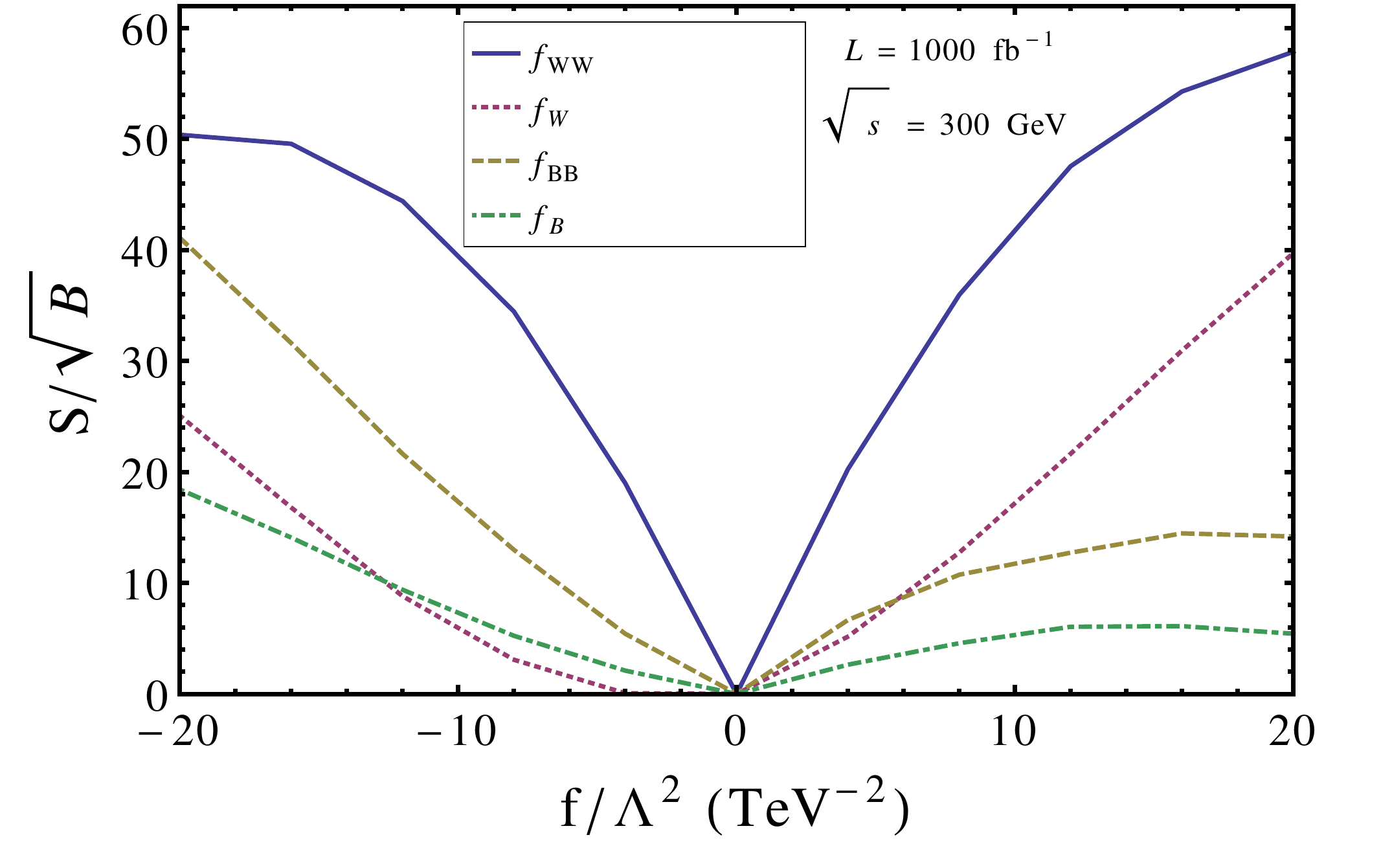}} \\
\hspace{0mm}(a)&&\hspace{4mm}(b)
\end{tabular}}
\caption{Significance ($\mathcal{S}/\sqrt{\mathcal{B}}$) as functions of $f_i/\Lambda^2$ for $\kappa=1$ at $\sqrt{s} = 300$ GeV for (a) $e^+e^-\to bb\ell\ell$ and (b) $e^+e^-\to bb+\cancel{E}$.}
\label{fig:signi}
\end{figure}

Alongside the issue of distinctness of the presence of
  the anomalous couplings, it is of interest to find out about the
  reach of a Higgs factory, or to know down to what
  strength the anomalous couplings can be detected. This information
 can be found in Fig.~\ref{fig:signi}. 
  There we have have plotted the quantities $\mathcal{S} =
  |\sigma_{BSM}^H - \sigma_{SM}^H|$ and $\mathcal{B} = \sigma_{SM}^H +
  \sigma_{SM}^{NH}$ for computing the significance. Here, $H$ ($NH$)
  signifies sub-processes  which involve (does not involve) the Higgs.

In Table~\ref{tab:bkg}, we show the cross-sections for both the signal
and background scenarios. For the signal we have considered two
benchmark points, {\textit{viz.}} SM and BP1 ($x_i\in
\{1,-3,8,-4,3\}$)). We show the cross-sections once after applying
just the trigger cuts (designated with the subscript $tc$) and next by
applying the channel-specific selection cuts (written with a subscript
$ac$) along with the basic trigger cuts. All the numbers have been
multiplied by $\eta_b^2$. We see that the effects of the invariant
mass selection cuts on the signal cross-sections are negligible whereas
these are very effective in reducing the backgrounds almost
completely. 

The study performed here is at parton level. Shower, hadronization and 
detector effects are expected to have an impact on the 
effective cross-sections reported in Table~\ref{tab:bkg}. That said,  
these effects will not change the conclusions of the paper.

\section{\label{sec:likelihood}Likelihood Analysis for $t$-channel}

The kinematics of the final state associated to the $s$-channel production has been studied extensively in the past. As pointed out in section~\ref{sec:intro}, the $t$-channel production provides limited phase-space because the momenta of the outgoing neutrinos cannot be disentangled experimentally. This leaves the Higgs boson kinematics as the only handle to explore the nature of the $HWW$ coupling. Studies are documented in the literature with the use of the Higgs boson momentum as a means to gain sensitivity. Here we  attempt to fully exploit the kinematics of the Higgs boson by means of a correlated two-dimensional likelihood analysis. The primary intent of this section is to shed light on the relative improvement of this two-dimensional approach, rather than determining absolute sensitivity to the size of anomalous couplings. The latter requires a detailed study that carefully incorporates experimental effects. This is beyond the scope of this paper.

We use a test-statistic (TS) to distinguish the BSM hypothesis from
its SM counterpart by defining the logarithm of a profile likelihood
ratio ($q_{ij}=\ln\lambda_{ij}$) for two different hypotheses $i$ and
$j$ defined as
\begin{equation}
q_{ij}=\ln\lambda_{ij}=\ln\frac{L(P_i|D_i)}{L(P_j|D_i)},
\end{equation}
where $\lambda_{ij}$ is the ratio of two likelihood functions
$L(P_i|D_i)$ and $L(P_j|D_i)$ describing two different
hypotheses \footnote{Alternatively, its reciprocal is also sometimes
  used, depending on the analysis required. It should be noted here
  that both likelihoods are constructed using the same $D_i$, but
  different $P_i$s.}, $D_i$ is the data set used and $P_{i,j}$ are the
probability density functions. Due to the discrete nature of the
probabilities in this analysis, the likelihood functions are defined
as products of binned Poisson probabilities over all channels and bins
\cite{ATLAS}. From the TS, a $p$-value can be calculated to quantify
the extent to which a hypothesis can be rejected. In general, a
$p$-value is a portion of the area under a normalised TS which, after
calculation, is the percentage confidence level (CL) by which a
hypothesis can be rejected.

In Monte Carlo (MC) studies, these TSs emerge as binned peaks which
show up on running pseudo-experiments, each of which returns a value
for the TS based on a randomly generated set of pseudo-data. The
number of pseudo-data points generated is fixed by the cross-section
of the process being studied. The TSs concerned in this analysis are
always produced in pairs, in order to discriminate between the SM and
BSM hypotheses. This pair of TSs is represented as
\begin{equation}
\label{eqn:ts1}
q_U=\ln\frac{L(P_{SM}|D_{SM})}{L(P_{BSM}|D_{SM})}~~~~\textrm{and}~~~~
q_L=\ln\frac{L(P_{SM}|D_{BSM})}{L(P_{BSM}|D_{BSM})}.
\end{equation}
The $q_U$ TS tends to have a more positive value due to its ordering,
and we refer to it as the \textit{upper} TS for our purposes, while
we refer to $q_L$ as the \textit{lower} TS.
A hypothesis can be rejected by calculating the associated $p$-value as follows 
\begin{equation}
\label{eqn:ts}
p=\int_{m_{q_U}}^{\infty}q_{L}(q) dq,
\end{equation}
where $m_{q_U}$ is the median of the upper TS, $q_{U}$.  The
confidence by which a hypothesis can be rejected, can alternatively be
quantified by knowing the \textit{significance} of the separation
between the two TSs. The median-significance, $Z_{med}$, is defined as
the number of standard deviations between the median of $q_L$ and the
left edge of the $p$-value area, that is, the median of $q_U$.
\begin{figure}[H]
\centering
\subfloat{
\begin{tabular}{ccc}
\resizebox{70mm}{!}{\includegraphics{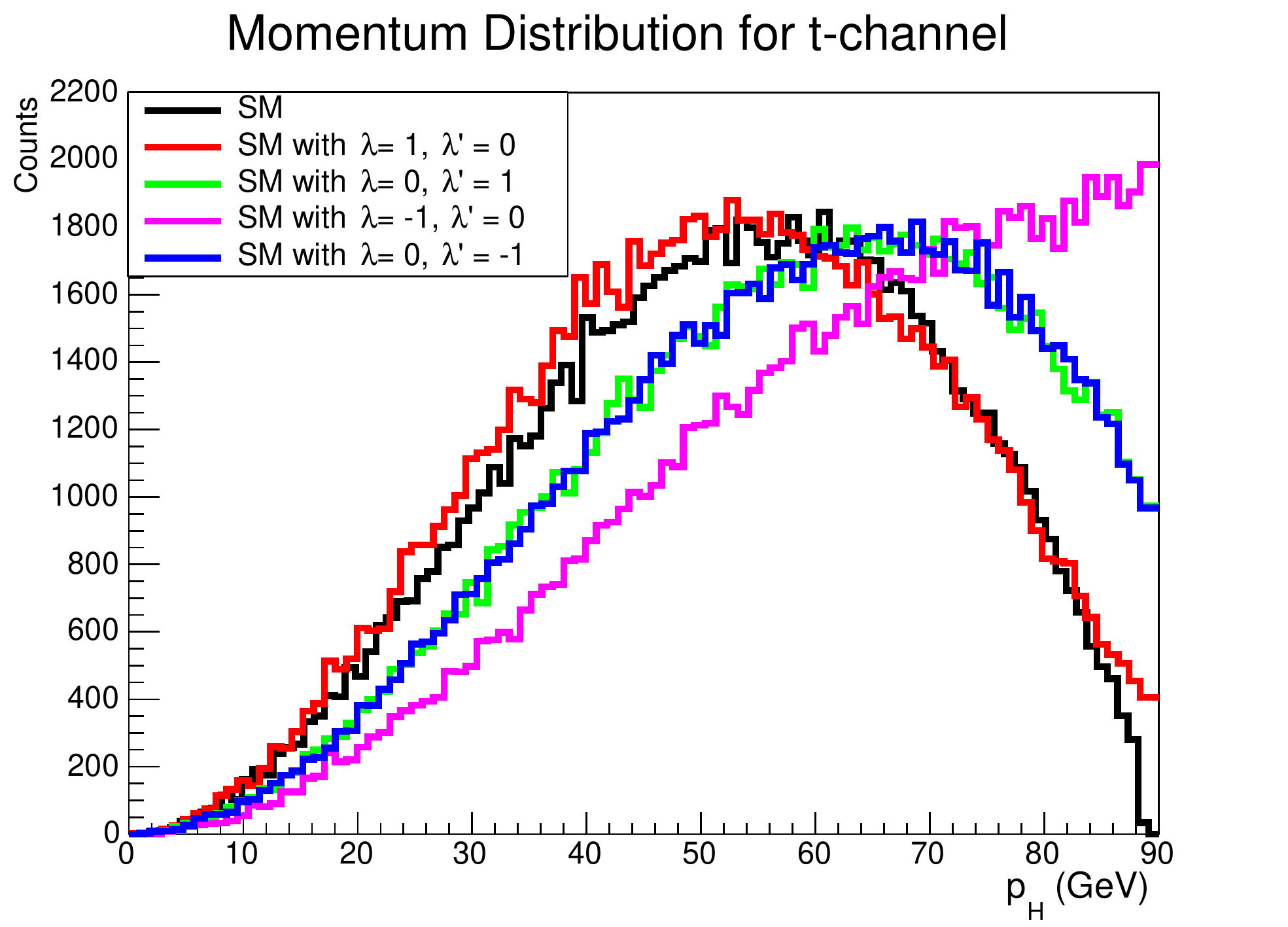}} &&
\resizebox{70mm}{!}{\includegraphics{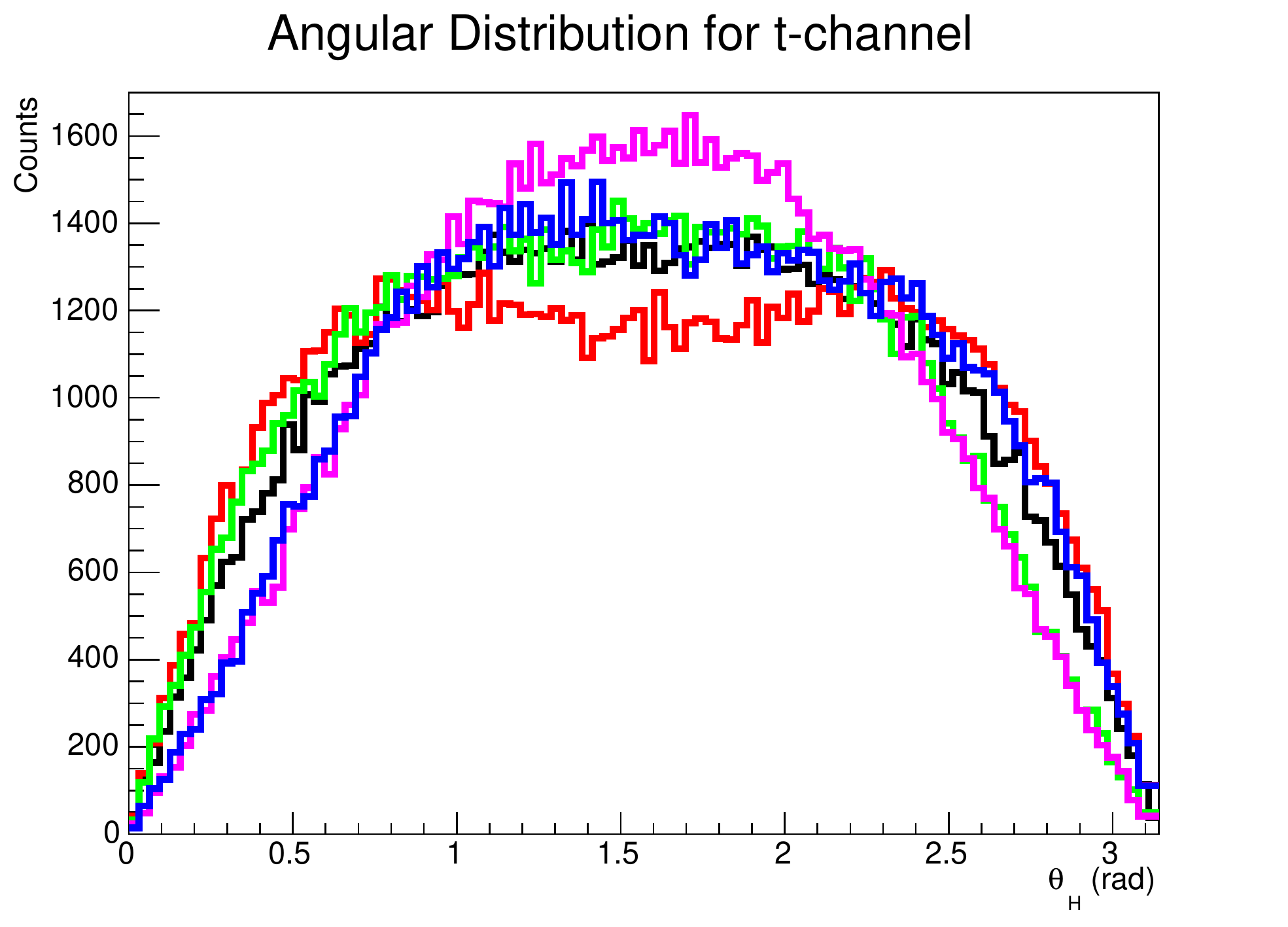}} \\
\hspace{8mm}(a)&&\hspace{10mm}(b)
\end{tabular}}
\caption{Normalised kinematic distributions of (a) Higgs momentum,
  $p_H$ and (b) the angle of the Higgs with the beam-axis, $\theta_H$
  for different benchmark points for the $t$-channel process at
  $\sqrt{s}=250$ GeV.}
\label{img:t}
\end{figure}

\begin{figure}[H]
\centering
\subfloat{
\begin{tabular}{ccc}
\resizebox{70mm}{!}{\includegraphics{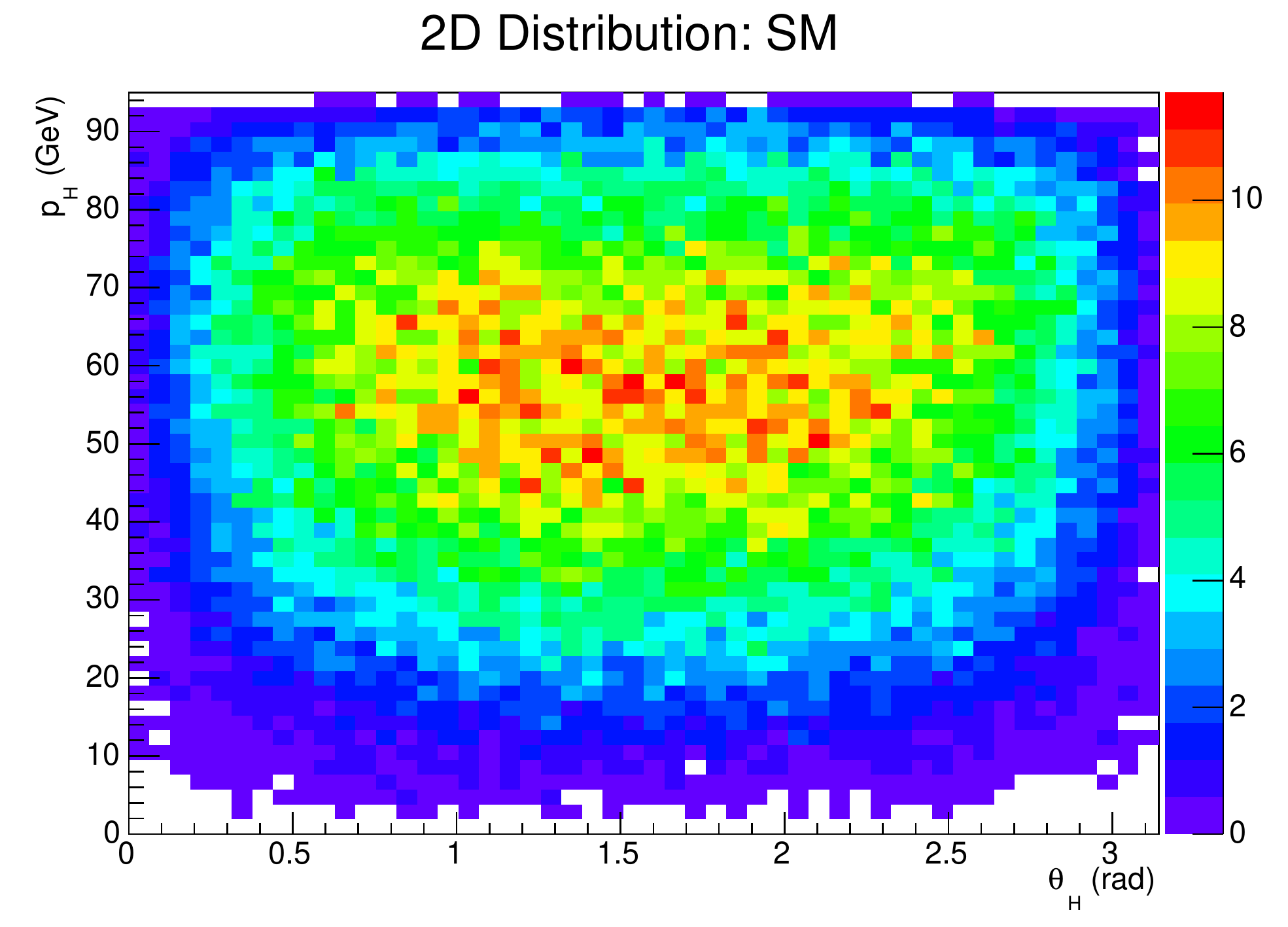}} &&
\resizebox{70mm}{!}{\includegraphics{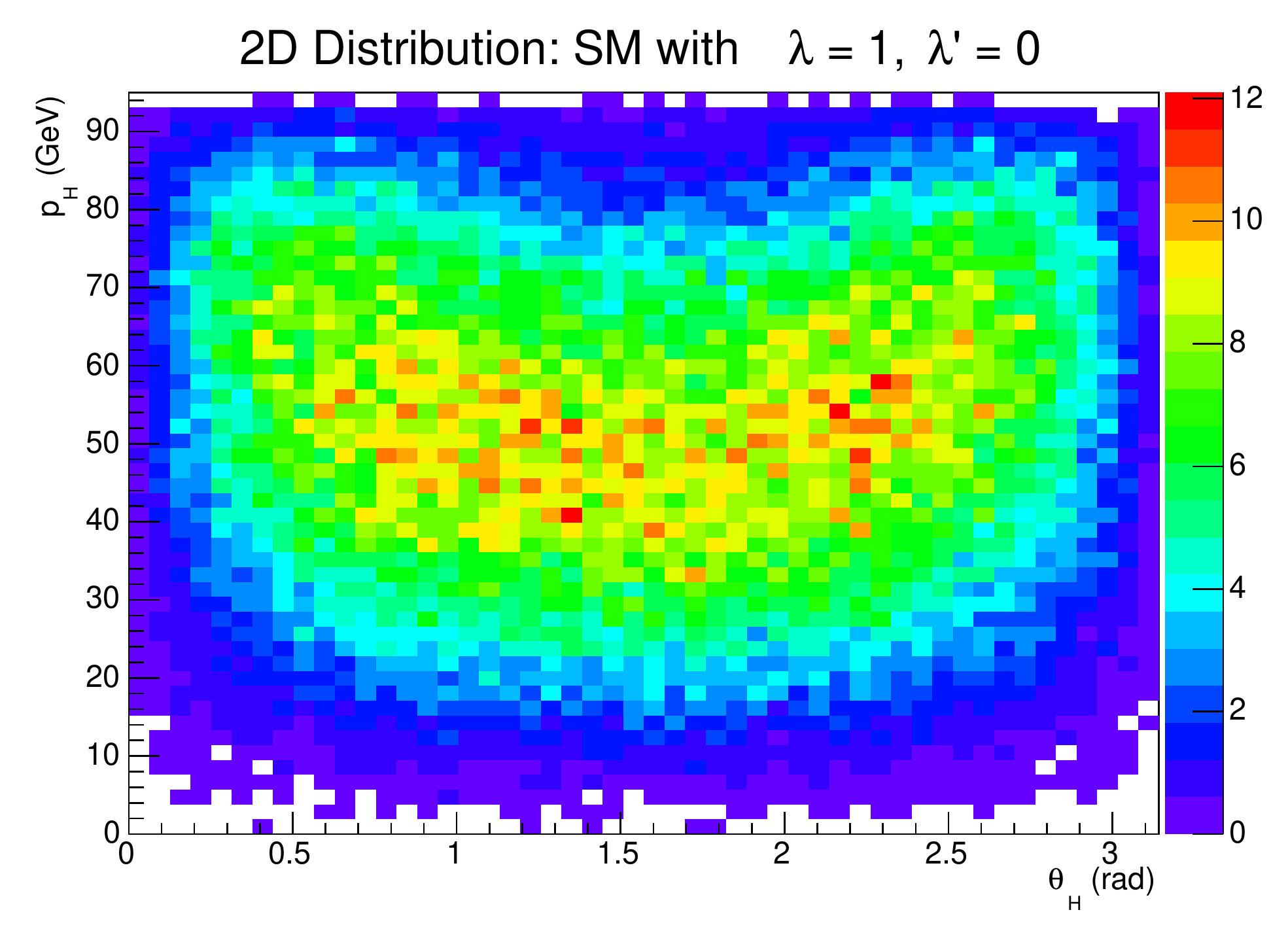}} \\
\hspace{8mm}(a)&&\hspace{10mm}(b)
\end{tabular}}
\caption{Two dimensional histograms showing the correlation of the
  $t$-channel Higgs momentum, $p_H$ and the angle of the Higgs with
  the beam-axis, $\theta_H$ at $\sqrt{s}=250$ GeV. The $z$-axis is an
  indication of the frequency of events, in arbitrary units. The
  effect of the correlation can be seen by noting how the BSM
  parameter $\lambda$ affects the distribution.}
\label{img:t-2D}
\end{figure}

As stated above, we focus on the $t$-channel process (in $e^+e^-\to \nu\bar{\nu} H$)
which has not been studied as extensively as the $s$-channel. The
$s$-channel ($t$-channel) contributions can be separated out from the
$\nu\bar{\nu}H$ events by applying the $E_H$-cut ($E_H^c$-cut) in
Eq.~\ref{eq:stcut}. For this purpose, we work with the
phenomenological parametrization of anomalous $HWW$ interaction
characterised by $\lambda$ and $\lambda'$, as defined in
Eq.~\ref{eq:LIP}.

In our analysis, the vertices for the Lagrangians in the SM and in BSM with
spin-0 bosons are calculated  in {\scshape{FeynRules}}~\cite{Alloul:2013bka} and
passed to the event-generator {\scshape{MadGraph}}~\cite{Alwall:2011uj}, which is used for the
generation of  the matrix elements for Higgs production in the $t$- and $s$-channels. MC samples are produced at parton level. Effects related to detector resolution are taken into account when defining requirements to suppress the contamination from the $s$-channel process (see Eq.~\ref{eq:stcut}).

We set the stage for the likelihood analysis by showing some plots for
distributions in terms of $\lambda$ and $\lambda'$. In
Figs.~\ref{img:t}(a) and (b), we show the $p_H$ (Higgs momentum) and
$\theta_H$ (the angle of the Higgs with the beam-axis) distributions
respectively for the $t$-channel at $\sqrt{s} = 250$ GeV. We see that
significant deviations from the SM can be seen. This is in contrast to
what was shown for the gauge invariant formulation (in
Fig.~\ref{fig:KD}) because there we stick to moderate values of the
parameter coefficients, whereas for example, here, $\{\lambda=
1,\lambda'=0\} \Rightarrow x_i \approx \{1,77,0,0,0\}$). In
Figs.~\ref{img:t-2D}(a) and (b), two dimensional histograms in
$p_H$-$\theta_H$ plane are shown for the SM and a BSM (SM with
$\lambda=1$, $\lambda'=0$) benchmark point respectively at
$\sqrt{s}=250$ GeV.

\begin{figure}[H]
\centering
\subfloat{
\begin{tabular}{ccc}
\resizebox{70mm}{!}{\includegraphics{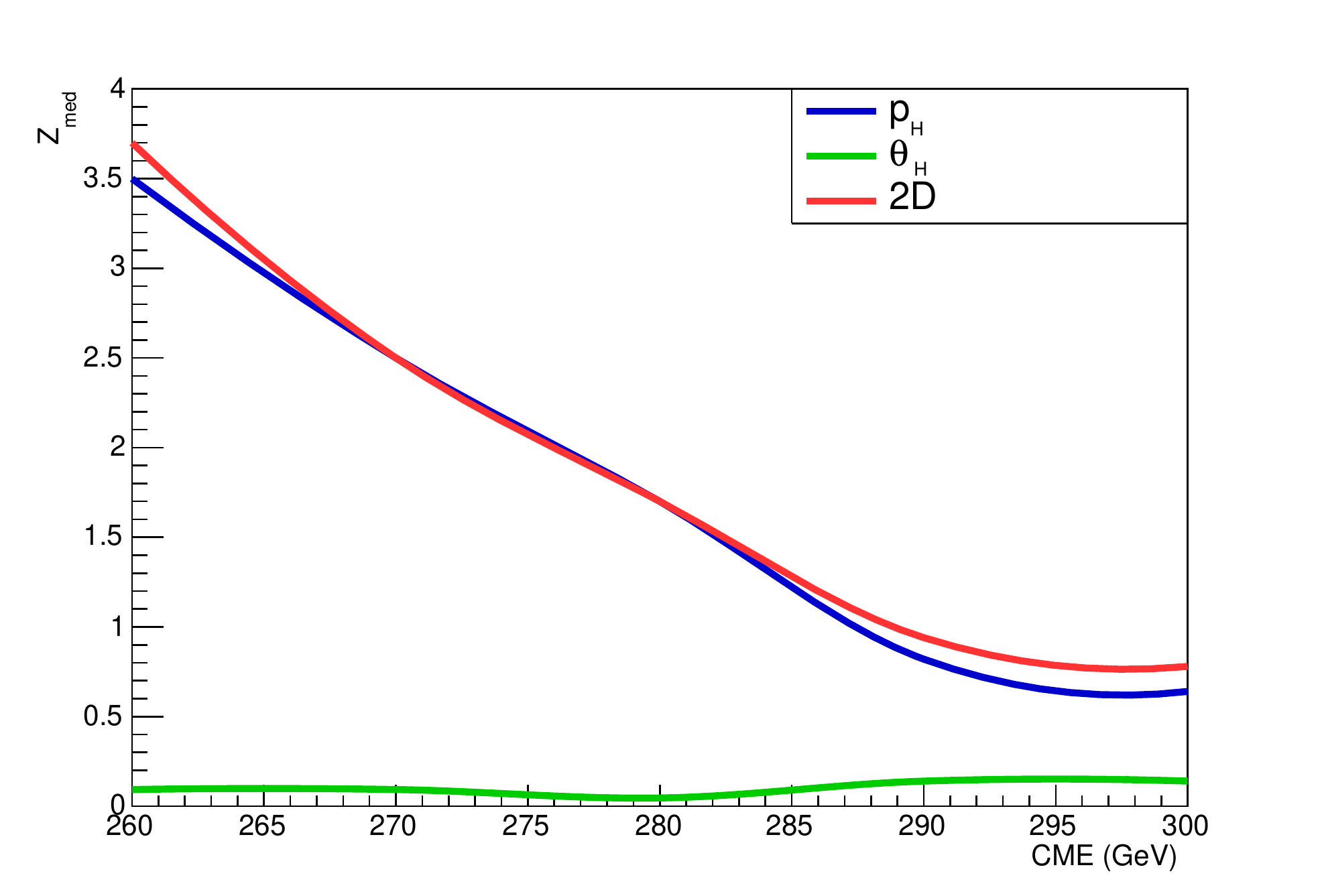}} &&
\resizebox{70mm}{!}{\includegraphics{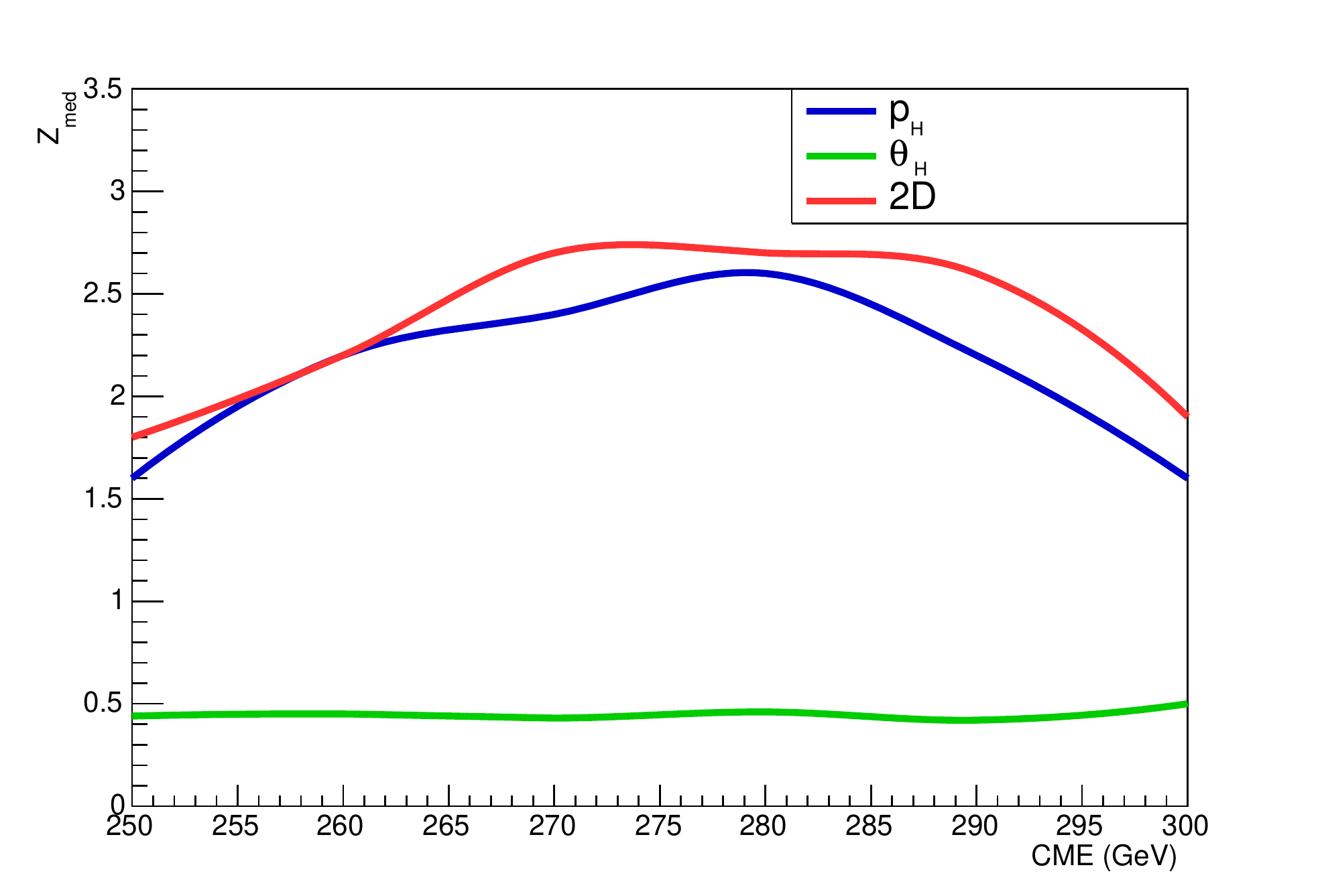}} \\
\hspace{0mm}(a)&&\hspace{4mm}(b) \\
\resizebox{70mm}{!}{\includegraphics{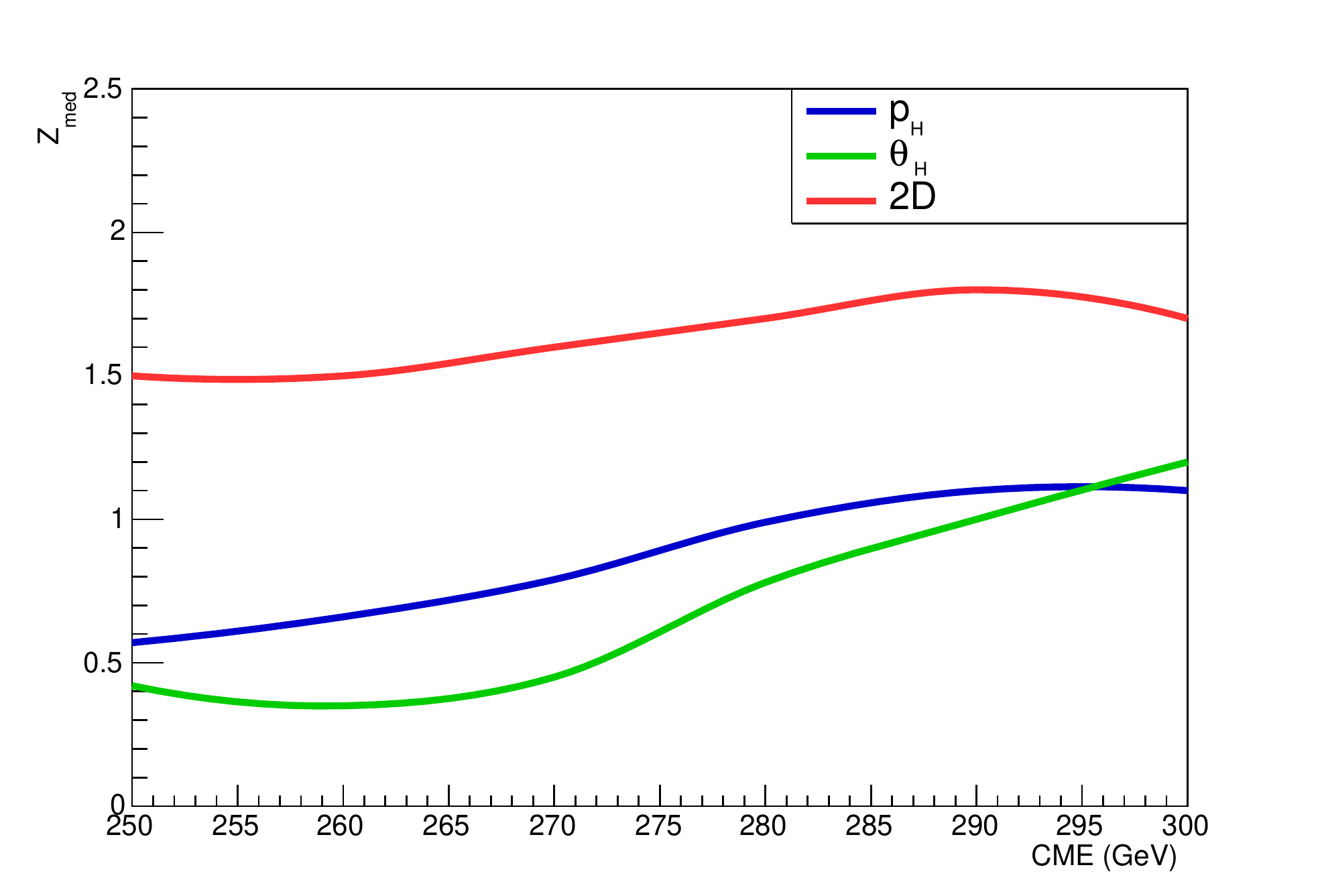}} &&
\resizebox{70mm}{!}{\includegraphics{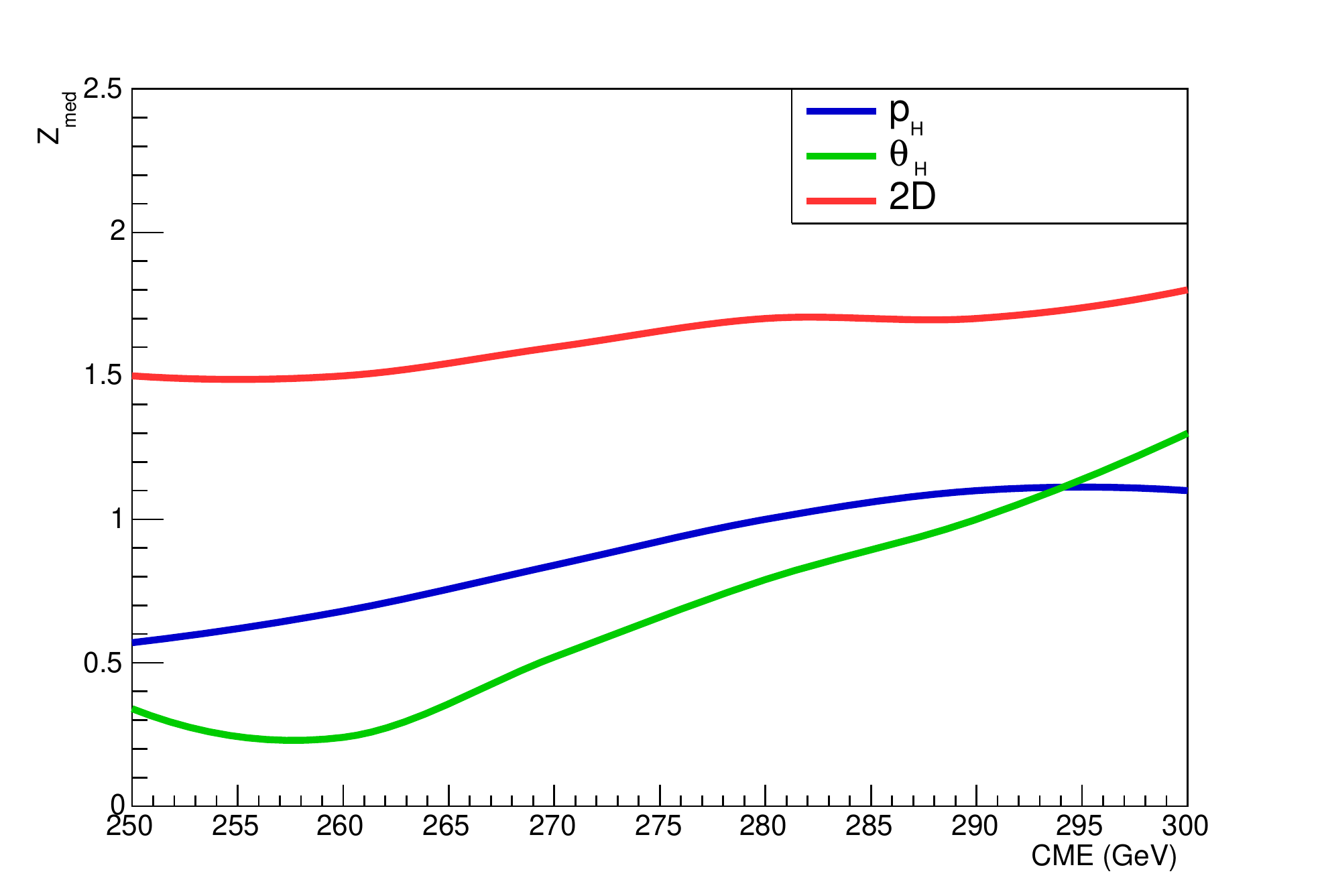}}\\
\hspace{0mm}(c)&&\hspace{4mm}(d)
\end{tabular}}
\caption{Median significance values for likelihood analyses done with
  both one dimensional and two dimensional distributions. (a) SM with
  $\lambda=1$, (b) SM with $\lambda=-1$, (c) SM with $\lambda'=1$ and
  (d) SM with $\lambda'=-1$. Results are obtained with 1\,fb$^{-1}$ of integrated luminosity.}
\label{img:likelihoods}
\end{figure}

A likelihood analysis for each BSM hypothesis is performed for
integrated luminosities of 1 fb$^{-1}$, 5 fb$^{-1}$ and 10
fb$^{-1}$. The number of pseudo-data points in each analysis is
determined from the SM cross section. The $Z_{med}$ for the 1
fb$^{-1}$ case are plotted as functions of the CME for each hypothesis
as shown in Fig.~\ref{img:likelihoods}. These plots show the power of
using two dimensional distributions in likelihood analysis. The
likelihood analysis is performed using a total number of 100,000
pseudo-experiments for each TS. The two dimensional distributions,
examples of which are shown in Fig.~\ref{img:t-2D}, are also included
in the likelihood analysis to demonstrate the effect of the
correlation between the two variables, $p_H$ and $\theta_H$.

Fig.~\ref{img:likelihoods} displays the significance for
one-dimensional analyses using the Higgs boson momentum and the polar
angle separately. Results are shown for illustration purposes for 1\,fb$^{-1}$ of integrated luminosity. Conclusions drawn here  are found not to depend on the integrated luminosity in the range studied here. The corresponding results for the combined 2D
likelihood are shown. The upper two plots correspond to admixtures
with the CP-even term. The sensitivity of the polar angle is
significantly less than that of the Higgs boson momentum. The lower
plots display the corresponding results for admixtures with the CP-odd
term. In this case the sensitivity of the polar angle is similar to
that of the momentum. As a result, the improvement from the 2D
analysis is significant, to the extent that the sensitivity can be enhanced by about a factor of two. The sensitivity of the angular variable grows with the CME.

The results provide a good motivation for the role of an electron
positron collider in understanding the nature of the $HVV$
couplings. The plots in Fig.~\ref{img:likelihoods} show the utility in
using two dimensional distributions in discerning the rejection of
hypotheses. That is, using the same accrued data from two separate one
dimensional distributions, one can enhance the confidence in rejecting
hypotheses. The correlation of the two dimensional distributions thus
carries vital information about the dynamics of the processes which
are studied in $e^+e^-$ collisions.

\section{\label{sec:disc}Summary and Conclusions}

We have attempted to demonstrate the efficacy as well as limitations
of an $e^+ e^-$ Higgs factory operating at $250 - 300$ GeV in probing
anomalous, higher-dimensional couplings of a Higgs to $W$-and
$Z$-pairs, suppressed by a scale $\mathcal{O}$(TeV). For this purpose,
we have mostly adhered to the set of gauge-invariant operators that
can lead to such interactions, since it is such terms that are
expected to emerge on integrating out physics above the electroweak
symmetry breaking scale.  We have utilised the consequent correlation
of the anomalous $HWW$, $HZZ$ and $HZ\gamma$ couplings, and also the
concomitant effect on $ZWW/\gamma WW$ interactions, as reflected in
gauge boson pair-production rates.

The general conclusion reached by this study is that the total rates
can be quite useful as probes of higher-dimensional operators.  Based
on this, we have performed a detailed analysis of the cross-sections
for $s$-and $t$-channel Higgs production, specifying event selection
criteria for minimising their mutual contamination. A general scheme
of computing the rates with more than one gauge-invariant operators
has been outlined.  Based on such an analysis, we conclude that, even
with the additional operators well within the erstwhile experimental
bounds (including those form the LHC), a number of observations can
probe them at a Higgs factory. These include not only the individual
total cross-sections but also their ratios at different values of
$\sqrt{s}$ and also the ratio of the $s$-and the $t$-channel Higgs
production rates at fixed energies. We also indicate the correlated
variation of $W$-pair production rates. The Higgs production rate
contours with more than one type of anomalous gauge-invariant
operators are also presented. Finally, using some illustrative values
of anomalous $HWW$ couplings in a more phenomenological
parametrization, we indicate the viability of a correlated two-dimensional likelihood analysis to fully exploit the kinematics of the Higgs boson. The latter is particularly relevant to disentangle the SM from CP-violating admixtures. On the whole, we thus conclude that a Higgs
factory can considerably improve our understanding of whether the
recently discovered scalar is the SM Higgs or not, as evinced from its
interactions with a pair of weak gauge bosons.

\section*{Acknowledgements}
  We thank Taushif Ahmed, Satyanarayan Mukhopadhyay and Narayan Rana for 
  helpful discussions. The work done by SvB was supported  by The Claude Leon
  Foundation. The work of S.B., T.M. and B. Mukhopadhyaya was partially 
  supported by funding available from the Department of Atomic Energy, Government
  of India for the Regional Centre for Accelerator-based Particle
  Physics (RECAPP), Harish-Chandra Research Institute. B. Mellado acknowledges
  the hospitality of RECAPP, Harish-Chandra Research Institute, during
  the collaboration.


\end{document}